\documentclass{article}
\usepackage{graphicx} 
\usepackage{color, amsmath}
\usepackage{mathrsfs}
\usepackage{appendix}
\usepackage{geometry}
\usepackage{comment}
\usepackage{url}
\usepackage{hyperref}
\usepackage{multirow}
\usepackage{multicol}
\usepackage{rotating}
\usepackage{pdflscape}
\usepackage{longtable}
\usepackage{array}
\usepackage{adjustbox}
\usepackage{afterpage}
\usepackage{float}
\usepackage{lineno}
\usepackage{tabularx}
\usepackage{cancel}
\usepackage{soul}
\usepackage{ulem}
\usepackage{diagbox}
\usepackage{tablefootnote}

\DeclareMathOperator*{\argmax}{arg\,max}

\def \sp {{\sigma^+}}
\def \sm {{\sigma^-}}
\def \spi {{\sigma^+_i}}
\def \smi {{\sigma^-_i}}
\def \AE #1 #2 #3 {{#1}^{+#2}_ {-#3}}
\def \EA #1 #2 {{}^{+#1}_ {-#2}}
\def \half {{1 \over 2}}

\def \N {{\cal N}}

\long\def \change #1 #2 {#1}

\def \boxit#1{ \centerline{\hbox {\hsize 12 cm \vrule \vbox{
\hrule  \vskip 2 pt \hbox{\  \vbox{#1} }\vskip 2 pt \hrule}\vrule  }}}

\newcommand{\sub}[1]{\ensuremath{_{\mbox{\scriptsize \,#1}}}}
\newcommand{\supers}[1]{\ensuremath{^{\mbox{\scriptsize #1}}}}
\newcolumntype{P}[1]{>{\centering\arraybackslash}p{#1}}

\title{Asymmetric Errors}

\author{
Roger Barlow \\ The University of Huddersfield \\Huddersfield, UK \\ {\tt Roger.Barlow@cern.ch}  
\and 
Alessandra Rosalba Brazzale \\ University of Padova \\ Padova, Italy
\and 
Igor Volobouev \\ Texas Tech University \\ Lubbock, Texas, USA 
}

\date{\today}

\begin{document}

\maketitle

\begin{abstract}
We present a procedure for handling asymmetric errors.  Many results in particle physics are presented as values with different positive and negative errors, and there is no consistent procedure for handling them.  We consider the difference between errors quoted, using pdfs and using likelihoods, and the difference between the rms spread of a measurement and the 68\% central confidence region.  We provide a comprehensive analysis of the possibilities, and software tools to enable their use.
\end{abstract}


\section{Introduction}

\subsection{Background and motivation}
\label{sec:intro}

Results in particle physics are often given with errors which are asymmetric, of the form \change {$\AE R {\sp} {\sm} $} {$\AE x {\sp} {\sm} $}.  For instance, taking some results from the latest (at the time of writing) EPS conference:
\begin{itemize}
    \item ATLAS quote their measurement of the Higgs width as $\Gamma_H=\AE 4.5 3.3 2.5 $ MeV \cite{ATLAShiggs}.
    \item CMS quote the same quantity as $\Gamma_H=\AE 3.2 2.4 1.7 $ MeV \cite{CMShiggs}.
    \item NOvA have measured the neutrino CP violating parameter as $\delta_{CP}= \AE 0.82 0.27 0.87 \,\pi$ \cite{NOvA}.
    \item Belle II quote the branching ratio for $B \to \rho^+ \rho^0$ as $(\AE 23.2 2.2 2.1 (stat.) \pm 2.7 (sys.)) \times 10^{-6}$ \cite{Belle}.
    \item LHCb give the difference in the decay width for the neutral $B$ mesons as $\Gamma_s-\Gamma_d= \linebreak \AE -0.0056 0.0013 0.0015 (stat.) \pm 0.0014 (sys.) \,{\rm ps}^{-1}$ \cite{LHCb}.
\end{itemize}
Despite their widespread use, their interpretation and their handling is unclear and discussion in the literature is limited 
\cite{Schmelling,Systematic,Statistical,dagostini,Possolo,Demortier}.
This paper explores the reasons why they appear, and gives methods for their consistent handling. It addresses the basic question of how to use results presented as $ \AE R {\sp} {\sm} $, in the absence of any further information.

\begin{figure}[t]
    \centerline{\includegraphics[width=0.9\textwidth]{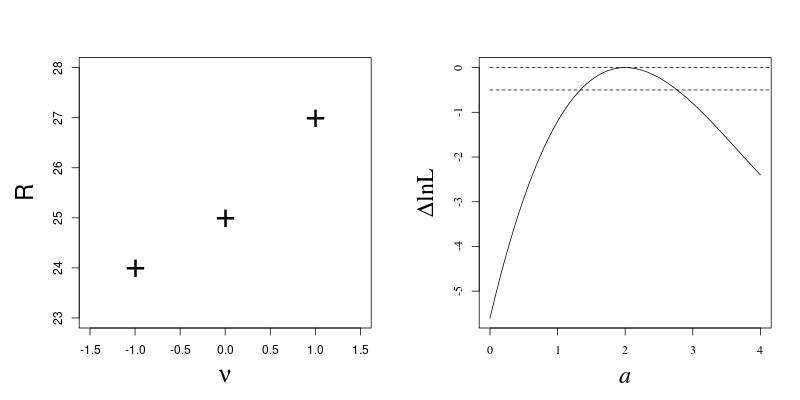}}
    \caption{\label{fig:typical}
        The two classes of asymmetric error, from OPAT systematic studies of changes to a result $R$ in response to a nuisance parameter $\nu$ (left) and maximum likelihood estimation for a parameter $a$ (right).}
\end{figure}

In practice such errors spring from three causes:
\begin{enumerate}
    \item when systematic errors are being studied, often using the OPAT (``One Parameter At a Time'') method when profiling out the nuisance parameters in the full likelihood is not feasible. In evaluating a systematic error due to the uncertainty (which can be Bayesian or frequentist) of a nuisance parameter $\nu$  one can change it by $\pm \sigma$ to observe the shift this produces in a result, and, as illustrated in Figure~\ref{fig:typical}, the upward and downward shifts are different.  Alternatively, when it is technically possible, many values can be generated according to a Gaussian distribution, to find and quantify the distribution of the results, and this may not be symmetric.
    \item from maximum likelihood estimation with errors given by the $\Delta \ln L=-\half$ method, and the log likelihood is not a symmetric parabola. This is also shown in Figure~\ref{fig:typical}.  (The $\Delta \ln L=-\half$ method is 
    an approximation, but its use is widespread, essentially universal, and in what follows this is what is meant by `likelihood errors'.) 
    \item when a random variable is distributed according to a non-Gaussian distribution, such as a Poisson distribution with small mean.  This includes as an important special case the replacement of a Gaussian random variable by another variable according to a non-trivial transformation, as happens on a plot with a logarithmic scale.
\end{enumerate}
The third case does not hold conceptual problems as one has, in principle, full knowledge of the underlying distribution function.  The challenges come with the first and second cases, where one has to handle the quoted value and errors with no further information about how they arose.  This prompts us to consider more closely some fundamental facts about the nature of `errors' which are normally hidden from view by the convenient properties of the Gaussian function.

\vspace{2ex}
\boxit{For the avoidance of confusion we point out that 
in this paper $\sm$ is 
positive, like $\sp$. And if $R$ falls with $\nu$ 
(rather than rising as shown in Figure~\ref{fig:typical}), then $\sp$ and $\sm$ are still the magnitudes of the positive and negative 
deviations in $R$, not in $\nu$. If (as occasionally happens)  both deviations are in the same direction then special care is needed, as described in the section on `flipped' distributions, Section~\ref{sec:flipped}.} 

\vspace{1ex}

\subsection{Beyond the Gaussian}
\label{sec:beyondgaussian}

The Gaussian (``Normal'') distribution is described by the density function
%
%
\begin{equation}
    \phi(x;a)={1 \over \sigma \sqrt{2 \pi}} e^{-\half{(x-a)^2 \over \sigma^2}}
    \label{eq:gauss}
\end{equation}
for a measurement $x$ of some value $a$.  This is what is encoded in any reported result of the form $\hat a=x\pm \sigma$.  When a result is reported as $\hat a= \AE x {\sp}  {\sm} $ then this is making a clear analogy with the Gaussian form.  But this similarity conceals two important questions which cannot be evaded.

The first is whether we are talking about a function which is asymmetric in $x$ or in $a$. 
Equation~\eqref{eq:gauss} is, of course, symmetric in both.  Sometimes we consider it as a function of $x$ for a given $a$, in which case it must be normalised to 1 and we call it a pdf (probability density function).  Sometimes we consider it as a function of $a$ for fixed data $x$, in which case we call it a likelihood and normalisation is irrelevant.  So the first question to consider in any discussion of `asymmetric errors' is: what is asymmetric, the pdf or the likelihood?

The second question is the meaning of 
the error.  Specifically, are we talking about the rms spread of the distribution, or 
the 68\% central confidence interval? These happen to be the same for a Gaussian distribution, but not in general.
 In Equation~\eqref{eq:gauss} the parameter $\sigma$ describes both the rms spread of the distribution about the mean, $\sigma=\sqrt{\left< x^2\right>-\left<x\right>^2}$, and the 68\% central confidence region for $a$: $[x-\sigma,x+\sigma]$, and we do not need to be concerned with the difference.  But for non-Gaussian distributions we have to know which definition of $\sigma$ we are using.

\begin{figure}[h]
\centerline{
\includegraphics[width=10 cm]{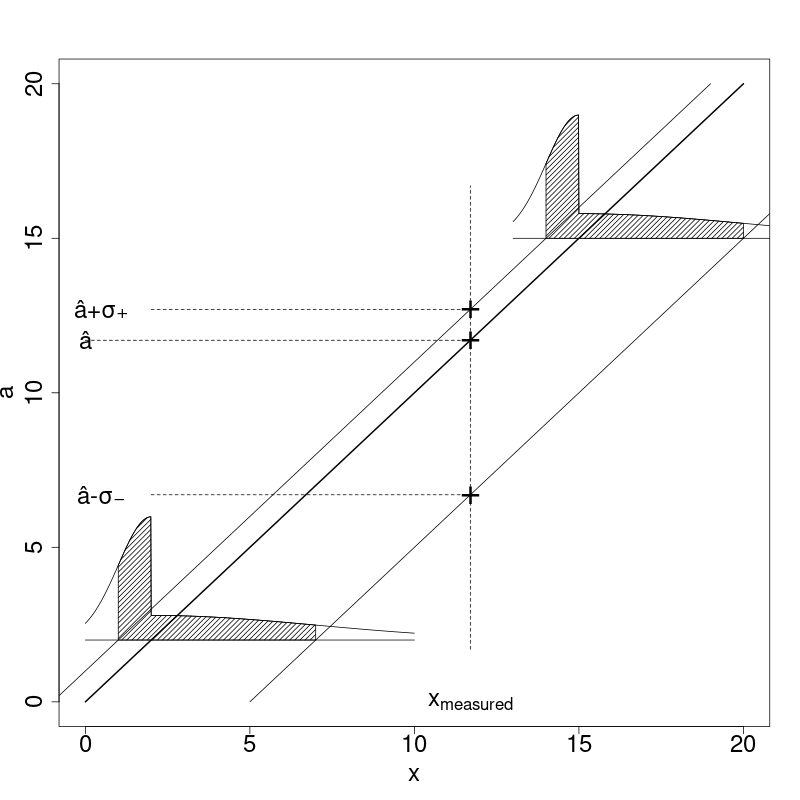}
}
\caption{\label{fig:confidencebelt} The Neyman confidence belt referred to in the text, showing how a positive pdf asymmetry $\AE 11.7 5.0 1.0 $  matches a negative likelihood asymmetry $\AE 11.7 1.0 5.0 $. The shaded areas display 68\% confidence regions of the pdf. This is an artificial example to make the point clear:  real life is more complicated. }
\end{figure}

The difference between rms spread `pdf errors' and confidence interval  `likelihood errors' requires clarification.  They both refer to uncertainties, the former from random probability and the latter from statistical inference. The Poisson process provides a simple illustration (though for a discrete probability distribution rather than a continuous pdf): the former describes the uncertainty of predicting results from a particular true mean, the latter the uncertainty of inferring the true value of the mean from a particular result.  
\\ \indent 
Pdf errors arise especially through measurements and their errors in experimental work. 
At high school and university the physicist is firmly taught never to quote an experimental result as a single number $x$, but {\it always} to give its error $\sigma$. That $\sigma$  is the standard deviation of the Gaussian distribution, as given in \eqref{eq:gauss}, from which this value is a sample.  If, for better accuracy, the mean of $N$ measurements is 
quoted then it is accompanied by the standard error $\sigma/\sqrt N$ which is the standard deviation of the pdf of $\overline x$.
The well-known formula for the combination of errors, given below as Equation~\ref{eq:combinationoferrors}, known to statisticians as the delta method, builds on this. It describes the combination of measurement errors: if $I$ is sampled from a Gaussian with standard deviation $\sigma_I$, and $V$ with $\sigma_V$, then the resistance measurement $R=V/I$ is sampled from a Gaussian for which (assuming the function can legitimately be linearised) $\sigma_R^2=\sigma_V^2/I^2 + V^2 \sigma_I^2 / I^4$. The quoted result, $V/I \pm \sigma_R$, is saying the value has been drawn from a Gaussian distribution of unknown mean and standard deviation $\sigma_R$. 
\\ \indent 
Such measurements can be converted to confidence intervals using the Neyman confidence belt, as described in Section 40 of the Particle Data Group handbook \cite{PDGhandbook}. For Gaussians this is trivial. 
In frequentist usage, to say ``I have measured $a$ to be $x\pm 3$." means ``I have a measurement $x$ from a device which returns values distributed about the true value $a$ according to a Gaussian distribution with a standard deviation of 3: I therefore assert that $x$ falls within the interval $[a-3, a+3]$ with probability 68\%, which in the Gaussian case can be turned into saying that, given $x$, the interval $[x-3, x+3]$ covers the true value $a$ with 68\% confidence.  (It does not mean ``The true value of $a$ lies within within $x-3$ and $x+3$ with 68\% probability.", that would be a Bayesian credible interval, though this subtle distinction may not always be appreciated.)
\\ \indent
If the pdf is not Gaussian --- and for asymmetric errors this must be the case --- then this trivial equivalence breaks. `Pdf errors' describe the standard deviation and `likelihood errors' the 68\% central confidence interval.
To take an artificially exaggerated but hopefully illuminating example, suppose
a measuring device has a 50\% chance of recording a measurement below the true value and does so according to a half-Gaussian with $\sigma=1$, and a 50\% chance of recording a measurement above the true value, according to a half-Gaussian with $\sigma=5$. (This is the dimidiated Gaussian, described later.) If it measures $x=11.7$ this would be written as $\AE 11.7 5.0 1.0 $, to describe the pdf. 
However the Neyman construction for the confidence belt would be $\AE 11.7 1.0 5.0 $: if 11.7 is a downward fluctuation it cannot have come very far from the true value, whereas an upward fluctuation may have been produced by a true value one distance away. This is shown in Figure~\ref{fig:confidencebelt}.
\\ \indent
It is tempting to suggest that any result presented with an asymmetric error should denote a confidence-level, and there is a case for doing so for final results.
However in an analysis there will be intermediate values used to construct a final result using addition-in-quadrature (with the aid of the combination-of-errors formula), and this is done in the framework of pdfs and their moments. Variances add under convolution, but 68\% central regions do not, requiring more complicated meta-analysis procedures.
In a typical particle physics analysis, there are many intermediate steps between the raw data and the final quoted figure, and typically for these the errors quoted are properties of the pdf.


If we are considering a pdf $p(x;a)$, then we can integrate it to obtain the spread (as given by the variance) of the measured $x$ for a given value of $a$ and the `errors' are expression of that spread.  It tells us nothing about $a$, as that is given.  If we are dealing with a likelihood function $L(a; x)$  we can get the `best' estimate $\hat a$ and take the 68\% CL central region from the $\Delta \ln L=-\half$ points,\footnote{This is only an approximation to the Neyman construction, but this point is generally not considered. For a discussion see Reference~\cite{deltalnL}.   } but we can say nothing more about $x$. 
It is true that a pdf can give 68\% central limits, but in the language of the Neyman confidence belt
they are the horizontal lines constructed on the plot shown in Figure~\ref{fig:confidencebelt}, 
not the vertical ones from which the confidence interval for a result is read off. So an `asymmetric error' may refer to the variance and skewness of a pdf, or to the confidence region for a likelihood.  Both cases arise in practice.  We will therefore consider asymmetric pdfs and the variances (and higher moments) of $x$, and also, separately,  asymmetric likelihoods and the confidence regions for $a$.

\boxit{
    Usage varies in the literature, but within this paper we use the variable names $x$ and sometimes $y$ to refer to results of measurements, and $a$ to refer to an ideal `true' value. Thus pdfs are considered as functions of $x$ and likelihoods as functions of $a$. Alternative names may be used in particular examples, as when $R$ and $\nu$ (for `Result' and `nuisance') are used in the OPAT plot of Figure~\ref{fig:typical} rather than $\hat a$ and $x$.
}

One then has to consider how the `errors' are going to be used.  Again there are two possibilities: Combination of Errors and Combination of Results. \change {} { This time the distinction is not so clear-cut. }

\subsubsection {Combination of Errors}

The simple formula for  combination of errors is how most physicists first meet statistics, in the context of experimental errors and uncertainties.  For some function $u=u(x,y)$ where \change { the errors on } {} $x$ and $y$ are statistically independent \change {and} {with errors that are } small (so that a linear approximation is valid), 
\begin{equation}
    \sigma_u^2=\left( {\partial u \over \partial x}\right)^2 \sigma_x^2 + \left( {\partial u \over \partial y}\right)^2 \sigma_y^2\ . 
    \label{eq:combinationoferrors}
\end{equation}
We need to extend it to cope with $\sigma^+_x,\sigma^-_x, \sigma^+_y, \sigma^-_y$ and thus $\sigma^+_u, \sigma^-_u$.

In introductory texts this formula is used in problems like determining the speed from the distance travelled in a given time using $v=x/t$, or the acceleration due to gravity from the length and period of a simple pendulum using $g=4\pi^2 L/T^2$.  In more advanced experimental work it is used in instances like calculating a branching ratio, $ Br={N-b \over \eta N_T}$, from some number $N$ of observed decays
into the channel of interest out of a total of $N_T$ decays, with a background $b$ and an efficiency $\eta$, or in fitting a straight line (or some more sophisticated function) to a set of values $\{x_i,  y_i \pm \sigma_i\}$.
 
\begin{figure}[t]
    \centerline{\includegraphics[width=0.9\textwidth]{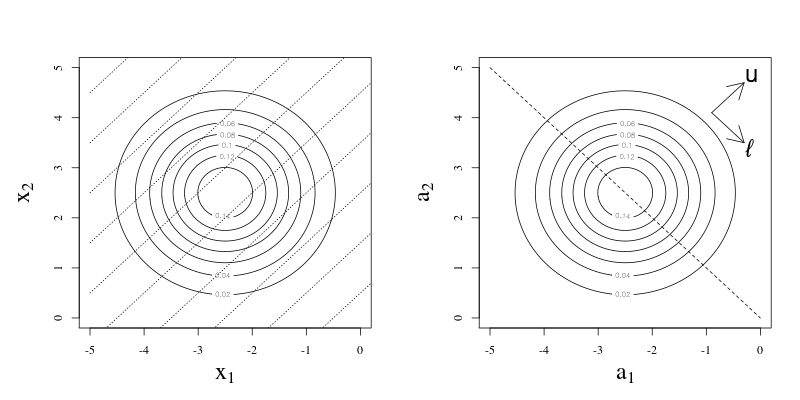}}
    \caption{\label{fig:coe}
        Combination of errors: $\ell =(2.5 \pm 1.0)- (-2.5\pm 1.0)$ using pdfs (left) and likelihoods (right). }
\end{figure}

As a very basic pedagogical example we consider the measurement of the length $\ell$ of a rod by measuring the positions $x_1$ and $x_2$ of the two ends, with $\ell=x_2-x_1$.  Working with pdfs, we have a two dimensional pdf $p(x_1,x_2)$.  The pdf for $\ell$ is given by the convolution of the two individual pdfs
\begin{equation}
    p(\ell)=
    \int \int p_1(x_1) p_2(x_2) \delta(\ell-x_2+x_1) \, dx_1 dx_2
    =\int p_1(x_1) p_2(\ell+x_1)\, dx_1 ,
    \label{eq:simpleconvolution}
\end{equation}
{where the two random variables $X_1$ and $X_2$ are assumed independent}. 
If the pdfs are Gaussian, as shown in the left hand plot of Figure~\ref{fig:coe}, where the dotted lines are the lines of constant $x_2-x_1$, then the integral 
gives a Gaussian for $\ell$.  This Gaussian has standard deviation $\sqrt{\sigma_1^2+\sigma_2^2}$, as variances add. This
is what is expressed by  Equation~\eqref{eq:combinationoferrors}.

In a likelihood framework, we suppose that $x_1$ and $x_2$ are measurements of (unknown) true values $a_1$ and $a_2$ and consider the joint likelihood, as shown in the right hand plot of Figure~\ref{fig:coe}.  For the (symmetric) Gaussian this is the same as the left hand plot, apart from the labels on the axes, but  in general it will be different.  This two-dimensional plot is reduced to one dimension by profiling: writing  $\ell=a_2-a_1$ and $u=a_2+a_1$, for each value of $\ell$ we find the maximum value of $L_1(a_1)L_2(a_2)$, which lies along the dashed line,
$$
    \hat{\hat u}={\ell (\sigma_2^2 - \sigma_1^2) + 2(\hat x_2 \sigma_2^2 + \hat x_1 \sigma_2^2) \over \sigma_1^2 + \sigma_2^2},
$$ 
and the resulting profile likelihood for $\ell$ is just 
$$
    -\half{\left(\ell - (\hat x_2 - \hat x_1)\right)^2 \over \sigma_1^2 + \sigma_2^2}.
$$
From this we read off the peak at $\ell=x_2-x_1$, and the $\Delta \ln L=-\half$ errors at $\pm\sqrt{\sigma_1^2+\sigma_2^2}$.  This is exactly the same result, for the best value and the `error', as obtained from the combination of errors formula.  So when dealing with Gaussians, the same combination of errors formula is found using pdfs (and convolution) or using likelihoods (and profiling), even though the meaning is subtly different in the two cases, {because in the first the resulting error is the root of a variance which happens to span a 68\% probability interval, and in the second it is the half-width of a 68\% confidence interval}. 
For non-Gaussians we have to consider them separately. 

Combination of errors is usually a matter for pdfs, because variances add, even for non-Gaussian pdfs.  This is dealt with in Section~\ref{subsec:PCOE}.  A typical analysis will consider many sources of error, usually labelled as systematic, and the combination of these is a major concern of this paper.  In such cases the Central Limit Theorem can be useful as when a large number of uncertainties is combined the overall distribution may be adequately described by a Gaussian, even though several of the contributions are asymmetric.  However, there are instances where combination of errors needs to be done using  likelihoods, and this is considered in Section~\ref{subsec:LCOE}.

\subsubsection {Combination of Results}

The second use we need to consider is the combination of results, also known as  meta-analysis.  Given a set of measurements of the same quantity one wishes to find the appropriate combined best value and its error --- and usually one would also want some goodness of fit statistic to describe whether the different results are compatible.  This can readily arise when the results are presented in the form of likelihoods (or the maximum likelihood estimator and the 68\% central interval obtained from the likelihood function).  It is conceptually simple to form the complete likelihood as the product of the individual ones, and find the location of the maximum and the 68\% confidence band; there are technical challenges, and Section~\ref{sec:likelihood} deals with those. And this is the usual form of the problem.  If these are all simple Gaussian measurements this requires the maximisation of $-\half \sum_i \left({a_i-\hat a \over \sigma_i}\right)^2$, which occurs at $\hat a={\sum a_i/\sigma_i^2 \over \sum 1/\sigma_i^2}$.  (The index $i$ runs over the results being combined; each result may involve many data points but that is not relevant here.)

But it is also possible that one has a set of results presented as pdfs: the $\hat a_i$ are estimators of the true $a$, they are functions of the data $x_i$ and are therefore described by pdfs.  If they are unbiased then any weighted sum $\sum_i w_i \hat a_i$ is an unbiased estimate \change {provided $\sum_i w_i=1$} {} , and one has the freedom to choose the $w_i$ to minimise the variance on this, using the combination of errors formula, Equation~\eqref{eq:combinationoferrors}.  For Gaussian errors this gives the well known prescription $w_i = {1 /\sigma_i^2 \over \sum_j 1/\sigma_j^2}$:
the weight is proportional to the inverse square of the error.  Hence for Gaussian errors the combination results using pdfs and using likelihoods gives the same answer, even though the meaning of the answer is subtly different. Again, for non-Gaussian errors this is not the case and we must consider both separately.

\subsection {Summary on combination-of-errors and combination-of-results}

The distinction between combination of errors and combination of results
turns out to require clarification, perhaps because both are considered as `the error analysis'.
The fundamental difference is that, in the first case, we want to quantify the precision of a combination, which involves different quantities, while in the second case, we compute the precision of an updated measurement of the same quantity, which we obtain by combining different measurements of the same quantity. 
 In the first case, precision goes down, because uncertainty adds.  In the second case, precision goes up because we measure something with more data.  (In statistics, the same name is kept for the precision, but different names are used for the outcome: measurement in the first case, estimate in the second.)
 \\ \indent 
 Combination-of-errors fits naturally into the framework of pdf errors, though 
 combination of results is perfectly possible where the result is a weighted sum of measured values. Conversely combination-of-results emerges very naturally from likelihoods, although combination of errors is perfectly possible through profiling.
 \\ \indent
 For clarification, the distinctions are shown in Table~\ref{tab:clarity}.
\begin{table}[p]
{\small
\begin{tabular}{ | p{0.18\linewidth} |
                   p{0.32\linewidth} |
                   p{0.28\linewidth} |
                   p{0.30\linewidth} | }
\hline
\diagbox[width=9.8em]{Combining\\[-2ex]~}{~\\[-2ex]Through}
      & {\bf PDF} (Section~2)
      & {\bf LIKELIHOOD} (Section~3) 
      & Notes \\
\hline
&&& \\[-2ex]
{\bf ERROR} \newline
(Sections~2.2 \newline and 3.4) \newline~\newline
Want to quantify variability of a measurement obtained by combining measurements of {\bf different} quantities. 
      & $X \sim p_x(x; a_x, \sigma_x^+, \sigma_x^-)$ \newline
        $Y \sim p_y(y; a_y, \sigma_y^+, \sigma_y^-)$ \newline
        ~\newline
        Parameter of interest: $\psi = a_x/a_y$, fitted by $\hat\psi = x/y$ (for example). \newline
        ~\newline
        Uncertainty of $\hat\psi$ is quantified through \newline
        $U = X/Y \sim p_u(u; \psi, \sigma_u^+, \sigma_u^-).$ \newline
        ~\newline
        ~\newline~\newline
        Bayesian view: \newline
        ~\newline
        $A_x \sim p_1(a_x \mid x, \sigma_x^+, \sigma_x^-)$ \newline
        $A_y \sim p_2(a_y \mid y, \sigma_y^+, \sigma_y^-)$ \newline
        ~\newline
        combined to yield \newline~\newline
        $U=A_x/A_y \sim p_u(u\mid x, y, \sigma_u^+, \sigma_u^-)$ 
      & $L(a_1 ; x) \quad \text{and} \quad L(a_2 ; y)$ \newline
        $ \Downarrow \newline
           L(a_1, a_2; x, y) = L(\psi, \lambda; x, y)$, \newline
        ~\newline   
        with $\psi = a_1/a_2$ parameter of interest (for example) and $\lambda$ any function of $(a_1,a_2)$ such that $(\psi,\lambda)$ is a reparametrization of $(a_1, a_2)$    
      & $\bullet$ The parameter of interest $\psi$ is a combination of other model parameters (say, $a_1$ and $a_2$). \newline
        $\bullet$ $n$ represents the number of random variables (here, $X$ and $Y$) which provide information (here on $a_1$ and $a_2$).  These are then combined to provide information on $\psi$. \newline
        $\bullet$ Uncertainty goes up! \newline
        $\bullet$ The procedure is usually based on PDFs (where $\sigma$ represents an rms error). \newline
        $\bullet$ If based on LIKELIHOODS ($\sigma$ then typically represents the spread of an 68\% CI), this requires profiling out the nuisance parameters $\lambda$ which complete the reparametrization from $(a_1,a_2)$ to $(\psi,\lambda)$. 
      \\
&&& \\[-2ex]
\hline
&&& \\[-2ex]
{\bf RESULT} \newline
(Sections~2.4\newline and 3.2) \newline~\newline
Want to quantify confidence of measurement obtained by combining measurements of {\bf the same} quantity. 
       & Weighted average (meta-analysis)
         $\hat a =\sum_i w_i \hat a_i$,
         with $\sum w_i=1$
       & Combine \newline
         ~\newline
         $\AE {\hat a(x_1)} {\sigma^+(x_1)} {\sigma^-(x_1)} $ \ from $L(a ; x_1)$ \newline 
         $\AE {\hat a(x_2)} {\sigma^+(x_2)} {\sigma^-(x_2)} $  \ from $L(a ; x_2) $ \newline   
         ~\newline 
         to yield\newline
         ~\newline
         $\AE {\hat a(x)} {\sigma^+(x)} {\sigma^-(x)} $ \ from $L(a ; x),$ \newline
         ~\newline
         with  $x=(x_1,x_2)$.
       & $\bullet$ There is an unique parameter of interest $a$. \newline
         $\bullet$ $n$ is the number of observations available on $a$. \newline
         $\bullet$ Uncertainty goes down! \newline
         $\bullet$ The procedure is usually based on LIKELIHOODS (where $\sigma$ represents the spread of an 68\% CI). \newline
         $\bullet$ It can be based on PDFs (where $\sigma$ represents an rms error) using meta-analysis. \newline 
         $\bullet$ Check for goodness-of-fit!
       \\
&&& \\[-2ex]
\hline
&&& \\[-2ex]
Notes  & $\bullet$ $\sigma$ measures {\bf variability} (e.g.\ precision of measurement device). \newline
         $\bullet$ Approximation needs be normalized (to yield a PDF).
       & $\bullet$ $\sigma$ measures the {\bf confidence } region (e.g. as expressed by Neyman confidence belt). \newline
         $\bullet$ Approximation needs be ``nearly'' parabolic (to work well).
       & 
       \\[15ex]
\hline
\end{tabular}
}\caption{\label{tab:clarity} Combination of errors and results using pdf errors and likelihood errors. }
\end{table}


\section{Pdf errors}

\subsection{Modelling pdf errors}

The question as to how a model function, whether considered as a pdf or as a likelihood, should best be written in a non-Gaussian way is a very open one. In this section we consider pdfs:  likelihoods will be dealt with in Section~\ref{sec:likelihood}.  If we restrict ourselves to distributions which look similar to a Gaussian, we can expect them to require an additional third parameter, expressing the asymmetry in some way: we do not consider extensions to more than 3 parameters, but the formalism is open should they be needed.


The three parameters may be specified in various ways, we may use:
\begin{enumerate}
    \item the moments, namely the mean $\mu= \left<x\right>$, the variance $V=\left<x^2\right>-\left< x \right> ^2$, and the unnormalised skewness, $\gamma=\left<x^3 \right> - 3 \left< x^2 \right> \left< x \right> +2 \left< x \right> ^3$.
    \item the quantiles. The median $M$ is the middle quantile $x_{0.50}$, and the 68\% (one sigma) central confidence probability region $[x_{0.16},x_{0.84}]$ can be written as $[M-\sm,M+\sp]$. This can be expressed in the form $\AE M {\sp} {\sm} $ and will be referred to as ``the quantile parameters" in what follows.
    (The region is actually bounded by the values of the cumulative Gaussian $\Phi(-1)$ and $\Phi(+1)$: though here and elsewhere only 2 significant figures are shown, the exact 1 sigma region is always meant.)   
    
    \item parameters appropriate to the particular model. 
\end{enumerate}

It might be thought that the mean could be used as an alternative to the median \label{sec:problems}
in combination with the bounds of the 68\% central confidence region, writing it as $[x_{0.16},x_{0.84}]=[\mu-\sm,\mu+\sp]$.  However, this parameterisation has an unhelpful behaviour in some models, as will be noted in Sections~\ref{sec:edgeworth} and \ref{sec:azzalini}, and it will not be pursued further. 

Any implementation must provide functions to convert any of the 3 parameterisations into any of the others. The algebra for doing so is given in Sections~\ref{sec:dimidiated} to \ref{sec:lognormal}, and the software in the appendices~\ref{sec:Cpp} and~\ref{sec:R}.

Two such models are readily obtained by considering the OPAT plot of Figure~\ref{fig:typical} and supposing that the dependence of $R$ on $\nu$ be described by two straight lines, or by a quadratic.  Under the first assumption, the Gaussian in $\nu$ becomes a dimidiated Gaussian: two equal-area half-Gaussians, under the second a somewhat distorted Gaussian.  These models have the advantage of being clearly motivated and giving simple algebraic relations between parameters (details are in Sections~\ref{sec:dimidiated} and \ref{sec:distorted}). 

\begin{figure}
    \centerline{\includegraphics[width=10 cm]{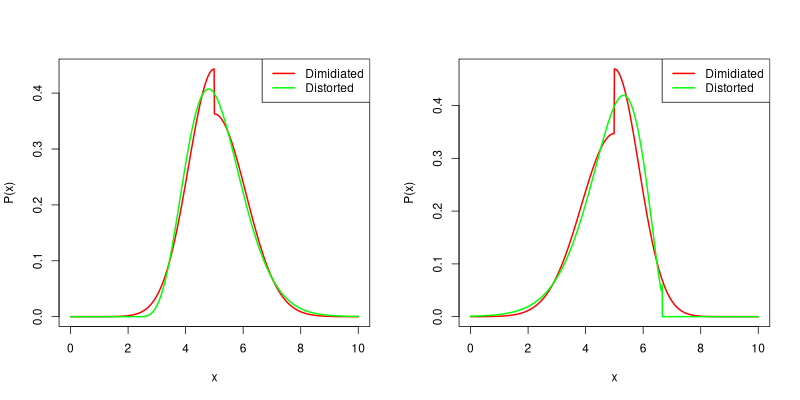}}
    \caption{\label{fig:simplepdfs}
        The dimidiated and distorted Gaussians for $\AE 5.00 1.0 0.9 $ and $\AE 5.00 0.85 1.15 $.}
\end{figure}

Typical examples are shown in Figure~\ref{fig:simplepdfs} for two cases, one with a small positive skewness and one with a slightly larger negative skewness.  The two shapes are broadly similar but differ in detail, as one would expect.  The dimidiated form has a discontinuity at the central value, but is well behaved elsewhere. The distorted form appears better behaved at the centre, but the turnover of the parabola can give a cutoff, as appears on the positive side of the second plot.  

These forms are not `correct': there can be no such guarantee.  They give reasonable answers in most practical cases, and we have introduced them here to aid the discussion of ideas.  Other models will be described in Section~\ref{sec:allthepdfmodels}. 



\subsubsection {Flipped distributions}

\label{sec:flipped} 

\begin{figure}[t]
    \includegraphics[width=0.9\textwidth]{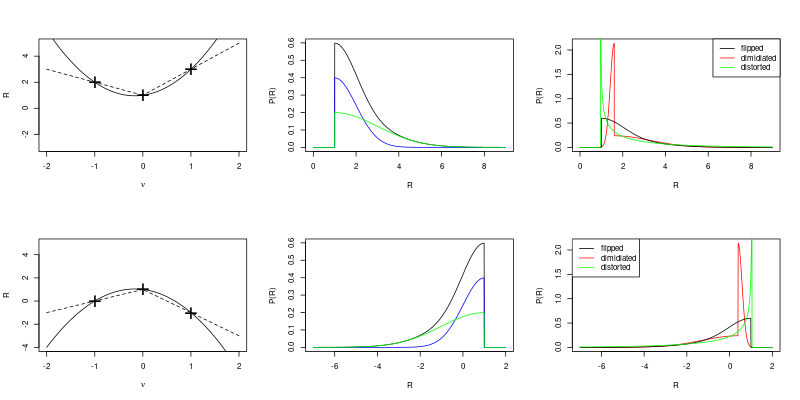}
    \caption{\label{fig:flipped} 
        The top row shows an  OPAT analysis where the deviations are both positive, 
        though of differing magnitude, and the dimidiated and distorted fits.  The second plot shows  the dimidiated distribution: the total (black) is the sum of the two half-Gaussians (green and blue).  The third plot also shows this pdf, together with the pdf obtained from the same 3 points using the distorted model, in green, and that from the equivalent \change{dimidiated} {} pdf in red.  The bottom row shows the \change {same} {equivalent} for two negative deviations.}
\end{figure}

It can happen that an OPAT analysis results in both deviations having the same sign, as illustrated in Figure~\ref{fig:flipped}.

This is implausible but not impossible.  If the error concerned is important then this presents a major problem: clearly something significant is going on, and the standard use of linear approximations in  
evaluating errors is inappropriate.  Fortunately such cases are rare. A more typical situation where this arises, and it does arise, is in the evaluation of the contribution of many systematic uncertainties to the total error. Among the many small contributions there may be one or two where both deviations go the same way, though this is often due to statistical fluctuations on a small and unimportant quantity whose treatment will make no difference to the final quoted result.  Nevertheless one needs a way of dealing gracefully with such cases.  The best advice to the user is to repeat the analysis with more points and more data --- but this extra effort may not be justifiable. 

The distorted model can handle such a situation. The minimum or maximum is now inside the 68\% central confidence region so the need to consider both \change{branches of the parabola for} {solutions in} the pdf is very significant and the Jacobian spike is
considerable, but all the algebra holds provided care is taken of the signs of $\sm$ and $\sp$.

For the dimidiated model the distribution in $R$ is the sum of two half-Gaussians, with the same nominal 
mean, which is actually the extreme or cut-off value ${\cal M}$.  If the `errors' are $\sigma_1$ and $\sigma_2$ \change{and are both positive, as in the upper row of Figure~\ref{fig:flipped},} {} then, from the properties of the Gaussian, the combined distribution has moments
\begin{equation}
    \label{eq:flipped}
    \mu={\cal M}+{\sigma_1+\sigma_2 \over \sqrt{2 \pi}}, \quad V={\sigma_1^2 + \sigma_2^2 \over 2}- {(\sigma_1+\sigma_2)^2 \over 2 \pi}, \quad \gamma=2{\sigma_1^3 + \sigma_2^3 \over \sqrt{2 \pi}} -3{\sigma_1^2 + \sigma_2^2 \over 2} {\sigma_1+\sigma_2 \over \sqrt{2 \pi}}+2\left({\sigma_2+\sigma_2 \over \sqrt{2 \pi}}\right)^3.
\end{equation}

It would be possible to introduce a specific flipped-dimidiated model using this algebra, but this seems an unnecessary complication to handle a situation which is either so serious that a simple 3-point analysis is inadequate, or so trivial that its impact will be small.  We suggest instead that, when a flipped distribution is encountered, it be replaced by a  conventional dimidiated distribution with the same moments as given by Equation~\eqref{eq:flipped}, and that the software (Appendices~\ref{sec:Cpp}--\ref{sec:R}) should provide a tool for doing this.  This, the red curve in Figure~\ref{fig:flipped}, has the same mean, variance and skewness as the black curve: they may appear different to the eye but this difference is not as large as it looks, and is similar to the differences between the dimidiated and distorted model (in green).

Another scenario in which this formalism might be applied can occur with a discrete nuisance parameter; for example one might need to consider both the (favoured) normal and the (disfavoured but possible) inverted neutrino mass hierarchy.  Suppose a result $R_1$ is obtained from the basic model or assumption, whereas a less-favoured alternative gives $R_2$.  If, say $R_2>R_1$, one might wish to quote this as \change { $\AE  R_1 |R_2-R_1| 0 $} { $\AE  R_1 R_2-R_1 0 $ } . We are not recommending this, but if someone chooses to do so we can accommodate it as a flipped pdf where
$\sm$ and $\sp$ are equal (or opposite: the signs require careful handling in the software)
and given by $|R_1-R_2|$.

For 
some models this cannot be handled, and they should not be deployed.  The relation between $\nu$ and $R$ is no longer monotonic, so the quantiles of $\nu$ do not match the quantiles of $R$.

\subsubsection{Comparison of the models}
\label{sec:allthepdfmodels}

Many different 3-parameter forms can be used to model asymmetric pdfs, and we have considered several.  These are briefly listed in Table~\ref{tab:pdfmodels} and fully described in Appendix~\ref{app:modelling-pdf-errors}.

The Asymmetry $A$ is here defined as ${\sp-\sm \over \sp+\sm}$.
Many models can handle arbitrary values of $\sp$ and $\sm$ and thus 
arbitrary $A$. Others, such as the Fechner and the skew normal, for which the extreme case is a half-Gaussian,
cannot model the largest $|A|$. The dimidiated Gaussian can handle arbitrary $A$ but not arbitrary $\gamma/V^{3/2}$. 

\begin{table}[tp]
\begin{tabular}{|p{0.25\textwidth}|p{0.25\textwidth}|p{0.15\textwidth}|p{0.1\textwidth}|p{0.20\textwidth}|}
\hline
&&&& \\[-1ex]
Name                    &  Description     
                        &  Range of $A$   
                        &  Handles flipping?    
                        &  Notes  \\[4ex]
\hline
&&&& \\[-1ex]
Dimidiated Gaussian     &  OPAT with 2 straight lines  
                        &  $\pm 1$  
                        &  Special case  
                        &  Discontinuity which can give problems when fitting \\[1ex]
&&&& \\[-2ex]
Distorted Gaussian      &  OPAT with a parabola  
                        &  $\pm 0.57382$  
                        &  Yes  
                        &  Limited range  \\[1ex]
&&&& \\[-2ex]
Railway Gaussian        &  OPAT with parabola morphing to straight lines 
                        &  $\pm 0.60467$  
                        &  Yes  
                        &  Arbitrary smoothing  \\[1ex]
&&&& \\[-2ex]
Double cubic Gaussian   &  OPAT with two cubics morphing to straight lines  
                        &  $\pm 0.74538$
                        &  Yes  
                        &  Arbitrary smoothing \\[1ex]
&&&& \\[-2ex]
Symmetric beta Gaussian\tablefootnote{Subsequent tables utilise the notation "Symmetric beta Gaussian($p$,$K$)" for this model. This notation supplies shape parameters $p$ and $h = 0.1 K$ detailed in Section~\ref{sec:symbetaGauss}.} &  OPAT with polynomial morphing to straight lines  
                        &  $\pm 1$
                        &  Yes  
                        &  Two arbitrary  tuning parameters \\[1ex]
&&&& \\[-2ex]
QVW Gaussian            &  Gaussian with $\sigma$ linear function of cumulant  
                        &  $\pm 0.68269$
                        &  No  
                        & 
 Messy numerically  \\[1ex]
&&&& \\[-2ex]
Fechner distribution    &  Two half Gaussians of same height 
                        &  $\pm 0.21564$  
                        &  No  
                        &  Little motivation  \\[1ex]
&&&& \\[-2ex]
Edgeworth expansion     &  Edgeworth expansion  
                        &  Very limited  
                        &  No  
                        &  Goes negative  \\[1ex]
&&&& \\[-2ex]
Skew Normal             &  Azzalini's form  
                        &  $\pm 0.21564$
                        &  No  
                        &  \\[1ex]
&&&& \\[-2ex]
Maximum entropy Johnson &  Johnson $S_U$  
                        &  $\pm 0.34719$
                        &  No  
                        &  Kurtosis from maximum entropy \\[1ex]
&&&& \\[-2ex]
Log normal              &  Takes exp of Gaussian distributed variable  
                        &  $\pm 1$
                        &  No  
                        &  Limited range  \\[4ex]
\hline
\end{tabular}
\caption{\label{tab:pdfmodels}
            Functions that can model asymmetric pdfs.}
\end{table}

\begin{figure}[t]
    \centerline{\includegraphics[width=.8\textwidth]{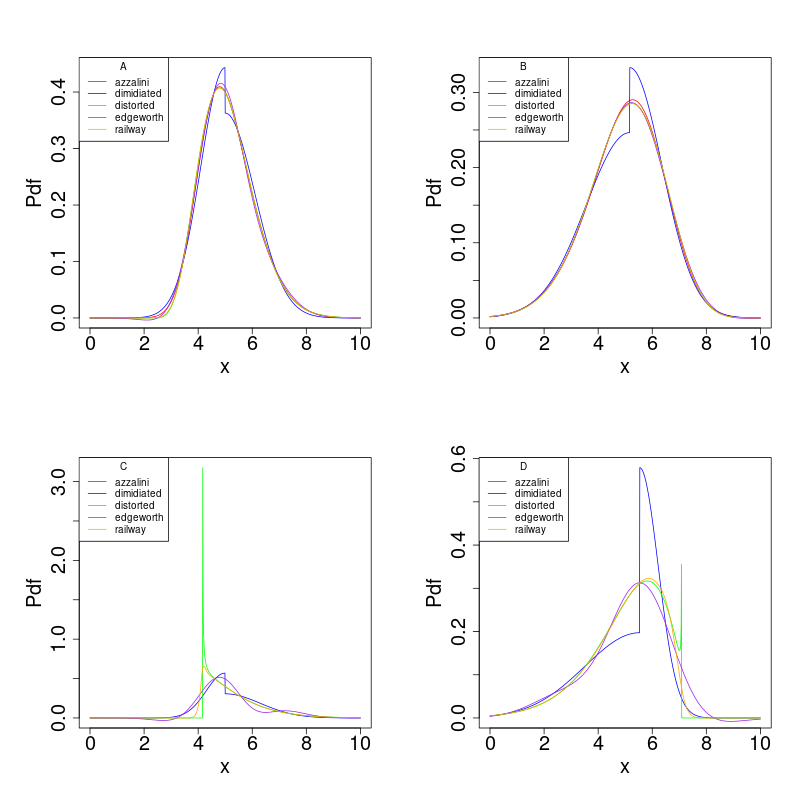}}
    \caption{\label{fig:pdf1}
        Pdfs for \change {a few} {the } different models\change {, listed in the legends.} {} Panel A is for $\AE 5.0 1.1 0.9 $,   B is for moments: $\mu=5.0, V=2.0, \gamma=-1.0$, C is for $\AE 5.0 1.3 0.7 $ and panel D for $\mu=5,V=2.0,\gamma=-3.0$.
        }
        
\end{figure}

The upper row of Figure~\ref{fig:pdf1} shows several models, with parameters chosen to match quantile values (on the left) and moments (on the right).  For these moderate asymmetries the behaviour appears sensible and there is generally little difference between the shapes \change {and the curves coincide,} {} except for the dimidiated Gaussian near the centre (where it probably doesn't matter very much).  

Two examples of larger asymmetry are shown in the lower row, and differences appear.  Azzalini's skew normal distribution cannot handle these values. The Edgeworth form goes negative and develops a second bump, which is
surely not physical. The distorted Gaussian shows a Jacobian spike which, again, is probably an undesired artifact. The railway Gaussian resembles the distorted but avoids the spike.

Other examples can be tried, and show the same behaviour: for moderate skewness the functions are well behaved and results are similar; for large skewness they need to be handled carefully. 

\begin{figure}[t]
    \centerline{\includegraphics[width=.7\textwidth]{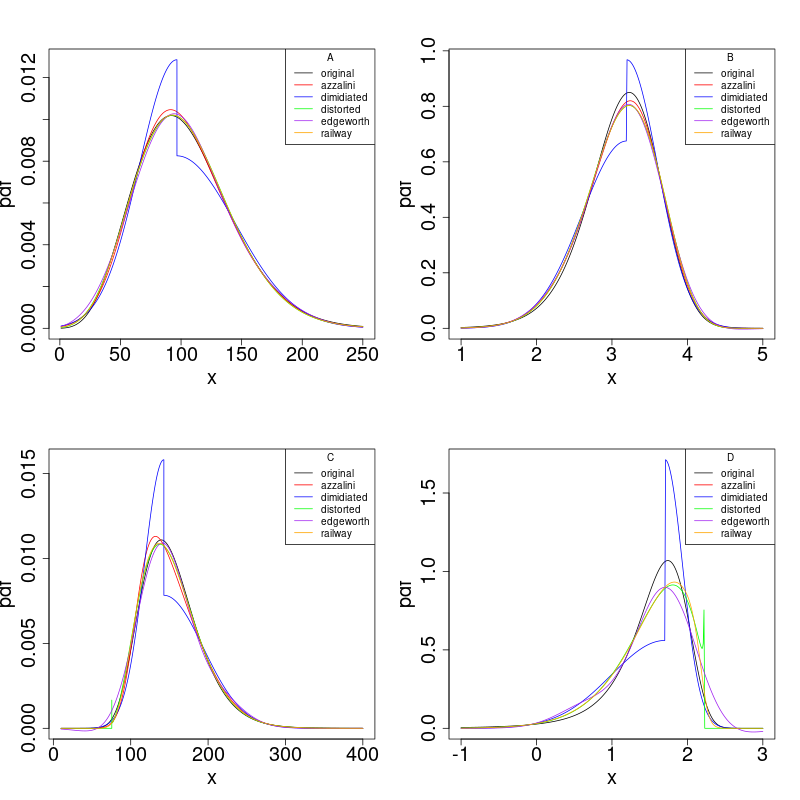}}
    \caption{\label{fig:pdf2}
        Pdfs from transformations of Gaussian distributions, and the model fits to them. Panel A is  the distribution for $x^2$, where $x$ is distributed according to $N(10.0,2.0)$.  B is for $\sqrt x$ with $x$ from $N(10.0,3.0)$.  C is $e^x$ with $x$ from $N(5.0,0.25)$ and panel $D$ is $\ln x$ with $x$ from $N(5.0,2.0)$.}
\end{figure}

As a further exploration, we can use some examples of the third original category of Section~\ref{sec:intro}, where a Gaussian-distributed variable is transformed to make it non-Gaussian.  We apply the various models to the moments of this distribution, and compare the model with the true original. Examples are shown in Figure~\ref{fig:pdf2}, taking the square of a Gaussian-distributed variable, the square root, the exponential and the logarithm.  (Different means and standard deviations were used, for presentational reasons; square roots and logarithms of negative numbers are ignored.)  For small to moderate asymmetries there is general agreement among the models (apart from the dimidiated model near the centre, as before) and they match the original.  The fourth example shown, panel D, considers the logarithm of a Gaussian variable, and  is more interesting
in that the models generally disagree with each other and  the original.

It is strongly recommended that any calculations with asymmetric errors use (at least) two models, as a check for robustness.  The dimidiated and distorted models have the advantage of simplicity, even though some aspects are clearly unrealistic.  The railway model is similar to the distorted and avoids the Jacobian spike.  On the other hand the Edgeworth model utilizing only three leading cumulants has an inbuilt problem in that it always gives a negative (unphysical) probability for some arguments.
Azzalini's skew normal is restricted to small asymmetries.  It seems to us that the Edgeworth and skew normal models are not suitable for general-purpose use, which have to handle a wide range of possible asymmetries: they can be used if there is a specific reason for favouring them, and it is known {\em a priori} that the asymmetries will not be large.

\subsection{Combination of pdf errors}
\label{subsec:PCOE}

The combination of two (or more) errors requires the convolution of the relevant pdfs: given $p_x(x)$ and $p_y(y)$ with $u=u(x,y)$ we require $p_u(u)$.  
\change{One  assumes that the errors are `small' so the linear approximation can be used. Near the nominal point $(x_0,y_0)$ we write $u(x,y)\approx u_0+\left({\partial u \over \partial x}\right) (x-x_0) +\left({\partial u \over \partial y}\right) (y-y_0)$.  Scaling to absorb the differentials we write $\tilde u=u-u_0,\tilde x=\left({\partial u \over \partial x}\right) (x-x_0),
\tilde y=\left({\partial u \over \partial y}\right) (y-y_0)$, which gives $\tilde u =\tilde x +\tilde y$.  So within the limits of the approximation we can express any general $u=u(x,y)$ as a simple sum
and for simplicity in the following algebra for error manipulations we will work with $u=x+y$, knowing that the general form can be reclaimed when and if desired.  Using this, $p_u(u)$ is given by Equation~\ref{eq:simpleconvolution}.} {In the linear approximation we assume, without loss of generality, that the values are shifted to have a nominal value of 0, $u(x,y)=u(x_0,y_0)+\tilde u(x-x_0,y-y_0)$ and then scaled to absorb the differentials, $x \to x_0+\left( {\partial u \over \partial x} \right) \tilde x$ and $y \to y_0+\left( {\partial u \over \partial y} \right) \tilde y$.  Then (dropping the tildes) we have $u=x+y$ and the pdf for $u$ is
the convolution of the two, as given in Equation~\ref{eq:simpleconvolution}. }
In some cases this can be evaluated algebraically, in others it needs to be done numerically.

This presents a problem, in that, although the convolution of two Gaussians gives another Gaussian,
this is not in general true for other functions.
The convolution of two pdfs, both belonging to one of the parameterisations in Table~\ref{tab:pdfmodels}, does not give a function described by that parameterisation.  See Appendix~\ref{sec:convolution} for a detailed discussion using the dimidiated Gaussian.

\vspace{2ex}
\boxit{
    To proceed, we note that the moments $\mu,V, \gamma$ add under convolution.  To provide a consistent procedure for `adding errors' we can evaluate the individual moments using Equations~\eqref{eq:dimidiate1}, \eqref{eq:distorted1}, or their equivalents for other  models, sum them to obtain the totals, and then use Equations~\eqref{eq:dimidiated2}, \eqref{eq:distorted2}, or their equivalents, to find the model parameters and, if desired, the quantile parameters that give a pdf that has these moments.  
}
\vspace{1ex}

\begin{sidewaystable}
\centering
\begin{tabular}{ | l | c c c | c c c | c c c | c c c | }
\hline
  & \multicolumn{12}{|c|}{} \\[-1ex]
  & \multicolumn{12}{|c|}{\bf Errors}  \\[2ex]
\hline
  & \multicolumn{3}{|c|}{} & \multicolumn{3}{|c|}{} & \multicolumn{3}{|c|}{} & \multicolumn{3}{|c|}{} \\[-1ex]
  & \multicolumn{3}{|c|}{$\sigma_x^- = 1.0$, $\sigma_x^+ = 1.0$} 
  & \multicolumn{3}{|c|}{$\sigma_x^- = 0.8$, $\sigma_x^+ = 1.2$} 
  & \multicolumn{3}{|c|}{$\sigma_x^- = 0.5$, $\sigma_x^+ = 1.5$}  
  & \multicolumn{3}{|c|}{$\sigma_x^- = 0.5$, $\sigma_x^+ = 1.5$}  \\[2ex]
  & \multicolumn{3}{|c|}{$\sigma_y^- = 0.8$, $\sigma_y^+ = 1.2$} 
  & \multicolumn{3}{|c|}{$\sigma_y^- = 0.8$, $\sigma_y^+ = 1.2$} 
  & \multicolumn{3}{|c|}{$\sigma_y^- = 0.8$, $\sigma_y^+ = 1.2$}  
  & \multicolumn{3}{|c|}{$\sigma_y^- = 0.5$, $\sigma_y^+ = 1.5$}  \\[2ex]
\hline
  & \multicolumn{3}{|c|}{} & \multicolumn{3}{|c|}{} & \multicolumn{3}{|c|}{} & \multicolumn{3}{|c|}{} \\[-1ex]  
  {\bf Models}
  & $\sigma^-$  &  $\sigma^+$  &  $\Delta$  &  $\sigma^-$  &  $\sigma^+$  &  $\Delta$  
  &  $\sigma^-$  &  $\sigma^+$  &  $\Delta $  & $\sigma^-$  &  $\sigma^+$  & $\Delta$ \\
    Dimidiated Gaussian     &  1.32  &  1.52  &  ~~0.080  &  1.22  &  1.62  &  0.160  &  1.09  &  1.78  &  0.284  &  0.97  &  1.93  &  0.413  \\  
    Distorted Gaussian      &  1.33  &  1.54  &  ~~0.098  &  1.25  &  1.64  &  0.203  &  1.17  &  1.88  &  0.349  &  1.12  &  2.07  &  0.534  \\ 
    Railway Gaussian        &  1.34  &  1.53  &  ~~0.098  &  1.25  &  1.64  &  0.199  &  1.18  &  1.86  &  0.325  &  1.13  &  2.04  &  0.484  \\  
    Double cubic Gaussian   &  1.32  &  1.52  &  ~~0.080  &  1.23  &  1.62  &  0.160  &  1.11  &  1.79  &  0.285  &  1.00  &  1.95  &  0.419  \\   
    Symmetric beta Gaussian(1,10)  &  1.32  &  1.52  &  ~~0.080  &  1.23  &  1.62  &  0.162  &  1.12  & 1.80  & 0.288  &  1.01  &  1.95  &  0.423  \\
    Symmetric beta Gaussian(1,30)  &  1.33  &  1.53  & ~~0.096  &  1.24  &  1.64  &  0.195  &  1.17  & 1.86  & 0.343  &  1.11  &  2.03  &  0.519  \\  
    Symmetric beta Gaussian(4,10)  &  1.32  &  1.52  &  ~~0.078  &  1.22  &  1.62  &  0.157  &  1.11  &  1.79  &  0.280   &  0.99  &  1.94  &  0.409  \\  
    Symmetric beta Gaussian(4,30)  &  1.33  &  1.53  &  ~~0.090  &  1.23  &  1.63  &  0.182  &  1.15  &  1.83  & 0.323  &  1.07  &  1.99  &  0.482  \\
    QVW Gaussian            &  1.32  &  1.52  & ~~0.083  &  1.23  &  1.62  &  0.167  &  1.12  &  1.80  &  0.298  &  1.03  &  1.96  &  0.440  \\
    Fechner distribution    &  1.30  &  1.52  &  ~~0.092  &  1.19  &  1.62  &  0.186  &  \multicolumn{3}{|c|}{--} &  \multicolumn{3}{|c|}{--}  \\  
    Edgeworth expansion     &  1.31  &  1.52  &  ~~0.095  &  1.21  &  1.62  &  0.195  &  \multicolumn{3}{|c|}{--} &  \multicolumn{3}{|c|}{--}  \\  
    Skew normal             &  1.33  &  1.49  &  ~~0.101  &  1.22  &  1.60  &  0.190  &  \multicolumn{3}{|c|}{--} &  \multicolumn{3}{|c|}{--}  \\  
    Johnson system          &  1.24  &  1.68  &  $-0.006$  &  1.28  &  1.83  &  0.156  &  \multicolumn{3}{|c|}{--} &  \multicolumn{3}{|c|}{--}  \\  
    Log normal              &  1.33  &  1.59  &  ~~0.075  &  1.26  &  1.70  &  0.188  &  0.73  &  1.93  &  0.115  &  0.91  &  2.42  &  0.351  \\[1ex]  
\hline
\end{tabular}
\caption{\label{tab:pdfcoe} Combining errors from various pdfs using various models. $\Delta$ is the combined median.}
\end{sidewaystable}

Table~\ref{tab:pdfcoe} illustrates the results of combining various typical errors, using different models.   The behaviour is much as one would expect. The combined errors increase in all cases, but the asymmetry falls. This is brought out in the rightmost entry, where
two very asymmetric errors ($\sp=3 \sm$) combine to give an error with only $\sp \approx 2 \sm$.  For moderate asymmetries, as in the first two columns, there is reasonable agreement between the models, to two significant figures, but not three, which gives an estimate of how far the precision of such calculations can be trusted.
For large asymmetries, as in the third and fourth columns, the agreement is lost, showing that in such cases the accuracy should not be considered as being definite.  (Several models fail to accommodate the examples with very high asymmetries, so the entries are blank). After all, any method will break down somewhere.

An important point to note is that in all cases the nominal value shifts: this is the number shown as $\Delta$.
If two asymmetric pdfs are convolved, then the median of the result is not the sum of the individual medians.  A procedure for combining errors using asymmetric distributions must, to be consistent, include a shift in the central value.  If practitioners are not prepared to make such an adjustment, then they should perhaps reconsider the use of asymmetric errors and revert to the usual symmetric form,  averaging their $\sp$ and $\sm$ values. The values of $\sp$ and $\sm$ in the table, and returned by the software, give, within the framework of the model concerned, the 68\% central region 
around the nominal point: when the $\Delta$ shift is applied the region boundaries should be adjusted accordingly.

\subsection{Why the `usual procedure' is wrong}

It is common practice (though we have not been able to find a reference) to combine asymmetric errors by combining all the positive errors in quadrature, and likewise all the negative errors, and then using the dimidiated (``bifurcated'') Gaussian.  This is obviously wrong.  Suppose one has $N$ sources of error, and that they all have the same positive error $\sp$ and negative error $\sm$.  The combined error given by this procedure will have positive error $\sqrt N \sp$ and negative error $\sqrt N \sm$.  The pdf of the combination has a width that increases like $\sqrt N$ but {\em does not change its shape.}  But this contradicts the Central Limit Theorem, which declares that at large $N$ \change {the distribution of a sum tends to a Gaussian (symmetric) shape, whatever the distributions of the components.} {all distributions tend towards a Gaussian (symmetric) shape.}

\change {}
{The flaw in the logic is as follows: when two random variables, described (for the sake of the example) by dimidiated Gaussians,  are combined, there is a 25\% chance that they will both fluctuate upwards.  That is indeed described, within the framework of this model, by a 2-D Gaussian for which the standard deviation of the sum is the sum in quadrature of the two positive $\sigma$.   
Likewise there is a 25\% chance that they will both go negative, and that is described by $({\sigma^-_1}^2 + {\sigma^-_2}^2)^{1/2}$.  But there is a 50\% chance that one will go up and the other will go down.   This is neglected by the `usual procedure', and it is this filling in of central values that lets the Central Limit Theorem do its work.}

Examples of how this can give wrong results are in Section~\ref{sec:lifetime}.

\subsection{Combination of pdf results}

As discussed in Section~\ref{sec:intro}, the use of likelihoods to combine results is a very standard procedure and their use to produce a combined error is less mainstream, whereas for pdfs the case is reversed.  In fact, for pdfs the combination of results can be considered as an instance of the combination of errors.  The measurements are all of the same quantity and combined by taking a weighted mean, rather than distinct quantities combined with some general function.
 
If several independent estimates $\hat r_i$ with pdfs of variances $V_i$ are obtained by different experiments for the same result $r$, then an obvious way to form a combined result is to take a weighted sum:
\begin{equation}
    \hat r =\sum_i w_i \hat r_i \qquad{\rm with } \qquad \sum w_i=1,
\end{equation}
where the requirement for the sum of the weights ensures that if the individual estimates are unbiased, so is the combination.  For an efficient estimator one wants to minimise the variance of $\hat r$. This is given by
\begin{equation}
    V_{\hat r} = \sum w_i^2 V_i,
\end{equation}
which is minimised by taking 
\begin{equation}
    w_i \propto {1 \over V_i},
\end{equation}
the familiar result that results should be weighted by the inverse of their variance: the point here is that it holds even for non-Gaussian distributions.

For $\hat r_i$ we must take the mean of the distribution (as $V$ is the variance about the mean).  If the median is given, it must be converted, according to the model being used, before averaging.  The skewness of the result is just $\sum_i w_i^3 \gamma_i$ and the resulting moments can be used to give the parameter set of whatever model is being used. 

\begin{table}[tbp]
{\footnotesize
\begin{tabular}{|c | c || p{0.25\linewidth} | p{0.25\linewidth} |}
\hline
\multicolumn{2}{|c||} {}  &  \multicolumn{2}{|c|}{} \\[-1ex]
\multicolumn{2}{|c||} {Input}  &  \multicolumn{2}{|c|}{Combined result using model} \\[1ex]
\hline
&&& \\[-1ex]
$r_1$  &  $r_2$  &  Dimidiated Gaussian  &  Distorted Gaussian   \\
$\AE 32.571 7.571 6.571 $  &  $\AE 18.429 7.571 6.571 $  
     &  $\AE 25.700 5.252 4.752 $  &  $\AE 25.750 5.262 4.763 $  \\[1ex]
  &  &  Railway Gaussian  &  Double Cubic Gaussian  \\
  &  &  $\AE 25.749 5.261 4.765 $  &  $\AE 25.699 5.254 4.754 $  \\[1ex] 
  &  &  QVW Gaussian  &  Fechner Distribution  \\
  &  &  $\AE 25.707 5.255 4.755 $  &  $\AE 25.712 5.252 4.749 $  \\[1ex]
  &  &  Edgeworth expansion  &  Skew normal  \\
  &  &  $\AE 25.749 5.251 4.749 $  &  $\AE 25.741 5.255 4.777 $  \\[1ex]
  &  &  Johnson System  &  Log normal  \\
  &  &  $\AE 25.742 5.333 4.809 $  & $\AE 25.748 5.282 4.776 $  \\[1ex]
  &  &  Sym.\ beta Gaussian(1,10)  &  Sym.\ beta Gaussian(1,30) \\
  &  &  $\AE 25.700 5.254 4.755 $  &  $\AE 25.741 5.260 4.761 $  \\[1ex]
  &  &  Sym.\ beta Gaussian(4,10)  &  Sym.\ beta Gaussian(4,30)  \\
  &  &  $\AE 25.695 5.253 4.754 $  &  $\AE 25.725 5.257 4.758 $  \\[1ex]
\hline 
&&& \\[-1ex]
$r_1$  &  $r_2$  &  Dimidiated Gaussian  &  Distorted Gaussian  \\
$\AE 41.142 7.571 6.571 $  &  $\AE 12.878 7.571 6.571 $  
     &  $\AE 27.200 5.252 4.752 $  &  $\AE 27.250 5.262 4.763 $  
     \\[1ex]
  &  &  Railway Gaussian  &  Double Cubic Gaussian  \\
  &  &  $\AE 27.249 5.261 4.765 $  &  $\AE 27.199 5.254 4.754 $  \\[1ex]
  &  &  QVW Gaussian  &  Fechner Distribution  \\
  &  &  $\AE 27.207 5.255 4.755 $  &  $\AE 27.212 5.252 4.749 $   \\[1ex]
  &  &  Edgeworth expansion  &  Skew normal  \\
  &  &  $\AE 27.249 5.251 4.74 $  &  $\AE 27.241 5.255 4.777 $  \\[1ex]
  &  &  Johnson System  &  Log normal  \\
  &  &  $\AE 27.242 5.333 4.809 $  &  $\AE 27.248 5.282 4.776 $  \\[1ex] 
  &  &  Sym.\ beta Gaussian(1,10)  &  Sym.\ beta Gaussian(1,30) \\
  &  &  $\AE 27.200 5.254 4.755 $  &  $\AE 27.241 5.260 4.761 $  \\[1ex]
  &  &  Sym.\ beta Gaussian(4,10)  &  Sym.\ beta Gaussian(4,30)  \\
  &  &  $\AE 27.195 5.253 4.754 $  &  $\AE 27.225 5.257 4.758 $  \\[1ex]
\hline 
\end{tabular}
}
\caption{\label{tab:coepdf} Examples of combination of errors with pdfs using different models.}
\end{table}

As an example, suppose that a variable $x$ is distributed according to a Gaussian distribution with mean $\mu=5$ and standard deviation $\sigma=1/\sqrt{2}$.  However the variable of interest is $r=x^2$, which accordingly has quantiles $M=25$, $\sp=7.571$ and $\sm=6.571$.  Suppose that $x$ is sampled twice and, as it happens, the two values are at $\mu-\sigma$ and $\mu+\sigma$.  They will be added with equal weight, and the parameters of the pdf of the result extracted using the combination of errors procedure, assuming one of the models for asymmetric pdfs.  The results of this are shown in Table~\ref{tab:coepdf}, together with the results if the two samples happen to be at $\mu\pm 2 \sigma$.  (Results are shown to more significant figures than the data would normally warrant, to enable comparison of small differences between models.)  This example illustrates the point that despite imposing $\sum w_i=1$, bias in the estimate is essentially inevitable.  Unbiased measurements of $x$ will not give unbiased estimates of $\mu$. Any non-trivial transformation of the variable of interest will entail a bias.

\begin{figure}[t]
    \centering
    \includegraphics[width=10 cm]{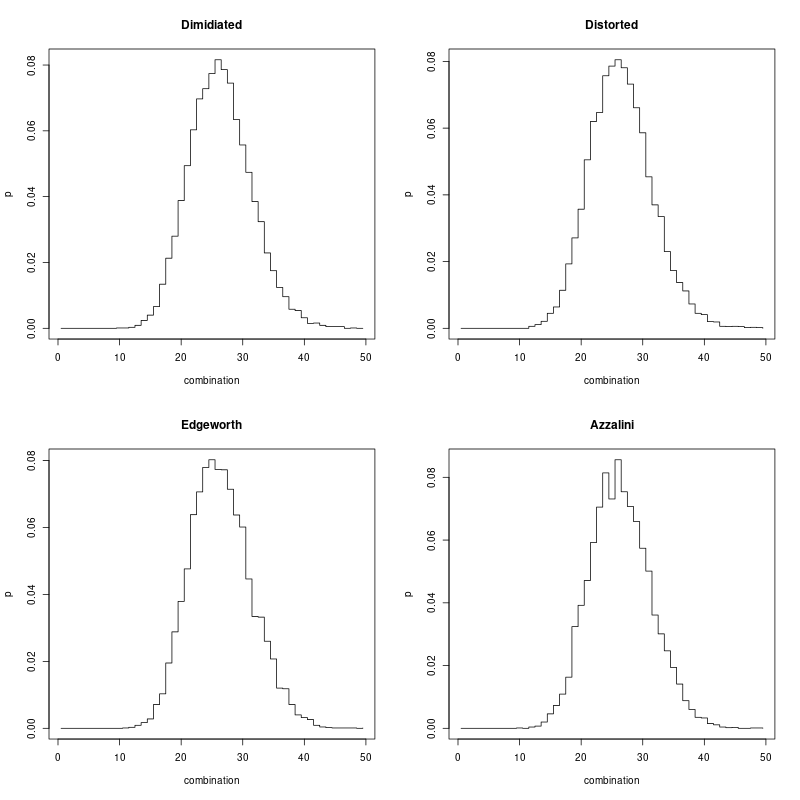}
    \caption{\label{fig:pdfcoe}
        \change{Example, using Monte Carlo simulations with 10,000 samples, of combination of results using pdfs,
        taking two random Gaussian numbers and squaring them, both being ascribed errors $\AE {} 7.57 6.57 $, and then combined using different pdf models. From such distributions the quantities $V_{MC}$ and $\gamma_{MC}$ in Table~\ref{tab:corcomp} are calculated.} {Combination of results using pdfs.} }
\end{figure}

Continuing with this example, we use a toy Monte Carlo to generate pairs of `samples' in $x$ according to a random Gaussian process \change { as described above,} {}  combine the two, and histogram the results.  This is shown in Figure~\ref{fig:pdfcoe}.  The mean, variance and skewness can be found from each histogram and compared with the values expected from the quoted results, shown in Table~\ref{tab:corcomp}. It can be seen that there is good agreement for the variances in all cases and fair agreement for the skewness, except that the dimidiated Gaussian underestimates the skewness of the pdf of the combination, because the dimidiated model is less influenced by extreme values than the others. 

%
\begin{table}[tp]
\centering
\begin{tabular}{| l | c c | c c| }
\hline
&&&& \\[-1ex]
Model & $V_{pred}$ & $\frac{V_{pred} - V_{MC}}{V_{MC}}$ & $\gamma_{pred}$ & $\frac{\gamma_{pred} - \gamma_{MC}}{\gamma_{MC}}$ \\[1ex]
\hline
&&&& \\[-1ex]
Dimidiated Gaussian            & 25.045 & -0.0096 & 14.967 & -0.610 \\
Distorted Gaussian             & 25.250 & -0.0015 & 37.750 & -0.015 \\
Railway Gaussian               & 25.249 & -0.0015 & 36.867 & -0.038 \\
Double cubic Gaussian          & 25.083 & -0.0081 & 20.765 & -0.458 \\
Symmetric beta Gaussian(1,10)  & 25.090 & -0.0078 & 21.804 & -0.431 \\
Symmetric beta Gaussian(1,30)  & 25.206 & -0.0032 & 34.141 & -0.109 \\
Symmetric beta Gaussian(4,10)  & 25.071 & -0.0086 & 19.007 & -0.504 \\
Symmetric beta Gaussian(4,30)  & 25.150 & -0.0055 & 28.741 & -0.250 \\
QVW Gaussian                   & 25.103 & -0.0073 & 23.277 & -0.393 \\
Fechner distribution           & 25.016 & -0.0107 & 25.157 & -0.344 \\
Edgeworth expansion            & 25.000 & -0.0114 & 37.751 & -0.015 \\
Skew normal                    & 25.583 &  0.0117 & 33.617 & -0.123 \\
Johnson system                 & 26.555 &  0.0501 & 44.119 & 0.151 \\
Log normal                     & 25.593 &  0.0121 & 39.367 & 0.027 \\[1ex]
\hline
\end{tabular}
\caption{\label{tab:corcomp} Comparison of quoted quantities and toy Monte Carlo results based upon $10^6$ toys. The toy results are not identical for all models but agree to five significant digits, $V_{MC} = 25.288$ and $\gamma_{MC}= 38.388$.  }
\end{table}

This framework may be unrealistic, in that it assumes that the (asymmetric) errors are independent of the measurement.  The quoted errors on the `input' column in Table~\ref{tab:coepdf} are all the same, because the true mean of $x$ is known.  This will be complicated by additional information from the experiments which does not appear in the quoted values and errors.  This is considered in detail by Schmelling~\cite{Schmelling} using a second order approximation (the ``distorted Gaussian" in this work). Rather than the 
ideal case where all true quantities are known, he considers the more realistic case where the errors are taken from measurements (as happens when $\sqrt N_{observed}$ errors are used for Poisson statistics), and shows that as well as the bias from the difference between median and mean there is a bias from the use of an estimator $\hat V$ for the variance, and a further significant bias from the correlation between $V$ and $\hat r$.  If the parameter of interest is, for example, the square of a Gaussian quantity, then upward fluctuations will be ascribed larger errors, and thus smaller weights. 

A combination of results generally brings with it the question of whether such a combination is valid: is it plausible that these are really measurements of the same quantity? This is generally expressed as a $\chi^2$ value, or some other goodness of fit measure.  Again, this is readily done when using likelihoods but is rather contrived when using pdfs.

For a given measurement $\AE x_i {\spi} {\smi}  $\ we can express its compatibility with some quoted $x_c$ using the p-value or its equivalent expression in terms of sigmas.  With the familiar Gaussian, we would say that a measurement of $12.7 \pm 0.1 $ was 5 sigma away from some nominal value of $12.2$, and accordingly treat it with strong reservation. A similar approach for asymmetric errors will use $\spi$ and $\smi$, but care must be taken over the direction of the deviation. Suppose the proposed 
true value is $12.2$ and the measurement is $\AE 12.7 0.1 0.2 $. Using the dimidiated model, this deviation is a 5 sigma, not 2.5 sigma, as the quoted measurement implies that a process which may go up a little or down a lot has given a result of 12.7.     For other models the arithmetic is not so simple, but the p-value 
can readily be determined from the pdf.

If we ask about the compatibility of $\AE x_i {\spi} {\smi} $ with some $\AE x_c {\sigma^+_c} \sigma^-_c $, as opposed to a single value, this can be obtained from the convolution of the two pdfs.  This can be done numerically, within the framework of the model adopted.

To consider the agreement of a whole set of $N$ measurements one can convert the individual p-values into their equivalent $\chi^2$ numbers and sum them to get a total $\chi^2$. If $x_c$ were imposed externally, this would have $N$ degrees of freedom; as $x_c$ has been obtained from the measurements it is probably reasonable to use $N-1$ in obtaining the overall p-value.


\section{Likelihood errors}
\label{sec:likelihood}

\subsection{Modelling non-parabolic likelihoods}

For the Gaussian the log likelihood function  $\ln L$ is a parabola with two parameters, the location and scale. (The parameter giving the value at the peak is irrelevant.)  For an asymmetric form a reasonable guess at the full likelihood function will be some curve with three parameters,  approximately parabolic, for which the maximum likelihood peak occurs at some $a=\hat a$ and $\Delta \ln L = - \half$ at $a=\hat a + \sp$ and $a=\hat a - \sm$.  Thus the curve must go through 3 points, having a maximum at the middle one. The 3 parameters will be obtainable from the values of $\hat a, \sp$ and $\sm$.
It should also have a `reasonable' behaviour elsewhere; in particular it should go to $-\infty$ at large positive and negative $a$.  Provision of such models is somewhat easier for log likelihoods than for pdfs, as there are only two sets of parameters deployed: these $\hat a,\sp,\sm$ and the 3 parameters of the specific model.  The software must convert between the two, but there is no third set corresponding to the moments of the pdfs.



Two models which are simple to use and turn out to be generally applicable arise from the suggestion by Bartlett that a likelihood function is described by a Gaussian whose width varies as a function of the argument $a$
\cite{bartlett1,bartlett2,deltalnL}.  With such an assumption one might suppose that the variation is linear, either for the standard deviation (`linear sigma model') or the variance (`linear variance model')
\begin{equation}
    \label{eq:bartlett}
    \ln L(a)=-\half \left( {a - \hat a \over \sigma+(a-\hat a) \sigma'}\right)^2
    \qquad {\rm or} \qquad -\half {(a-\hat a)^2 \over V+(a-\hat a)V'}.
\end{equation}
The parameters, $\sigma, \sigma'$ or $V,V'$, which are readily given by the requirement that the curve go through the two $-\half$ points, are
\begin{equation}
    \sigma={2 \sp \sm \over \sp+\sm}, \qquad \sigma'={\sp-\sm \over \sp+\sm},
    \qquad V=\sp \sm, \qquad V'=\sp-\sm.
\end{equation}
More details are in Sections~\ref{sec:linearsigma} and \ref{sec:linearvariance}.

\begin{figure}[t]
    \centerline{\includegraphics[width=12 cm]{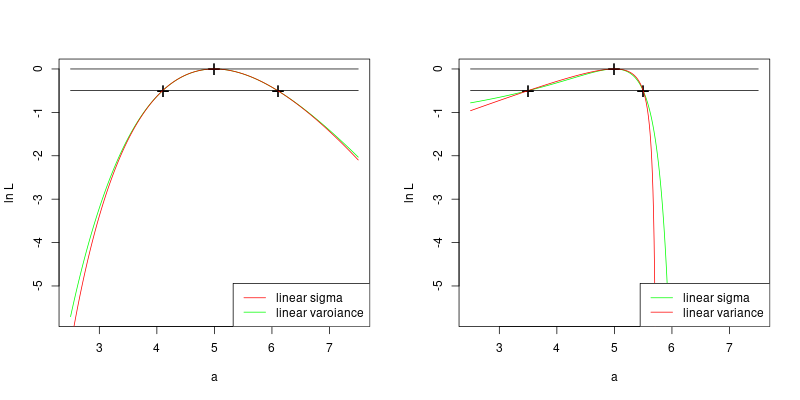}}
    \caption{\label{fig:twolike} 
        Using the linear sigma and linear variance models of the log likelihood arising from  $\AE 5.0 1.1 0.9  $ (left) and $\AE 5.0 0.5 1.5 $ (right).}
\end{figure}

The use of these two models  is illustrated in Figure~\ref{fig:twolike} which shows the likelihood curves resulting from a small positive asymmetry, $\AE 5.0 1.1 0.9 $ and a large negative  one, $\AE 5.0 0.5 1.5 $.  All curves go through the 3 defining points, by construction.  In the central region there is little difference, even in the very asymmetric instance. Differences do appear further from the peak.  As a general rule, the choice of model makes little difference for small asymmetries, but can become significant at large excursions for large asymmetries.

If any information is provided, or can be deduced, about the form of the likelihood then that should be used to guide the choice of model --- there is no general `best model', but some models do better for particular classes of result.  By the same token, if a result with an asymmetric error is quoted, any further available information should be provided. If all the information available to a meta-analyst is the 
3 numbers of the reported measurement, then the wide range of available models can potentially give a wide range of outcomes.

\subsubsection{Comparison of models and recommendations}

As well as these two models, many others may be used as 3-parameter descriptions of not-quite parabolic log likelihoods.  We list some in Table~\ref{tab:lnLmodels}, their full descriptions are in Appendix~\ref{app:loglikelihoods}

\begin{table}[p]
\begin{tabular}{|p{0.25\textwidth}|p{0.40\textwidth}|p{0.35\textwidth}|}
\hline
&& \\[-1ex]
Name                    & Description    
                        &  Notes \\[1ex]
\hline
&& \\[-1ex]
Cubic                   & $-\half (\alpha a^2+\beta a^3)$ 
                        & Badly behaved, $\ln L$ turns over. \\
Broken parabola         & $-\half {a^2 \over \sigma_\pm^2} $  
                        & Simple, but $L''$ discontinuity at peak \\
Constrained quartic     & $-\half(\alpha + \beta a)^2$  
                        & Quartic constrained at single peak \\
Molded quartic          & $-\half (\alpha a^4+\beta a^3+\gamma a^2)$  
                        & Quartic is best match to broken parabola in central region. \\
Matched quintic         & Quintic centrally, $\sigma_\pm$ outside  
                        & Quintic coeffts from matching $L$ and $L''$ at $\sigma_\pm$ \\
Interpolated 7th degree & 7th order centrally, $\sigma_\pm$ outside  
                        & Coeffts from matching $L,L',L''$ \\
Simple double quartic   & Central quartics, quadratics outside $\sigma_0=\sqrt{\sm\sp}$  
                        & Coeffts from matching\\
Molded~double quartic   & Central quartics, quadratics outside $\sigma_0=\sqrt{(\sm^4+\sp^4 )/(\sm^2+\sp^2}$ 
                        & Coeffts from matching \\
Simple double quintic   & Central quintics, quadratics outside & \\
Conservative spline     & Cubic between spline points, quadratics outside 
                        & Spline points need to be chosen in central region. \\
Logistic beta           & $-\alpha \ln(1+e^{-x})-\beta \ln(1+e^x)-\ln B(\alpha,\beta)$ 
                        & Log of Type IV generalised logistic\\
Logarithmic             & $-\half({\ln(1+\gamma a) \over \ln \beta})^2$ 
                        & Parabola subject to scaled expansion. Limited domain\\
Generalised Poisson     & $-\alpha a + \N(1+{\alpha a \over \N})$ & Poisson, but $\N$ need not be integer. \\
Linear sigma            & $\sigma(a)=\sigma+a\sigma'$  
                        &  Good general-purpose use \\&&Finite valid domain \\
Linear variance         & $V(a)=V+aV'$   
                        &  Good general-purpose use.  Finite domain \\
Linear log sigma        & $\sigma = e^{\alpha Q(a)+\beta}$  
                        & $Q(a)$ is the cdf of the equivalent Fechner distribution.\\
Double cubic log sigma  & $\sigma=e^{cubics}$ centrally, $\sigma_\pm$ outside 
                        & Different cubics for $a\le 0$ and $a\ge 0$ \\
Quintic log sigma       &$\sigma=e^{quintic}$ centrally, $\sigma_\sp$ outside 
                        & Quintic by matching $L,L',L''$ \\
PDG method              &  $\sigma(a)=\sigma+a\sigma'$ or $\sigma_\pm$  
                        & Linear $\sigma$ centrally, $\sp$ and $\sm$ above and below. Discontinuous $L'$\\
Edgeworth expansion     & Log of Equation~(\ref{eq:edgeworth1}) 
                        & Numerically messy \\
Skew normal & Log of Equation~(\ref{eq:azzalini}) 
                        & Numerically messy \\[1ex]
\hline
\end{tabular}
\caption{\label{tab:lnLmodels}
            Functions for modelling asymmetric log likelihoods.  For simplicity, formul\ae\  assume $\hat a=0$.  The `central region' is between $\hat a-\sm$ and $\hat a+\sp$.  The symbol $\sigma_\pm$ is used to denote ``$\sp$ for positive values and $\sm$ for negative values".}
\end{table}


\begin{figure}[t]
    \centerline{\includegraphics[width=12 cm]{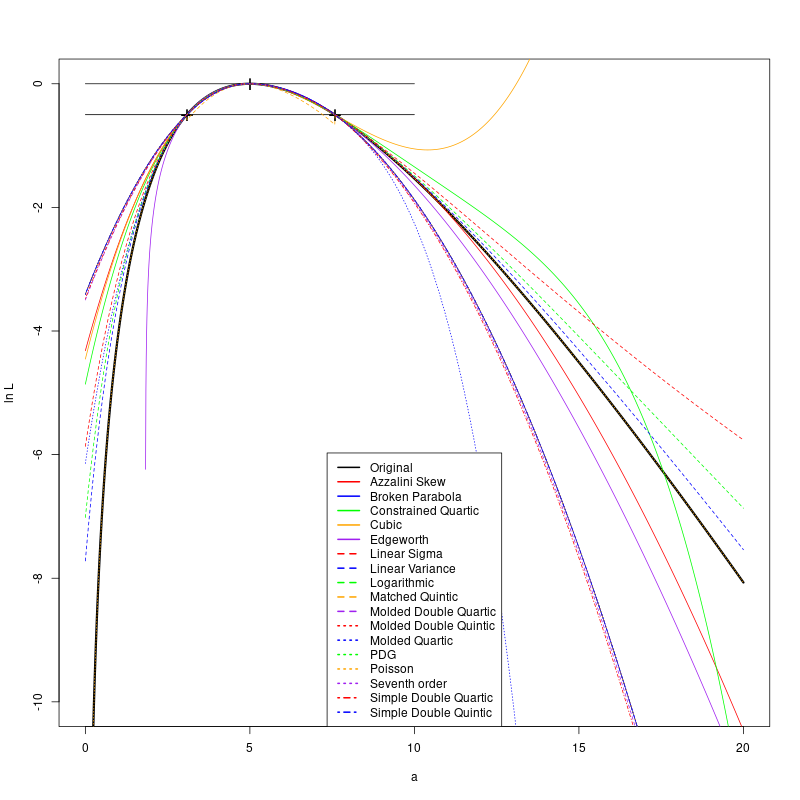}}
    \caption{\label{fig:Like1}
        Approximations to a Poisson likelihood.}
\end{figure}

As an illustration we take the result $a=\AE 5.0000 2.5811 1.9159 $, which would be reported from a Poisson measurement of 5 events.  This is shown in Figure~\ref{fig:Like1}.  The true likelihood is $L(a)=e^{-a}a^5/5!$, and we compare this with the results of the likelihoods given by the models.  All curves, by construction, peak at $a=5$ and  go through the $\Delta \ln L=\pm \half$ points, and their interpolations in the central region are very similar.  At larger values their behaviours are very different.  The cubic turns over, which is unacceptable. The quartic provides a better match to the original Poisson likelihood, especially on the negative side. The logarithmic form does fairly well, the Poisson does perfectly, as it should in this particular example. 
The two Bartlett forms, labelled Linear Variance and Linear Sigma, both do well, the linear variance is somewhat better than the linear sigma.  The PDG form does badly, as it assumes a Gaussian behaviour outside the central interval.  The Edgeworth form does badly at both small and large values, and is troubled by logarithms of negative numbers at large values and/or asymmetries. The skew normal is not as bad as Edgeworth's approximation, but it does not do very well.  These comparisons assume the fundamental $a=\AE 5.0000 2.5811 1.9159 $ measurement comes from a Poisson measurement: if it does not, then conclusions might be different.

\begin{figure}[t]
    \center{\includegraphics[width=12 cm]{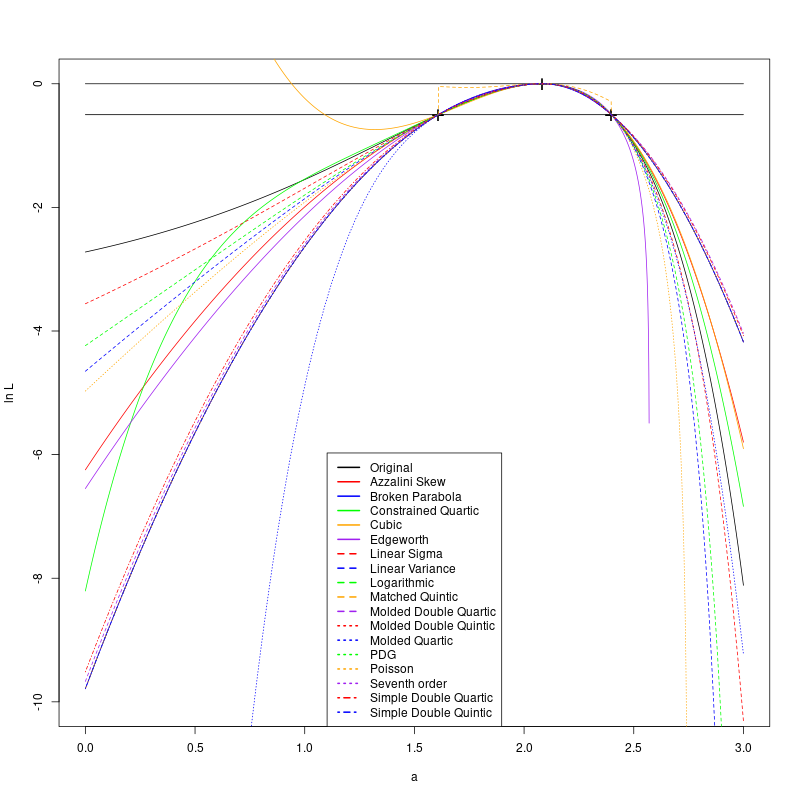}}
    \caption{\label{fig:Like2}
        Possible models of log likelihoods for the measurement $\AE 2.08 0.32 .047 $, which can arise from $\ln (8 \pm 3)$.  }
\end{figure}

As another example, this time with a negative skewness, we take the logarithm of a parameter with a likelihood described by a Gaussian with $\mu=8.0$ and $\sigma=3.0$.  This is shown in Figure~\ref{fig:Like2}. The reported result is $a=\AE 2.08 0.32 0.47 $.  Again the cubic does predictably badly, and the quartic surprisingly well.  The logarithmic does quite well, and the Poisson is not good.  The PDG form is again bad, but the Bartlett forms are good: this time the linear sigma does a bit better than the linear variance. The Edgeworth form does badly and the skew normal form does not do particularly well.

\begin{figure}[t]
    \center{\includegraphics[width=12 cm]{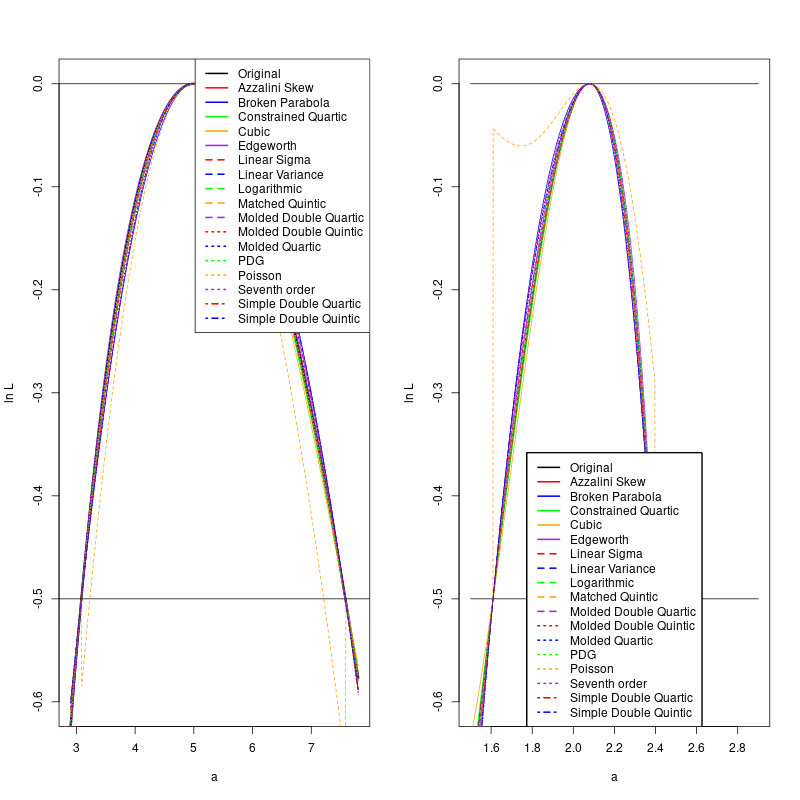}}
    \caption{\label{fig:manylikelihoodscloseup}
        The central regions for Figures \ref{fig:Like1} and \ref{fig:Like2}.  }
\end{figure}

All models appear to agree closely with each other and with the originals in the central region, and this is borne out  by closer examination, in Figure~\ref{fig:manylikelihoodscloseup}.  Outside the central region the models fall into two classes: those like the PDG model, the molded quartic and the matched quintic, which assume quadratic behaviour outside, and those which do not.  Considering the general form as a Gaussian with varying sigma, it is very counterintuitive to assume that sigma varies in the central region but is constant outside it. On the other hand, if it is varying, one has to guess correctly whether it is increasing or decreasing.

On the basis of these two examples it looks as if the two Bartlett forms generally behave well.  There is not a lot to choose between them, but one will not go far wrong using the linear variance with the linear sigma as a sanity check, though the other way round would be quite acceptable --- particularly if the measurement is basically a Poisson process. 
Other models may be useful in particular applications.

\subsection{Combination of results using likelihood}

Likelihood-based errors are greatly used in the combination of results. The combination of errors will be dealt with in the Section~\ref{subsec:LCOE}.

If one has several results measuring the same quantity, then the total likelihood is the product of the individual likelihoods, and the log likelihood is just the sum of the individual log likelihoods.  From that, one can read off the overall $\hat a$ and the $\Delta \ln L=-\half$ errors.  If we do not know the full forms for the likelihoods, but are only given the peak values and errors, then we can use the same approach, but modelling the likelihoods by one of the methods above.  In general a numerical maximisation can be used to give a solution.  For the two Bartlett models this can be done efficiently: the likelihood is
\begin{equation}
    \ln L(a) = - \half \sum_i \left({a-\hat a_i \over \sigma_i(a)}\right)^2,
\end{equation}
where $\hat a_i$ are the individual maximum likelihood estimates of $a$.  For the linear sigma form, differentiating and setting to zero leads to
\begin{equation}
    \hat a \sum_i w_i = \sum_i \hat a_i w_i \qquad {\rm with} \qquad w_i={\sigma_i \over (\sigma_i+\sigma'_i(a-\hat a_i))^3}.
\end{equation}
The equivalent for the linear variance model is
\begin{equation}
    \hat a \sum_i w_i = \sum_i w_i (\hat a_i-{V'_i \over 2 V_i}(\hat a-\hat a_i)^2)
    \qquad {\rm with} \qquad w_i={V_i \over (V_i+V'_i(a-\hat a_i))^2}.
\end{equation}
Solutions can be found by iteration. In all cases one can then find the errors on the final result by numerical root finding, to discover where the log likelihood falls by $\half$ from its peak value.

\begin{figure}[t]
    \centerline{\includegraphics[width=8 cm]{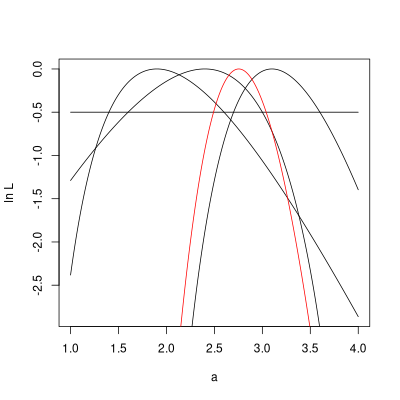}}
    \caption{\label{fig:comb1}
        Combining three asymmetric errors (black solid curves) to yield a result (red solid curve).}
\end{figure}

An illustration is shown in Figure~\ref{fig:comb1}.  The 3 values $\AE 1.9 0.7 0.5 $, $\AE 2.4 0.6 0.8 $ and $\AE 3.1 0.5 0.4 $ are combined (using the linear variance model) to give $\AE 2.754 0.286 0.263 $.  For the linear sigma model the result is $\AE 2.758 0.293 0.272 $ and the plot is identical to the eye.  Indeed other models give very similar results, as shown in Table~\ref{tab:combineresults}. The likelihood plots all look very similar to Figure~\ref{fig:comb1}, apart from that using the cubic model.

\begin{table}[p]
\centering
\begin{tabular}{|c|c||c|c|}
\hline 
& & & \\[-1ex]
Model & Result & Model & Result \\[1ex]
\hline 
& & & \\[-1ex]
{Cubic} & fails 
 & Truncated cubic & $ \AE 2.792  0.342  0.306 $ \\[1ex]
Broken parabola  & $ \AE 2.703 0.301 0.301 $ 
 & Symmetrized parabola & $ \AE 2.666  0.321  0.321 $  \\[1ex]
Constrained quartic & $ \AE 2.765  0.303  0.285 $ 
  & Molded quartic & $ \AE 2.721  0.246  0.240 $ \\[1ex]
Matched quintic & $ \AE 2.728  0.290  0.300 $ 
  & Interpolated 7th degree & $ \AE 2.702  0.301  0.296 $   \\[1ex]
Simple double quartic & $ \AE 2.720  0.289  0.299 $ 
  & Molded double quartic & $ \AE 2.718  0.286  0.298 $ \\[1ex]
Simple double quintic & $ \AE 2.702  0.301  0.298 $ 
  & Molded double quintic & $ \AE 2.701  0.302  0.297 $ \\[1ex]
Conservative spline(0.05) & $ \AE 2.720  0.291  0.292 $ 
  & Conservative spline(0.1) & $ \AE 2.732  0.285  0.287 $   \\[1ex]
Conservative spline(0.15) & $ \AE 2.738  0.285  0.281 $ 
  & Conservative spline(0.20) & $ \AE 2.742  0.287  0.281 $   \\[1ex]
Conservative spline(max) & $ \AE 2.742  0.287  0.282 $ 
  & Log logistic beta & $ \AE 2.763  0.336  0.366 $ \\[1ex]
Logarithmic & $ \AE 2.755  0.288  0.266 $ 
  & Generalised Poisson & $ \AE 2.753  0.283  0.258 $ \\[1ex]
Variable sigma & $ \AE 2.758  0.293  0.272 $ 
  & Variable variance & $ \AE 2.754  0.286  0.263 $   \\[1ex]
Variable log sigma & $ \AE 2.743  0.289  0.291 $ 
  & Molded cubic log sigma & $ \AE 2.702  0.301  0.297 $   \\[1ex]
Quintic log sigma & $ \AE 2.702  0.301  0.296 $ 
  & PDG & $ \AE 2.726  0.273  0.309 $   \\[1ex]
Skew normal & $ \AE 2.749  0.293  0.285 $ & & \\[1ex]
\hline
&&& \\[-1ex]
Railway Gaussian & $ \AE 2.753  0.277  0.250 $ 
  & Double cubic Gaussian & $ \AE 2.731  0.294  0.316 $ \\[1ex]
Symmetric beta Gaussian(1,10) & $ \AE 2.745  0.282  0.314 $ 
  & Symmetric beta Gaussian(1,30) & $ \AE 2.751  0.281  0.260 $   \\[1ex]
Symmetric beta Gaussian(4,10) & $ \AE 2.701  0.309  0.309 $ 
& Symmetric beta Gaussian(4,30) &  $ \AE 2.748  0.286  0.278 $ \\[1ex]
QVW Gaussian & $ \AE 2.741  0.290  0.297 $ 
  & Johnson system & --   \\[1ex]
Log normal & $ \AE 2.755  0.288  0.266 $ & & \\[1ex]
\hline
\end{tabular}
\caption{\label{tab:combineresults} 
            Results of combining the same 3 results\change{, $\AE 1.9 0.7 0.5 $, $\AE 2.4 0.6 0.8 $ and $\AE 3.1 0.5 0.4 $,} {} according to different models.}
\end{table}

It is worth pointing out how the combined error depends on the values being combined, as well as their quoted errors. If one of the values in the combination is changed, this affects not only the combined central value but also the quoted error. In  the symmetric case the quoted errors are independent of the values.

\subsection{Goodness of fit with likelihood}

According to Wilks' theorem, under the null hypothesis the improvement in likelihood due to the introduction of $n$ extra parameters is such that $-2 \Delta \ln L$ is distributed according to a $\chi^2$ distribution.  On that basis, the change in the log likelihood when using only 1 free parameter (the combined result) rather then $n$ fitted values (the individual results) gives a measure of goodness of fit such that $-2 \Delta \ln L$  is $\chi^2$ with $n-1$ degrees of freedom.

Wilks' theorem holds only in the \change {asymptotic limit,} {limit of large $N$ }  and provided the first model \change{is contained by} {contains } the second.  The first point depends clearly on the number and precision of the results being combined.  So this needs to be investigated empirically.

\begin{figure}[t]
    \centerline{\includegraphics[width=8 cm]{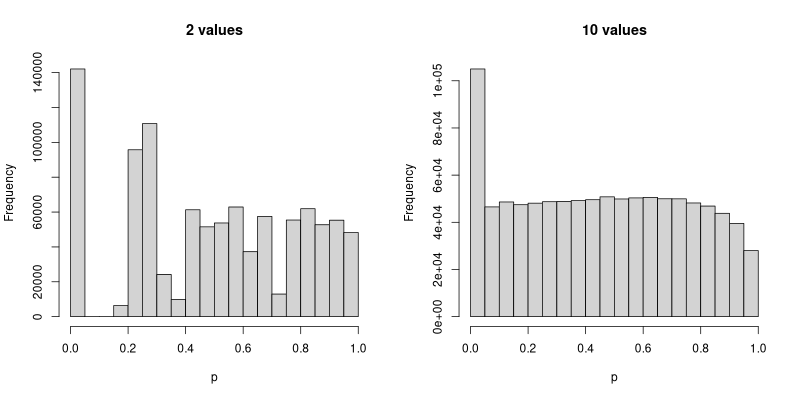}}
    \caption{\label{fig:chisq}
        Goodness of fit distributions for the simulations described in the text from combining 2 and 10 similar results.}
\end{figure}

As an illustration, we have performed a simulation in which two values were drawn from a Poisson distribution of mean 5.0.  These were encoded as two values with asymmetric errors (using the $\Delta \ln L=-\half$ prescription), which were then combined with the linear variance model.  The total likelihood at the combined best value was converted to a $\chi^2$ using Wilks' theorem.  This was repeated many times.  If the assumptions are valid this quantity will be distributed according to a $\chi^2$ distribution with one degree of freedom.  We actually show, in the left hand plot of Figure~\ref{fig:chisq}, the histogram of the corresponding p-value, as this is easier to interpret: for a true $\chi^2$ distribution this histogram would be flat.  It is not flat but it does its best, given that the space of possible results is discrete (the observations must be integer).  The right hand plot shows the result of a similar simulation, combining 10 such values at a time, and  the outcome has almost (but not quite) the ideal flat shape. 

On examination, the fact that the shape is not quite flat is caused by a small overall positive shift in the $-2\Delta \ln L$ distribution, as compared to the ideal $\chi^2$: this produces the spike at small values and the dip at large ones. This can be ascribed to the fact that a Poisson of mean 5 is not Gaussian.  If the simulations are repeated for means of 10, 20 and 30 then the distribution improves, becoming apparently  perfectly flat. 


It therefore appears that $\Delta \ln L$ --- which emerges as part of the combination procedure --- can be used as a criterion for the acceptability of the combination, though it should not be treated as an exact quantity as the conditions for the validity of Wilks' theorem  may not be met.

\subsection{Combination of errors with likelihood}
\label{subsec:LCOE}

As discussed earlier, one expects that most manipulation of errors from likelihoods will be concerned with the  combination of results. However there will be cases where the combination of errors is required.  If we had full knowledge of the likelihood we would use the profiling technique to obtain the error on the combination. 
Instead we do this as best we can using one of the models for the shapes of the asymmetric likelihoods.

\begin{figure}[t]
    \centerline{\includegraphics[width=12 cm]{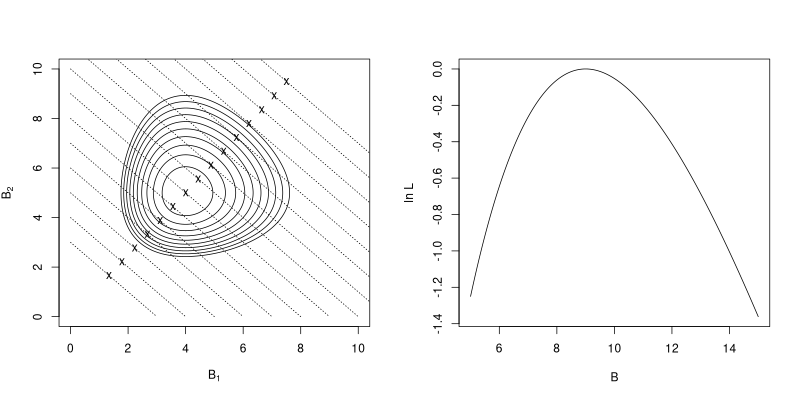}}
    \caption{\label{fig:lnLerrors}
        Errors from profiled likelihoods, from 2D (left) to 1D (right). }
\end{figure}

The concept is shown in an example in Figure~\ref{fig:lnLerrors}.  Some total background $B$ is the sum of $B_1$ and $B_2$ which were measured as 4 and 5 Poisson counts.  The dotted lines show contours of constant $B=B_1+B_2$, and for each value of $B$ the maximum $\ln L$ is found, indicated by the crosses. These points give the combined likelihood curve, shown on the right, from which the $\Delta \ln L=-\half$ errors may be read off.

If we have many variables, $\{a_1 \dots a_n\}$ and are concerned with the form of the maximum $\ln L(u)=\sum_i \ln L_i(a_i)$ for a given $u=a_1+a_2\cdots +a_n$, this can be found using Lagrange's method of undetermined multipliers 
$\sum {\partial \ln L_i \over \partial a_i} + \lambda {\partial u \over \partial a_i}=0$.  In our simple case 
${\partial u \over \partial a_i}$ is just 1, and we can write ${\partial \ln L_i \over \partial a_i}=
-{a_i \over w_i}$, where $w_i$ is approximately constant, as the form is roughly parabolic. This gives $a_i=\lambda w_i$ and the constraint reads $u=\lambda \sum w_i$,
\begin{equation} 
    a_i = u { w_i \over \sum_i w_i}  \qquad w_i = - \left( { 1 \over a_i} {\partial \ln L_i \over \partial a_i}\right)^{-1}.
    \label{eq:Lcoe}
\end{equation}
%
For the linear sigma model, $w_i={\left(\sigma_i + \sigma'_i a_i\right)^3 \over  \sigma_i}$, for the linear variance model $w_i={\left( V_i+a_iV'_i \right)^2 \over 2 V_i + V'_i a_i}$, and similarly for other models. If the cancellation \change{of the $a_i$ factor} {} cannot be done algebraically and numerical evaluation gives 0/0 at the peak, one can use $w=-\left( {\partial^2 L \over \partial a^2}\right)^{-1}$ for small $a$.

This is needed to read off the $\Delta \ln L=-\half$ errors. This can be done by starting at the peak, $u=0$, for which all $a_i$ are zero, and increasing $u$ in small steps, evaluating the $w_i$, and iterating Equation~\ref{eq:Lcoe}  to convergence, and continuing until $\Delta \ln L =-\half$ to find $\sp$. The process is then repeated for $\sm$.

\thispagestyle{empty}

\begin{table}[p]

{\scriptsize
\vspace{-3cm}
\begin{tabular}{| l | p{0.13\linewidth} p{0.13\linewidth} p{0.13\linewidth} 
                      p{0.13\linewidth} p{0.13\linewidth} p{0.13\linewidth} |} 
\hline 
&&&&&& \\[-1ex]
Inputs  &  Broken parabola  &  Truncated cubic  &  Constrained quartic  &  Molded quartic  &  Matched quintic  &  Interpolated 7th degree  \\
\vbox to 13 pt{} 4 5                &  $\AE 9 3.488 2.549 $  &   $\AE 9 3.218  2.620 $   &  $\AE 9 3.272  2.635 $  &  $\AE 9 3.439  2.678 $  &  $\AE 9 3.283  2.590 $  &  $\AE 9 3.425  2.558 $  \\
\vbox to 13 pt{} 3 6                &  $\AE 9 3.483 2.556 $  &  $\AE 9 3.224  2.620 $   &  $\AE 9 3.274  2.635 $   &  $\AE 9 3.433  2.676 $  &  $\AE 9 3.275  2.594 $  &  $\AE 9 3.402  2.567 $  \\
\vbox to 13 pt{} 2 7                &  $\AE 9 3.471 2.572 $  &   $\AE 9 3.238  2.622 $ &  $\AE 9 3.278  2.636 $  &  $\AE 9 3.417  2.670 $  &  $\AE 9 3.264  2.602 $  &  $\AE 9 3.345  2.583 $  \\
\vbox to 13 pt{} 1 8                &  $\AE 9 3.450 2.601 $  &  $\AE 9 3.264  2.628 $   &  $ \AE 9 3.290  2.640 $  &  $\AE 9 3.368  2.659 $  &  $\AE 9 3.269  2.617 $  &  $\AE 9 3.279  2.608 $  \\
\vbox to 13 pt{} 3 3 3              &  $\AE 9 3.603 2.453 $  &  $\AE 9 3.105  2.566 $  &  $\AE 9 3.206  2.594 $  &  $\AE 9 3.500  2.669 $  &  $\AE 9 3.161  2.523 $  &  $\AE 9 3.356  2.477 $  \\
\vbox to 13 pt{} 1 1 1 1 1 1 1 1 1  &  $\AE 9 4.073 2.095 $  & $\AE 9 2.605  2.284 $ &  $\AE 9 2.894  2.386 $  &  $\AE 9 3.389  2.508 $  &  $\AE 9 2.552  2.217 $  &  $\AE 9 2.576  2.157 $  \\[1ex]
\hline
&&&&&& \\[-1ex]
Inputs  &  Double quartic  &  Molded~double quartic  &  Double quintic  &  Molded double quintic  &  Conservative spline (0.05)  &  Conservative spline (0.10) \\
\vbox to 13 pt{} 4 5                &  $\AE 9 3.383  2.600 $  &  $\AE 9 3.435  2.617 $  &  $\AE 9 3.471  2.558 $  &  $\AE 9 3.479  2.561 $  &  $\AE 9 3.426 2.592 $  &  $\AE 9 3.379  2.618 $  \\
\vbox to 13 pt{} 3 6                &  $\AE 9 3.376  2.606 $  &  $\AE 9 3.432  2.623 $  &  $\AE 9 3.461  2.568 $  &  $\AE 9 3.473  2.573 $  &  $\AE 9 3.423 2.596 $  &  $\AE 9 3.377  2.620 $  \\
\vbox to 13 pt{} 2 7                &  $\AE 9 3.362  2.617 $  &  $\AE 9 3.428  2.635 $  &  $\AE 9 3.438  2.588 $  &  $\AE 9 3.460  2.595 $  &  $\AE 9 3.416 2.606 $  &  $\AE 9 3.374  2.626 $  \\ 
\vbox to 13 pt{} 1 8                &  $\AE 9 3.342  2.636 $  &  $\AE 9 3.421  2.651 $  &  $\AE 9 3.393  2.619 $  &  $\AE 9 3.439  2.627 $  &  $\AE 9 3.403 2.624 $  &  $\AE 9 3.367  2.636 $  \\ 
\vbox to 13 pt{} 3 3 3              &  $\AE 9 3.378  2.552 $  &  $\AE 9 3.506  2.589 $  &  $\AE 9 3.536  2.481 $  &  $\AE 9 3.575  2.492 $  &  $\AE 9 3.489 2.526 $  &  $\AE 9 3.405  2.569 $  \\  
\vbox to 13 pt{} 1 1 1 1 1 1 1 1 1  &  $\AE 9 3.250  2.380 $  &  $\AE 9 3.888  2.509 $  &  $\AE 9 3.517  2.245 $  &  $\AE 9 3.963  2.306 $  &  $\AE 9 3.727 2.263 $  &  $\AE 9 3.482  2.355 $  \\[1ex]
\hline
&&&&&& \\[-1ex]
Inputs  &  Conservative spline (0.15)  &  Conservative spline (0.20)  &  Conservative spline (max)  &  Log logistic-beta  &  Logarithmic  &  Generalized Poisson  \\
\vbox to 13 pt{} 4 5                &  $\AE 9 3.342  2.628 $  &  $\AE 9 3.316  2.627 $  &  $\AE 9 3.315  2.627 $  &  $\AE 9 3.177  2.545 $  &  $\AE 9 3.325  2.663 $  &  $\AE 9 3.342  2.676 $  \\
\vbox to 13 pt{} 3 6                &  $\AE 9 3.342  2.629 $  &  $\AE 9 3.319  2.629 $  &  $\AE 9 3.313  2.628 $  &  $\AE 9 3.192  2.555 $  &  $\AE 9 3.325  2.663 $  &  $\AE 9 3.342  2.676 $  \\
\vbox to 13 pt{} 2 7                &  $\AE 9 3.342  2.632 $  &  $\AE 9 3.322  2.632 $  &  $\AE 9 3.309  2.631 $  &  $\AE 9 3.221  2.576 $  &  $\AE 9 3.325  2.663 $  &   $\AE 9 3.345  2.583 $  \\ 
\vbox to 13 pt{} 1 8                &  $\AE 9 3.341  2.640 $  &  $\AE 9 3.325  2.640 $  &  $\AE 9 3.306  2.638 $  &  $\AE 9 3.264  2.609 $  &  $\AE 9 3.326  2.663 $  &  $\AE 9 3.342  2.676 $  \\ 
\vbox to 13 pt{} 3 3 3              &  $\AE 9 3.341  2.585 $  &  $\AE 9 3.292  2.584 $  &  $\AE 9 3.272  2.582 $  &  $\AE 9 3.071  2.459 $  &  $\AE 9 3.308  2.649 $  &  $\AE 9 3.342  2.676 $  \\  
\vbox to 13 pt{} 1 1 1 1 1 1 1 1 1  &  $\AE 9 3.310  2.383 $  &  $\AE 9 3.187  2.386 $  &  $\AE 9 3.025  2.370 $  &  $\AE 9 2.731  2.178 $  &  $\AE 9 3.211  2.572 $  &  $\AE 9 3.342  2.676 $   \\[1ex]
\hline
&&&&&& \\[-1ex]
Inputs  &  Linear sigma &  Linear variance  &  Linear log sigma &  Molded cubic log sigma  &  Quintic log sigma &  PDG method  \\
\vbox to 13 pt{} 4 5                &  $\AE 9 3.310  2.653 $  &  $\AE 9 3.333  2.668 $  &  $\AE 9 3.322  2.624 $  &  $\AE 9 3.475  2.561 $  &  $\AE 9 3.443  2.562 $  &  $\AE 9 3.310  2.653 $  \\
\vbox to 13 pt{} 3 6                &  $\AE 9 3.310  2.653 $  &  $\AE 9 3.333  2.668 $  &  $\AE 9 3.318  2.626 $  &  $\AE 9 3.467  2.572 $  &  $\AE 9 3.427  2.571 $  &  $\AE 9 3.311  2.653 $ \\
\vbox to 13 pt{} 2 7                &  $\AE 9 3.311  2.653 $  &  $\AE 9 3.333  2.668 $  &  $\AE 9 3.311  2.630 $  &  $\AE 9 3.449  2.595 $  &  $\AE 9 3.390  2.588 $  &  $\AE 9 3.311  2.653 $  \\ 
\vbox to 13 pt{} 1 8                &  $\AE 9 3.313  2.654 $  &  $\AE 9 3.333  2.668 $  &  $\AE 9 3.304  2.638 $  &  $\AE 9 3.416  2.628 $  &  $\AE 9 3.315  2.613 $  &  $\AE 9 3.313  2.654 $  \\ 
\vbox to 13 pt{} 3 3 3              &  $\AE 9 3.278  2.630 $  &  $\AE 9 3.323  2.659 $  &  $\AE 9 3.278  2.582 $  &  $\AE 9 3.555  2.491 $  &  $\AE 9 3.437  2.487 $  &  $\AE 9 3.278  2.630 $  \\  
\vbox to 13 pt{} 1 1 1 1 1 1 1 1 1  &  $\AE 9 3.098  2.500 $  &  $\AE 9 3.269  2.610 $  &  $\AE 9 2.971  2.381 $  &  $\AE 9 3.735  2.320 $  &  $\AE 9 2.910  2.202 $  &  $\AE 9 3.098  2.500 $  \\[1ex]
\hline
&&&&&& \\[-1ex]
Inputs  &  Skew normal  &  Railway Gaussian  &  Double cubic Gaussian  &  Symmetric beta Gaussian(1,10)  &  Symmetric beta Gaussian(1,30)  &  Symmetric beta Gaussian(4,10)  \\
\vbox to 13 pt{} 4 5                &  $\AE 9 3.302  2.628 $  &  $\AE 9 3.355  2.705 $  &  $\AE 9 3.312  2.579 $  &  $\AE 9 3.330  2.626 $  &  $\AE 9 3.350  2.678 $  &  $\AE 9 3.347  2.518 $  \\
\vbox to 13 pt{} 3 6                &  $\AE 9 3.304  2.630 $  &  $\AE 9 3.356  2.707 $  &  $\AE 9 3.291  2.584 $  &  $\AE 9 3.328  2.627 $  &  $\AE 9 3.350  2.679 $  &  $\AE 9 3.319  2.528 $  \\
\vbox to 13 pt{} 2 7                &  $\AE 9 3.308  2.635 $  &  $\AE 9 3.359  2.711 $  &  $\AE 9 3.254  2.597 $  &  $\AE 9 3.322  2.630 $  &  $\AE 9 3.350  2.680 $  &  $\AE 9 3.261  2.548 $  \\ 
\vbox to 13 pt{} 1 8                &  $\AE 9 3.318  2.645 $  &  $\AE 9 3.366  2.720 $  &  --                  &  $\AE 9 3.310  2.634 $  &  $\AE 9 3.351  2.683 $  &  --  \\ 
\vbox to 13 pt{} 3 3 3              &  $\AE 9 3.271  2.591 $  &  $\AE 9 3.373  2.740 $  &  $\AE 9 3.174  2.509 $  &  $\AE 9 3.303  2.587 $  &  $\AE 9 3.357  2.683 $  &  $\AE 9 3.195  2.399 $  \\  
\vbox to 13 pt{} 1 1 1 1 1 1 1 1 1  &  $\AE 9 3.166  2.444 $  &  $\AE 9 3.543  3.032 $  &  --                     &  $\AE 9 3.039  2.350 $  &  $\AE 9 3.409  2.740 $  &  --  \\[1ex]
\hline
&&&&&& \\[-1ex]
Inputs  &  Symmetric beta Gaussian(4,30)  &  QVW Gaussian  &  Johnson system  &   Log normal  & &  \\
\vbox to 13 pt{} 4 5                &  $\AE 9 3.341  2.652 $  &  $\AE 9 3.330  2.620 $  &  --  &  $\AE 9 3.325  2.663 $  & &  \\
\vbox to 13 pt{} 3 6                &  $\AE 9 3.339  2.653 $  &  $\AE 9 3.326  2.623 $  &  --  &  $\AE 9 3.325  2.663 $  &  &  \\
\vbox to 13 pt{} 2 7                &  $\AE 9 3.336  2.655 $  &  $\AE 9 3.318  2.628 $  &  --  &  $\AE 9 3.325  2.663 $  &   &  \\ 
\vbox to 13 pt{} 1 8                &  $\AE 9 3.332  2.659 $  &  $\AE 9 3.308  2.637 $  &  --  &  $\AE 9 3.326  2.663 $  & &  \\ 
\vbox to 13 pt{} 3 3 3              &  $\AE 9 3.330  2.634 $  &  $\AE 9 3.293  2.578 $  &  --  &  $\AE 9 3.308  2.649 $  &  &  \\  
\vbox to 13 pt{} 1 1 1 1 1 1 1 1 1  &  $\AE 9 3.238  2.546 $  &  $\AE 9 3.013  2.374 $  &  --  &  $\AE 9 3.211  2.572 $  &   &  \\[1ex]
\hline
\end{tabular}}
\caption{\label{tab:coelike} Examples of combination of errors using likelihoods. The `correct' answer is ${9}^{+3.342}_{-2.676}$.  Where models cannot accommodate the asymmetry (the  $N=1$ Poisson is strongly asymmetric and breaks several models, and the Johnson System cannot handle even moderate asymmetries) no result can be given.}
\end{table} 


\section{Examples and case studies}
\subsection {Examples}

Here we present some relevant examples where the true behaviour behind the simple three numbers of the result(s) is known, and see how well the various models do when their outcome is compared with the exact behaviour.

\subsubsection {A signal with several backgrounds: combination of errors using likelihoods}

Suppose an experiment counts some number of events.  To extract the signal size, the number of background events is to be subtracted.  There are several such background sources, and for simplicity we can suppose that each has been
measured by an ancillary experiment, running the apparatus over the same time, so they do not need to be scaled.

If there are two such backgrounds and the ancillary experiments give 4 and 5 counts, then we could, using $\Delta \ln L=-\half$ errors, report these as $\AE 4 2.346 1.682 $ and $\AE 5 2.581 1.916 $.  In this case we happen to know that the two results can be combined to give a total background count of 9, with errors ${}^{+3.342}_{-2.676}$.  But  we might be  unaware of this fact,  and just been given the raw numbers. We could then combine the uncertainties using, for example, the linear variance method, and obtain errors of ${}^{+3.333}_{-2.668}$, in excellent agreement with the true value. 

The same total background measurement of 9 might be obtained in various ways, such as 3+6 or 3+3+3 or even 9 separate measurements of 1 event. The same combination procedure also gives $\EA 3.333 2.668 $$ $ for the 3+6 case, and changes only slightly to $\EA 3.323 2.659 $ for 3+3+3 and $\EA 3.269 2.610 $ for 1+1+1+1+1+1+1+1+1. Outcomes for other models and for other inputs are shown in Table~\ref{tab:coelike}, and show remarkable agreement with the true values for the errors.

\subsubsection{A lifetime measurement: combination of results using likelihoods}
\label{sec:lifetime}

Suppose an experiment measures a lifetime $\tau$ from 3 decays. The values happen to be   1.241, 0.592, and 0.988, in some time units.  (These numbers were generated from an exponential distribution \change{with lifetime 1.0} {} , ensuring that this example is realistic.)  Maximising $\ln L$, which is \change{$\ln \prod  {\rm exp}(-t_i/\tau)/\tau$} {$\ln  {\rm exp}(-t/\tau)/\tau$} ,  and using $\Delta \ln L=-\half$  gives  a result $\tau=\AE 0.940 0.841 0.385 $.   
 
The experiment then measures 3 more lifetimes, which happen to be 0.834,  2.964, and  0.176, which combined on their own give $\AE 1.325 1.184 0.542 $.

If we combine all 6 values we get the combined result $\AE  1.1325  0.6225 0.3598 $.  
The likelihood used is the product of 6 separate exponential likelihoods, so we are using the full information.

Now suppose we take the two partial results separately (i.e.~just the value and $\pm$ errors, as quoted above) and combine them using the linear-sigma model.  That gives $\AE 1.1323 0.6213 0.3604 $.  The linear-variance model gives $\AE 1.1318 0.6249 0.3577 $.  Both these agree well with the full information  value, both in the central value and the quoted errors.  Both models (which know nothing about the fact that this was a lifetime measurement with an exponential likelihood) give a result very close to the full-information one.

\begin{table}[t] 
\begin{center} 
\begin{tabular}{ l c c c } 
\hline 
&&& \\[-1ex]
Model & Value & $\sp$ & $\sm $ \\ 
Full information &  1.1325  &  0.6225  &  0.3598 \\[1ex] 
\hline 
&&& \\[-1ex]
Poisson  &  1.1285  &  0.6282  &  0.3533 \\ 
Cubic  &  1.1074  &  NA  &  0.3433 \\ 
Quartic  &  1.1335  &  0.6243  &  0.3637 \\ 
Logarithmic  &  1.1319  &  0.6237  &  0.3586 \\ 
Linear sigma  &  1.1323  &  0.6213  &  0.3604 \\ 
Linear variance  &  1.1318  &  0.6249  &  0.3577 \\ 
PDG method  &  1.1318  &  0.6249  &  0.3613 \\ 
Skew normal  &  1.1494  &  0.6146  &  0.3764 \\[1ex] 
\hline 
&&& \\[-1ex]
Wrong & 1.1325  &  0.7261  &  0.3326 \\[1ex] 
\hline 
\end{tabular} 
\end{center} 
\caption{\label{tab:life} Results from combining two lifetime experiments.} \end{table} 
These results are shown in Table~\ref{tab:life} together with those from some other models. The Edgeworth model cannot cope with asymmetries this large, and the cubic does not find a positive error due to the turnover in the curve. It can be seen that the logarithmic and PDG models also do well, the quartic is less good, and the Poisson and 
skew normal models  perform relatively poorly. The final row, labelled ``wrong'', takes the mean of the two results and combines the positive and negative errors separately in quadrature, and it can be seen that its error estimates are seriously discrepant.

\subsubsection{Combining Poisson counts: combination of results using likelihoods}

Suppose a counting experiment sees 5 events in an hour. The result is quoted, using $\Delta \ln L=-\half$ errors even though this is a case where the full Neyman errors could be given, as $\AE 5 2.581 1.916 $.  This continues for another hour and, as it happens, 5 events are again seen.  The total gives a result $\AE 10 3.504 2.838 $ and with the knowledge we have of the way the experiment has been done, we can estimate the number of events per hour by dividing this by 2 to get $\AE 5 1.752 1.419 $.

But it could be that this knowledge is suppressed, and we are just presented with two estimates (which happen to be the same) and we have to combine them as best we can.  If we do this using the linear variance method, the result is $\AE 5.000 1.747 1.415 $.  This is an excellent match to the ideal value, with the errors differing only in the 4th significant figure. Using linear sigma we would get $\AE 5.000 1.737 1.408 $ which is also very good.

\begin{table}[b]
\begin{centering}
\begin{tabular}{ c c c c c c }
$\hat a_1$ & $\hat a_2 $  & Linear $\sigma$ & Linear $V$ & Skew normal & Quartic\\
\\ \hline 
&&&&& \\[-1ex]
$ \AE  5 2.581 1.916 $  & $ \AE  5 2.581 1.916  $  & $ \AE  5.000 1.737 1.408  $  & $ \AE  5.000 1.748 1.415  $  & $ \AE  5.000 1.732 1.395  $  & $ \AE  5.000 1.719 1.399  $  \\[1ex] 
  $ \AE  6 2.794 2.128 $  & $ \AE  4 2.346 1.682  $  & $ \AE  4.998 1.778 1.432  $  & $ \AE  5.000 1.759 1.425  $  & $ \AE  5.009 1.766 1.460  $  & $ \AE  4.992 1.815 1.440  $  \\[1ex] 
  $ \AE  7 2.989 2.323 $  & $ \AE  3 2.080 1.416  $  & $ \AE  5.038 1.937 1.530  $  & $ \AE  5.009 1.794 1.456  $  & $ \AE  5.002 1.744 1.594  $  & $ \AE  5.107 2.147 1.651  $  \\[1ex] 
  $ \AE  8 3.171 2.505 $  & $ \AE  2 1.765 1.102  $  & $ \AE  5.401 2.368 1.826  $  & $ \AE  5.054 1.856 1.516  $  & $ \AE  4.689 1.635 1.589  $  & $ \AE  5.282 1.277 1.707  $  \\[1ex] 
  $ \AE  9 3.342 2.676 $  & $ \AE  1 1.358 0.698  $  & $ \AE  7.348 3.149 2.549  $  & $ \AE  5.201 1.942 1.605  $  & $ \AE  3.633 1.426 1.414  $  & $ \AE  3.209 0.594 0.828  $  \\[1ex] 
   \hline 
 
\end{tabular}
\caption{\label{tab:errors} 
            Combining results from two samples from the same Poisson distribution.  The exact result is $\AE 5.000 1.752 1.419 $.}
\end{centering}
\end{table}

Table~\ref{tab:errors} shows this and also the results obtained from other possible pairs of results with the same sum and thus the same ideal answer.  It can be seen that the technique, especially for the linear variance model, works very well.  It is worth pointing out that the slightly larger discrepancies in the final two rows arise from rather unlikely experimental circumstances -- the probability of 10 events being split 9:1 or even 8:2 is small, and this shows up in a poor goodness of fit.

\subsubsection{Strength of a known mass peak: models of a known likelihood}

\begin{figure}[t]
    \centerline{\includegraphics[width=12 cm]{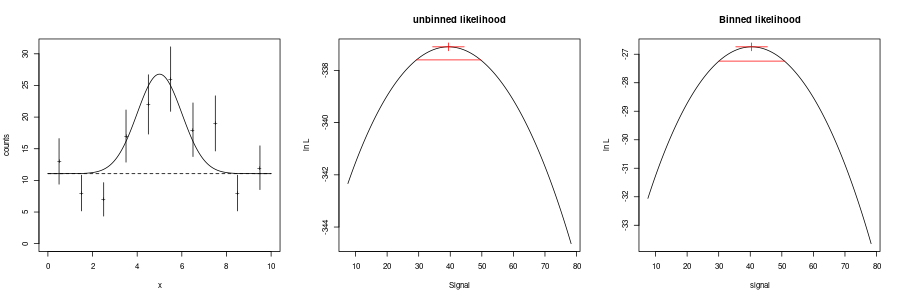}}
    \caption{\label{fig:example1}
        Fitting a peak with known shapes: the data (first panel) and the likelihoods (second and third panels), unbinned and binned, as a function of supposed signal strength, showing the peak and the $\Delta \ln L=-\half$ errors. }
\end{figure}

A common analysis method is to take measurements of a reconstructed particle mass, and fit them to a total of signal and background shapes where all parameters are known except for the normalisations.  Figure~\ref{fig:example1} shows such a situation, where the background is known to be flat and the signal is a Gaussian of mean 5.0 and standard deviation 1.0.  One may then, as appropriate, work with the raw (unbinned) likelihood $\ln L = \sum_i \ln{(N_B B(x_i)+N_S S(x_i))}$, where the sum is over all events, or the binned likelihood $\ln L=\sum_j (F_j \ln n_j - F_j)$, where $n_j$ is the contents of bin $j$ and $F_j$ is the prediction.  The likelihood curves are shown in the second and third panels of Figure~\ref{fig:example1}.  In this example the unbinned likelihood gives a result $N_S=\AE 39.41 10.20 10.13  $, and the binned likelihood gives $N_S=\AE 40.43 10.40 10.35  $.

\begin{figure}[t]
    \centerline{\includegraphics[width=11 cm]{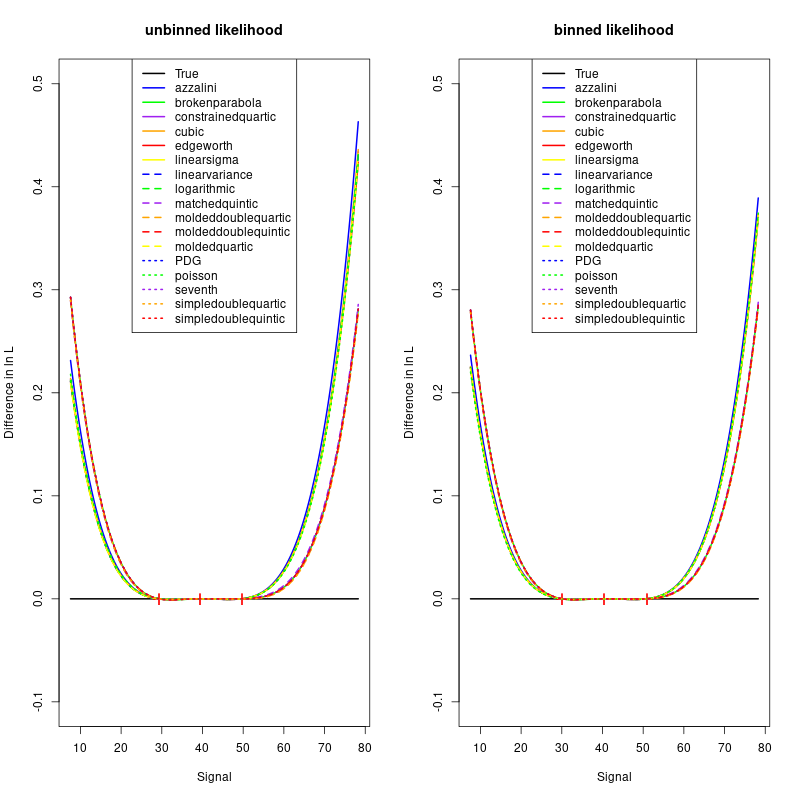}}
    \caption{\label{fig:example1a}
        Modelling the likelihoods in the case shown in Figure~\ref{fig:example1}, plotting the differences from the true curve. \change {The different models give very similar functional forms, 
        especially in the central region,
        making the point that results will be robust under choice of model, at least in this example. If the actual forms are shown, rather than the differences, the distinctions are even less apparent.} {} }
\end{figure}

From the quoted asymmetries one can model the log likelihoods and compare with the known true form. With these small asymmetries the modelling is good and the curves lie close to each other; it is more revealing to plot the difference between the modelled curve and the true one, as shown in Figure~\ref{fig:example1a}. 

Within the central region the modelling is excellent for all types except the Edgeworth form. Above and below the central region the agreement is not so good. Generally the modelled curves are too high, showing that they do not fall off fast enough, all with a very similar pattern. The exception (apart from the Edgeworth) is the PDG form, which is much too high below the peak, and too low above it.  The PDG form is Gaussian outside the central region: the signal form clearly falls off increasingly fast below the peak and more gently above.

\subsubsection{A product with asymmetric uncertainties}

The expected number of particle decays in some channel can be written as $N={\cal L} \sigma F$, where ${\cal L}$ is the integrated luminosity (or equivalent), $\sigma$ is the cross section\footnote{It is regrettable to use the same symbol for cross-sections and for standard deviations, but the use is entrenched.}, and $F$ is the branching fraction. $N$ is the ideal number and the actual number observed will be Poisson-distributed with mean $N$.  In a typical instance ${\cal L}$ will be accurately known, but $\sigma$ and $F$ may have asymmetric errors.  What should we quote as the error for the expected $N$ given, say, ${\cal L}=1000 {\rm pb}^{-1}, \sigma=\AE 12.3 0.4 0.5 {\rm pb}$, and $F=\AE 0.12 0.01 0.02 $?

The central value is $1000 \times 12.3\times 0.12=1476$ events. Writing $\sigma=\sigma_0+\Delta_\sigma, F=F_0+\Delta_F$ then, to leading order in a Taylor expansion, $N=N_0+({\cal L} F_0) \Delta_\sigma + ({\cal L} \sigma_0 ) \Delta_F$, and the appropriate function (pdf or $\ln L$) is found by combining the errors, scaled by ${\cal L} F_0$ and ${\cal L}\sigma_0 $ respectively.  For the linear sigma model this gives $\AE 1476 136 250 $. Using the linear variance model it is $\AE 1476 137 251 $. 

If, instead, one were calculating the cross section from some observed number of events using $\sigma={N  \over {\cal L} F}$ one would scale the errors on $N$ by ${1\over {\cal L}_0 F_0}$ and on $F$ by ${1 \over {\cal L}_0 F_0^2}$ and combine them.


\section{Likelihood or pdf? Variances or confidence?}

\subsection{Framework }

An error quoted on a measurement may denote the rms of a pdf or a 68\% central confidence region obtained from a likelihood. Previous sections have discussed the appropriate handling of these two separate cases. Now we consider the consequences if they are,
inadvertently or deliberately, confused.

As mentioned in Section~\ref{sec:beyondgaussian}, a pdf can give 68\% central limits, but they relate to the horizontal lines on the Neyman confidence belt, not the vertical ones, and the relationship is subtle.  In the example discussed in Figure~\ref{fig:confidencebelt} a pdf with positive skewness leads to a likelihood with negative asymmetry. 
Essentially the same problem occurs in the evaluation of confidence regions using the bootstrap.  A pdf is approximated by taking repeated samples from the data, and these can be used to give a confidence region for some quantity. However this is not the confidence region for the true quantity, and to use it as such ``amounts to looking up the wrong tables backwards'', as Peter Hall puts it \cite[p36]{Edgeworth} and  \cite[p928]{hall}.
The  proportional Gaussian  has a symmetric pdf, $p(x)={1 \over \sqrt{2 \pi} \alpha a} e^{-\half\left({x-a \over \alpha a}\right)^2}$, but a likelihood with positive asymmetry.
The Poisson has a positive asymmetry for both probability and likelihood, as its variance increases with the mean. So there is no easy equivalence between the 68\% confidence region obtained from the pdf and that from the likelihood. 

\subsection{What happens if you use the wrong formalism}

There may be occasions where one has to work with asymmetric errors quoted with no indication of whether they are obtained 
from pdfs or from likelihoods. 

As an example we consider the combination of errors from two results, both with error $\sp=2,\sm=1$. These are shown in Table~\ref{tab:mixedup} for several models.  Not surprisingly --- this is a large asymmetry --- the different models give a spread of different values for the combined $\sp$, and likewise for $\sm$.  But the pdf models are clustered separately from the $\ln L$ models.  The values of $\sp$ and $\sm$ for the combination under the assumption that these come from pdfs are both larger than those under the assumption that these come from likelihoods.  This discrepancy is not enormous, but it is larger than the differences between individual models.  (The difference between any of these and the `wrong' method of summing separately in quadrature is somewhat larger.)  The conclusion to be tentatively drawn \change {in this instance } { } is that if a pdf error is (wrongly) treated as a likelihood error, or vice versa, this makes a meaningful, though not enormous, difference. \change{But this should not be taken as being true in general, and individual cases will need to be justified.} {}


\begin{table}[t] 
 \begin{center} 
 \begin{tabular}{ l c c } 
 \hline 
 &  $\sp$ & $\sm $  \\ 
 Pdf models \\ \hline \\ 
Dimidiated &  2.636 & 1.651 \\ 
Distorted &  2.727 & 1.760 \\ 
Edgeworth &  2.656 & 1.586 \\ 
Railway &  2.715 & 1.775 \\ 
 \hline \\ ln$L$ models \\ \hline \\ 
Linear sigma &  2.467 & 1.526 \\ 
Linear variance &  2.562 & 1.562 \\ 
Cubic &  2.000 & 1.464 \\ 
Logarithmic &  2.530 & 1.550 \\ 
Constrained quartic &  2.363 & 1.492 \\ 
\hline \\ Wrong   
 & 2.828 & 1.414 \\ 
\hline 
 \end{tabular}
\end{center}
\caption { \label{tab:mixedup} Different outcomes from combining errors $ \EA 2 1 $ and $\EA 2 1 $. }  
\end{table}

Similarly Table~\ref{tab:mixedup2} shows the outcomes from combining two results: $\AE 1.0 2.0 1.0 $ and $\AE 2.0 2.0 1.0 $. 
Again, the result to be quoted depends on the choice of model, which is arbitrary (though the practitioner may choose to reject some models as being inappropriate for their use) so there is a spread.  There are also overall shifts between the pdf treatment and the $\ln L$ treatment, which is not arbitrary.  If you use the wrong treatment this will give wrong results, by an amount which is somewhat larger than the inescapable differences owing to the choice of model.


\begin{table}[ht] 
 \begin{center} 
 \begin{tabular}{ l c c c } 
 \hline 
 & {\rm Result} &  $\sp$ & $\sm $  \\ 
 \hline \\ ln$L$  models \\ \hline \\ 
Azzalini & 1.703 & 1.232 & 0.796 \\ 
Broken parabola & 1.800 & 1.166 & 0.892 \\ 
Constrained quartic & 1.676 & 1.239 & 0.784 \\ 
Cubic & 1.634 & 1.193 & 0.771 \\ 
Linear sigma & 1.673 & 1.244 & 0.760 \\ 
Linear variance & 1.668 & 1.255 & 0.738 \\ 
Logarithmic & 1.670 & 1.251 & 0.745 \\ 
Matched quintic & 1.655 & 1.232 & 0.809 \\ 
Molded double quartic & 1.729 & 1.217 & 0.808 \\ 
Molded double quintic & 1.767 & 1.188 & 0.856 \\ 
Molded quartic & 1.729 & 1.234 & 0.809 \\ 
PDG & 1.673 & 1.244 & 0.791 \\ 
Poisson & 1.661 & 1.262 & 0.720 \\ 
Seventh & 1.730 & 1.226 & 0.881 \\ 
Simple double quartic & 1.715 & 1.222 & 0.822 \\ 
Simple double quintic & 1.775 & 1.186 & 0.878 \\ 
\hline \\ Pdf models \\ \hline \\ 
Dimidiated & 1.703 & 1.318 & 0.825 \\ 
Distorted & 1.758 & 1.363 & 0.880 \\ 
Edgeworth & 1.732 & 1.328 & 0.793 \\ 
Railway & 1.744 & 1.357 & 0.888 \\ 
\hline 
 \end{tabular}
\end{center}
\caption { \label{tab:mixedup2} Different outcomes from combining results $ \AE 1.0 2.0 1.0 $ and $ \AE 2.0 2.0 1.0 $.}  
\end{table}

\subsection {A final result}
\label{final.result}

A typical final result is presented in the form $x^{+\sigma^+_1}_{-\sigma^-_1} {\ }^{+\sigma^+_2}_{-\sigma^-_2}
$ where ``the first error is statistical and the second is systematic". The first is thus, presumably, an error obtained from the likelihood to some final fit, and the second from some pdf-based uncertainty obtained from combining many separate sources.  In such cases one may still want to present a combined result: $ \AE x {\sp} {\sm}  $.

As a pedagogical example we consider the determination of the total strength (in Becquerel, or counts per second) of an isotropic  point source using a counter of intrinsic efficiency $\eta$.  If $n$ counts are measured in time $t$ then the strength $R$ is obtained from $n$ through a conversion factor $A$
\begin{equation}
    R=A n={{4 \pi \over S} {1 \over \eta t} n},
\end{equation}
where $S$ is the solid angle presented by the counter to the source.  

We suppose that $\eta$ and $S$ are well known, but that there is a Gaussian error $\sigma_t=0.1 s$ on the calibration of the time measurement $t$ which is systematic, i.e. described by a pdf, and gives an asymmetric error on $1/t$.   

Considering an example where $A=100, \eta=1, n=50$ and $t=1 s$, we would write $R={5000}^{+741}_{-674}{}^{+556}_{-455}$ where the first, `statistical', error comes from the Poisson errors on $n=50$ and the second `systematic' from the error on $1/t$.



If we, wrongly, treat the pdf error as if it were a likelihood error and combine the two errors this gives a combined result of $\AE 5000 906 828  $
(using the linear-sigma model: other models give near-identical values). 
If we, wrongly, treat the ln$L$ error as if it were a pdf and combine them,
this gives  $\AE 5000 909 831  $ (using the dimidiated model: other models give near-identical values.
There is little difference between the two outcomes.

As a validation we performed $10^7$ toy experiments,
generating values of $n$ from a Poisson distribution of mean 50 and values of $t$ from a Gaussian distribution of mean 1 and standard deviation 0.1, and applying the two prescriptions.
The pdf method gave a coverage of $69.074\pm 0.015$\%,
only slightly above the nominal 68.269\%, and
the lnL method was even closer with a coverage of $68.387\pm {0.015}$\%

It thus appears justifiable to combine (with caution) asymmetric
 `systematic' and `statistical' errors on a final result, treating them both either as one or the other.


\section{Conclusions}

We trust we have convinced any reader of this paper that asymmetric errors are not to be taken up lightly: if differences appear between the $\sp$ and $\sm$ values in some part of an analysis, they will introduce considerable complication both technically and conceptually. If these differences are small and can legitimately be ignored then our advice would be to do so, rather than to retain them out of misplaced diligence.  If the differences are large then the best course 
would be to report the full likelihood, at least for the parameter(s) of interest, or perhaps to provide future meta-analysts with some fuller description of the behaviour of the likelihood function.

If asymmetric errors cannot be avoided then they can be handled by the methods given.  You need to be clear whether they are likelihood errors or pdf errors, and whether you are combining results or combining errors.  You also need to choose models from among those given here,  or ones of your own, always taking more than one model to check the robustness of the result.  Experience should show which model(s) are appropriate for a particular type of problem.

This treatment assumes that the errors being considered are independent, and that the final result is a single quantity.
We hope to consider its extension to correlated inputs and outputs in a future publication.


\section{Acknowledgements}

We would like to thank our colleagues for many discussions and suggestions, (including a PHYSTAT informal review),  especially the PHYSTAT organisers: Sara Algeri, Olaf Behnke, Lydia Brenner, Louis Lyons (who suggested some of the examples) and Nick Wardle.  This paper arose out of a workshop at the Banff Center for Arts and Creativity in Alberta, Canada, and we are very happy to thank the Banff International Research Station (BIRS)  and its staff for their organisation and hospitality.  We are grateful to Hans Dembinski for helpful comments. 

ARB acknowledges that the publication was produced with funding from the Italian Ministry of University and Research under the Call for Proposals related to the scrolling of the final rankings of the PRIN 2022 call (Project title ``Latent variable models and dimensionality reduction methods for complex data'', Project No. 20224CRB9E, CUP C53C24000730006, PI Prof.~Paolo Giordani, Grant Assignment Decree No.~1401 adopted on 18.9.2024 by MUR).

IV acknowledges partial support by the US Department of Energy grant DE-SC0015592.


\appendix

\section{Modelling pdf errors}
\label{app:modelling-pdf-errors}

We present the details of possible models for near-Gaussian pdf functions.  The list is not exhaustive, and others, such as the Generalised Extreme Value distribution, can also be suggested \cite{Possolo}. Their motivations, and their technical challenges, vary widely. There is no universal `best' model and the choice is ultimately down to the user. Use of more than one model in any application  is strongly recommended as the differing results will give a feeling for the robustness of the whole procedure.

\subsection{The dimidiated Gaussian}
\label{sec:dimidiated}

\begin{figure}[ht]
    \centerline{\includegraphics[width=12 cm]{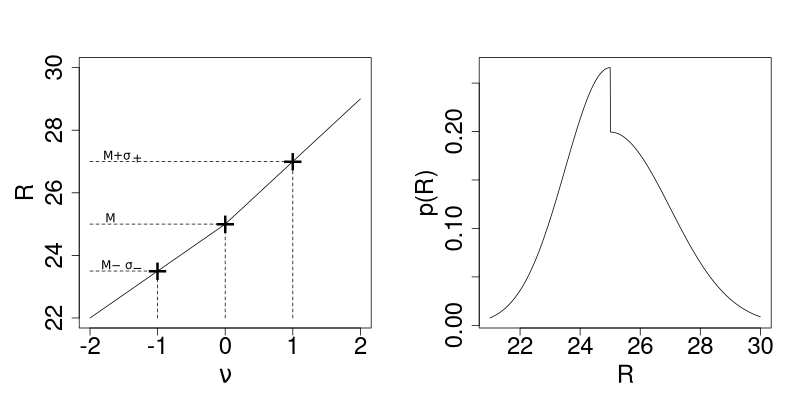}}
    \caption{\label{fig:OPAT1}
        The Dimidiated Gaussian. The left hand plot shows the results of a simple OPAT analysis that gave a result $\AE 25.0 2.0 1.5 $, and the proposed approximation.  The right hand plot shows the resulting pdf projected on the vertical axis that would come from a Gaussian 
        distribution for $\nu$ on the horizontal axis.
}
\end{figure}

Given the results of an OPAT analysis one can hypothesize that the dependence of $R$ on $\nu$ can be described by two straight lines, as shown in the left of Figure~\ref{fig:OPAT1}.  This can be expressed as
\begin{equation}
    R= \begin{cases}
    M + \sp \nu, & \nu \ge 0\\
    M + \sm \nu, & \nu \le 0\\ 
    \end{cases}.
    \label{eq:Rdim}
\end{equation}
The pdf for the variable $\nu$ is a unit Gaussian, so for $R$ the pdf is a dimidiated (or split) Gaussian with two equal halves, as shown on the right of Figure~\ref{eq:Rdim}.  This is sometimes referred to as a bifurcated Gaussian, but that is certainly inaccurate. The dimidiated Gaussian for $R$ is the probability transform of the unit Gaussian for $\nu$, under the model that they are related by Equation~\eqref{eq:Rdim}

Although the model is clearly unrealistic, it has some nice simple features.  The quantile parameters provide the parameters of the pdf  directly, as given by Equation~\eqref{eq:pdim}
\begin{equation}
    \label{eq:pdim}
    p(x)= \begin{cases}
    {1 \over \sm \sqrt{2 \pi}} e^{-\half \left( {x-M \over \sm}\right)^2} & \text{for } x < M \\
    {1 \over \sp \sqrt{2 \pi}} e^{-\half \left( {x-M \over \sp}\right)^2}
    & \text{for } x > M\\
    \end{cases}.
\end{equation}
The moments are given by
\begin{eqnarray}
    \mu &=& M+{\sp-\sm  \over \sqrt{2 \pi}}, \nonumber \\
    V &=&\half(\sp^2+\sm^2)- {1 \over 2 \pi} (\sp-\sm)^2, \label{eq:dimidiate1}\\
    \gamma &=&{1 \over \sqrt{2 \pi}}\left[ 2(\sp^3-\sm^3) - {3 \over 2} (\sp-\sm)(\sp^2+\sm^2) + {1 \over \pi} (\sp-\sm)^3\right]. \nonumber
\end{eqnarray}

To determine the quantile parameters from the moments these equations have to be solved numerically.  If we write $D=\sp-\sm$ and $S=\sp^2+\sm^2$, the equations
\begin{eqnarray}
    S=& 2V+{D^2 \over \pi}, \\ \nonumber
    D=& {2 \over 3 S} \left[ \sqrt{2 \pi} \gamma -D^3({1 \over \pi} -1)\right] 
\end{eqnarray}
can be solved as a cubic for $D$, using Cardano's formula. Having thus determined $S$ and $D$ the quantile parameters are given by
\begin{equation}
    M=\mu-{D \over \sqrt{2 \pi}},\qquad \sp= \half(\sqrt{2S-D^2}+D),\qquad \sm=\half(\sqrt{2S-D^2}-D).
    \label{eq:dimidiated2}
\end{equation}

Even though the model can handle arbitrarily large asymmetries,
in the sense of $\sp-\sm \over \sp + \sm$, there is a limit in terms of moments.
 If one of the two half-Gaussians has zero width, then 
\begin{equation}
    {|\gamma |\over \sqrt{V^3} }= {\pi+1 \over \sqrt{(\pi-1)^3}}\approx 1.641,
\end{equation}
so asymmetries, in the $\gamma \over {V^{3/2}}$ sense, larger than this cannot be
handled. But such cases are extreme.

\subsubsection{Convolution of two dimidiated Gaussians}
\label{sec:convolution}

\begin{figure}[t]
    \center{\includegraphics[width=0.7\textwidth]{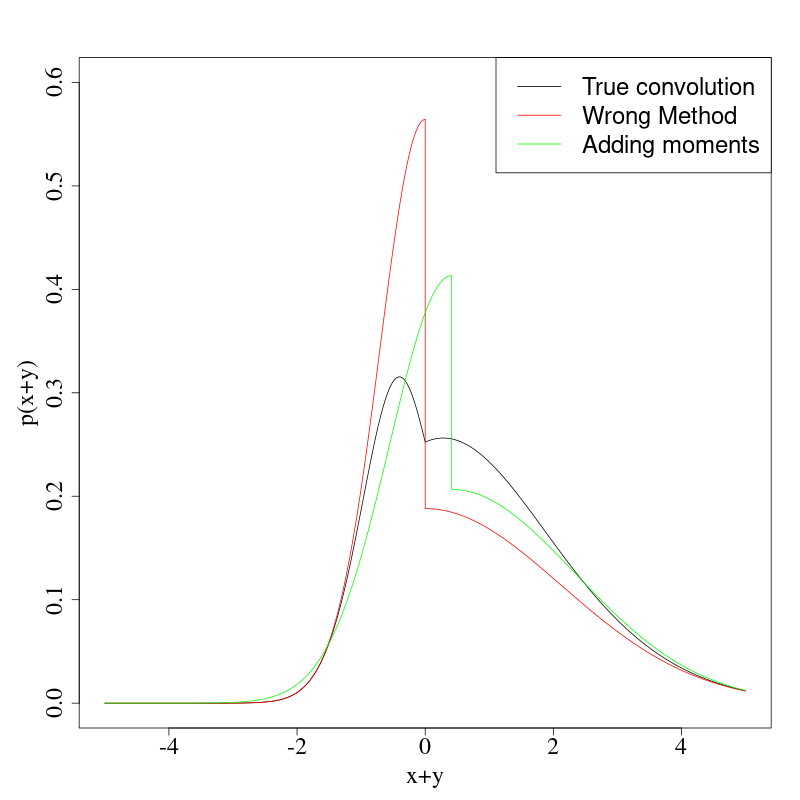}}
    \caption{\label{fig:convolute} 
        Combination of errors using the dimidiated Gaussian. Here $x$ and $y$ both have positive skewness: $\sp=1.5$ and $\sm=0.5$.}
\end{figure}

For the dimidiated Gaussian model, the convolution of two pdfs can be done analytically.  If $z=x+y$ then its pdf is  $p_z(z)=\int p_x(x) p_y(z-y)\, dx$.  The integral covers the top left and bottom right quadrants of the $(x,y)$ plane and either the bottom left quadrant (if $z$ is negative) or the top right (if $z$ is positive).  Writing, for clarity:

\begin{equation}
    \sigma^2_+={\sigma_x^+}^2+{\sigma_y^+}^2, \qquad 
    \sigma_\pm^x={\sigma_x^+}^2+{\sigma_y^-}^2, \qquad
    \sigma_\mp^2={\sigma_x^-}^2+{\sigma_y^+}^2, \qquad
    \sigma_-^2={\sigma_x^-}^2 + {\sigma_y^-}^2,
\end{equation}
the Gaussian integrals give, for $z>0$,
\begin{equation}
    p_z(z)=
    \phi\left({z \over \sigma_\mp}\right)
    \Phi\left({-z \sigma_x^- \over \sigma_y^+ \sigma_\mp}\right)
    +\phi\left({z \over \sigma_+}\right) \left[\Phi\left({z \sigma_y^+ \over \sigma_x^+ \sigma_+}\right)-\Phi\left({-z \sigma_x^+ \over \sigma_y^+ \sigma_+}\right)\right]
    +\phi\left({z \over \sigma_\pm}\right)\left[1-\Phi\left({z \sigma_y^- \over \sigma_x^+ \sigma_\pm}\right)\right],
\end{equation}
where $\phi$ is the Gaussian pdf and $\Phi$ is the corresponding cumulative density function. For $z<0$ the expression is
\begin{equation}
    p_z(z)=\hfill 
    \phi({z \over \sigma_\mp})
    \Phi\left({z \sigma_y^+ \over \sigma_x^- \sigma_\mp}
    \right)
    +\phi({z \over \sigma_-})\left[\Phi\left({-z \sigma_x^- \over \sigma_y^- \sigma_-}\right)-\Phi\left({z \sigma_y^- \over \sigma_x^- \sigma_-}\right)\right]
    +\phi\left( {z \over \sigma_\pm}\right) \left[1-\Phi\left({-z \sigma_x^+ \over \sigma_y^- \sigma_\pm}\right)\right].
\end{equation}
An example is shown in Figure~\ref{fig:convolute}.  Two quantities $x$ and $y$, both with large positive skewness, $\sp=1.5,\sm=0.5$, are combined.  The black curve shows the convolution of the two dimidiated Gaussians, using the above equations: if two dimidiated Gaussians are combined this is the `correct' pdf.  The red curve shows the curve from the `wrong method' of combining positive and negative sigmas separately and agreement is poor.  The green curve shows the dimidiated Gaussian whose first 3 moments match that of the convolution, and gives much better agreement. Note that the central value (the median) has shifted, also that the asymmetry is reduced by the combination, in accordance with the Central Limit Theorem: the green curve has $\sp=1.93, \sm=0.97$.

\subsection{The distorted Gaussian}
\label{sec:distorted}

\begin{figure}[ht]
    \centerline{\includegraphics[width=12 cm]{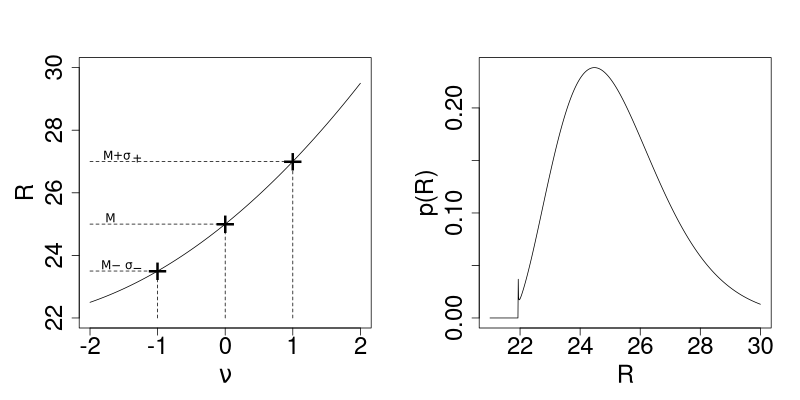}}
    \caption{\label{fig:OPAT2} 
        The Distorted Gaussian.  The left hand plot shows the results of a simple OPAT analysis that gave a result $\AE 25.0 2.0 1.5 $, as before, and the proposed approximation. The right hand plot shows the pdf projected on the vertical axis that would come from a Gaussian on the horizontal axis.}
\end{figure}

A more sophisticated procedure to handle the results of an OPAT analysis is to draw a parabola through the three points, as in Figure~\ref{fig:OPAT2},
\begin{equation}
    R=M+a\nu + b \nu^2 .
    \label{eq:distorted}
\end{equation}
%
The parameters $a$ and $b$ are given very simply from the quantile parameters: $a=\half(\sp+\sm)$ and $b=\half(\sp-\sm)$.
The pdf is again the probability transform of the unit Gaussian, but using a single quadratic transform rather than two piecewise linear ones.  Although this is more plausible than the two straight lines of the dimidiated Gaussian, care needs to be taken to include both arms of the parabola in  evaluating the pdf.   The support for $p(R)$ is limited by the minimum (or, for a negative skewness, the maximum) of the parabola. Where this occurs, there is a Jacobian spike, and this appears in Figure~\ref{fig:OPAT2} around $R=22$.

If $\nu$ is distributed according to the standard Gaussian with $\mu=0$ and $\sigma=1$
and $x(\nu)=x_0+ a \nu + b \nu^2$, then the pdf for 
$x$ is $p(x)={{\phi}(\nu) \over |x'(\nu)|}={{\phi}({\nu}) \over |a+2b\nu|}$, where $a$ and $b$ are as given in Equation~\eqref{eq:distorted}, and $\phi(\nu)$ is the unit Gaussian.  As this is a quadratic, there are two values of $\nu$ for a given $x$:  $\nu={\sqrt{a^2+4bx}-a \over 2b}$ and ${-\sqrt{a^2+4bx}-a \over 2b}$.  Except for small asymmetries and/or small deviations, where the contribution from the second can be neglected, the pdf for $x$ is the sum of the two.

The moments are given by
\begin{equation}
    \mu = M+b,   \qquad
    V = a^2 + 2 b^2, \qquad 
    \gamma  =  2b(3 a^2 + 4 b^2).  
    \label{eq:distorted1}
\end{equation}
%
%
Again, the parameters can be determined from the moments numerically, 
\begin{equation}
    b  =  {\gamma \over V-4 b^2}, \qquad
    a  =  \sqrt{V-4 b^2},\qquad
    M  =  \mu - b.
    \label{eq:distorted2}
\end{equation}
From these, if desired, the other parameters are just given by $\sp=a+b, \sm=a-b$.

Because of the second arm of the parabola, the parameters $M,\sp,\sm$ do not give the exact quantiles for $R$.  Because of this ambiguity the distorted Gaussian is treated slightly differently in the software packages: the R implementation uses the parameters described here and the Python/C++  
implementation uses the parameters giving the
exact quantiles.

\subsection{The ``railway'' Gaussian}

\begin{figure}
    \begin{center}
    \includegraphics[width=0.49\textwidth]{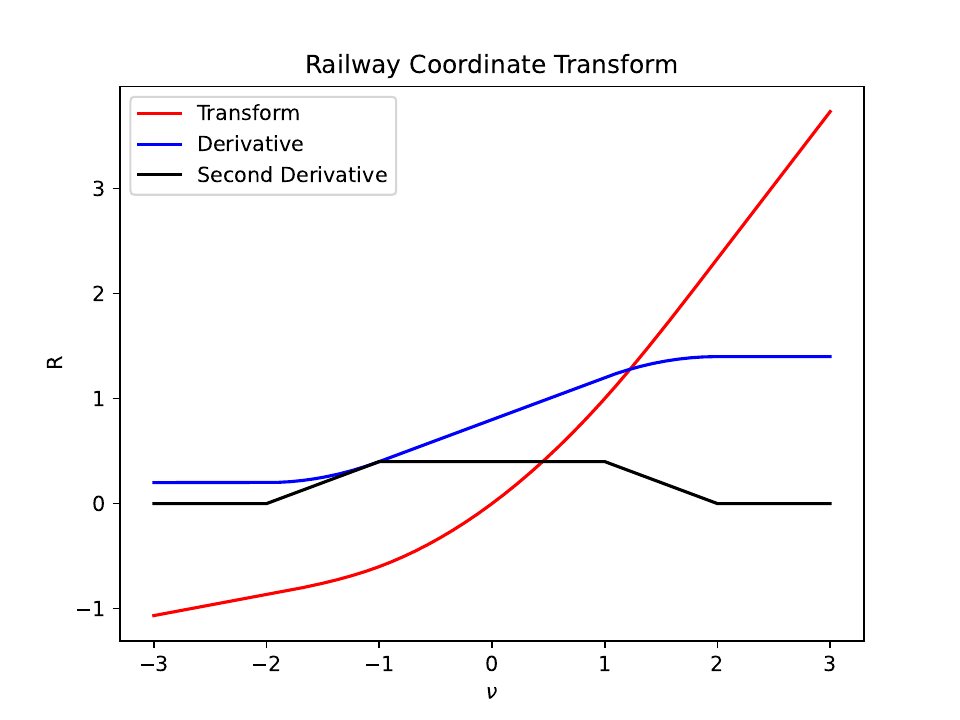}
    \includegraphics[width=0.49\textwidth]{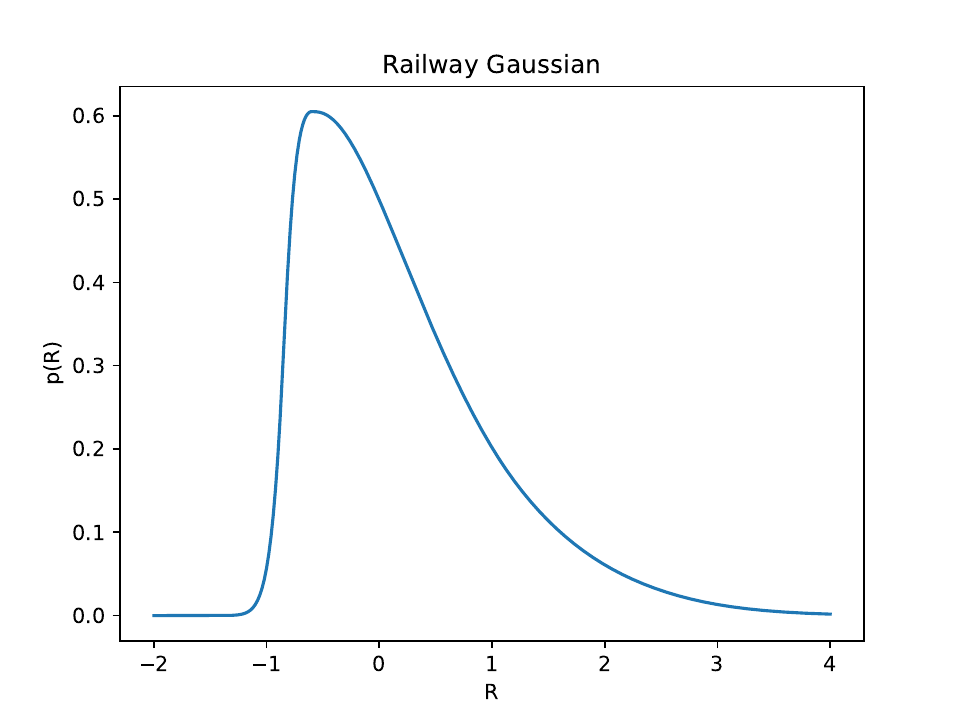}
    \caption{\label{fig:railwayGaussian} 
        The Railway Gaussian. The panel on the left illustrates the coordinate transformation used to construct a railway Gaussian with parameters $M = 0$, $\sigma^+ = 1$, $\sigma^- = 0.6$, $h_l = h_r = 1$.  The panel on the right shows the corresponding density.}
    \end{center}
\end{figure}

The railway Gaussian density attempts to mitigate the drawbacks of the dimidiated Gaussian (the discontinuity in the middle) and
of the distorted Gaussian (the Jacobian spike). This density is also based on a coordinate transformation. The curve drawn through the OPAT analysis points is a parabola in the $[-1, 1]$ region which smoothly transitions into straight lines on both left and right sides of the region. These transition regions are highly reminiscent of the track transition curves in railroad tracks, and this similarity gives the density its name. The coordinate transformation is defined by the equation
\begin{equation}
    \label{eq:railwaytransform}
    R = f(\nu)=\begin{cases}
    L(\nu, -1 - h_l) & \text{for } \nu \le -\!1 - h_l \\
    T(\nu, -1, -h_l) & \text{for } -\!1 - h_l  \le \nu \le -1 \\
    M + a \nu + b \nu^2 & \text{for } -\!1 \le \nu \le 1 \\
    T(\nu, 1, h_r) & \text{for } 1 \le \nu \le 1 + h_r \\
    L(\nu, 1 + h_r) & \text{for } \nu \ge 1 + h_r
    \end{cases},
\end{equation}
where $L(\nu, \nu_0)$ is a straight line defined by its value and its derivative at $\nu_0$, while $T(\nu, \nu_0, h)$ is a cubic transition curve. The coefficients of the cubic are determined from the value of the cubic and of its first and second derivatives at $\nu_0$ as well as from the requirement that the second derivative decays linearly to 0 at $\nu_0 + h$.  The explicit expression is
\begin{equation}
    T(\nu, \nu_0, h) = \left[\frac{f''(\nu_0)}{2} \left(1 - \frac{\nu - \nu_0}{3 h}\right) (\nu - \nu_0) + f'(\nu_0)\right] (\nu - \nu_0) + f(\nu_0).
\end{equation}
According to Equation~\ref{eq:railwaytransform}, for the right transition region $[1, 1 + h_r]$ with $\nu_0 = 1$, we must set $f(\nu_0) = M + 1 + b$, $f'(\nu_0) = a + 2 b$, $f''(\nu_0) = 2 b$. The transition curve in the $[-\!1 - h_l, -1]$ left transition region is similarly parameterized by the function and its derivatives at $\nu_0 = -1$.  An example railway coordinate transformation and the corresponding density is shown in Figure~\ref{fig:railwayGaussian}.

In comparison with the dimidiated and distorted Gaussian distributions, the railway Gaussian has two additional parameters, $h_l$ and $h_r$ (the widths of the left and right transition regions, respectively).  A reasonable default choice for them is given by $h_l = \left| \frac{f'(-1)}{f''(-1)} \right|$ and $h_r = \left| \frac{f'(1)}{f''(1)} \right|$.

The R code (Appendix~\ref{sec:R}) evaluates the moments of the railway Gaussian algebraically, whereas in the python/C++ code (Appendix~\ref{sec:Cpp}) the moments  are evaluated numerically and, for numerical reasons, the values of $h$ used are restricted to the range $[0.1, 10.0]$. 

Note that, in general, the railway coordinate transformation is allowed to be non-monotonic (for example, when $\sigma^+$ and $\sigma^-$ have different signs).  In this case, similar to the distorted Gaussian, the railway Gaussian will have a semi-infinite support and will exhibit a~Jacobian spike.

\subsection{The double cubic Gaussian}

\begin{figure}
    \begin{center}
    \includegraphics[width=0.49\textwidth]{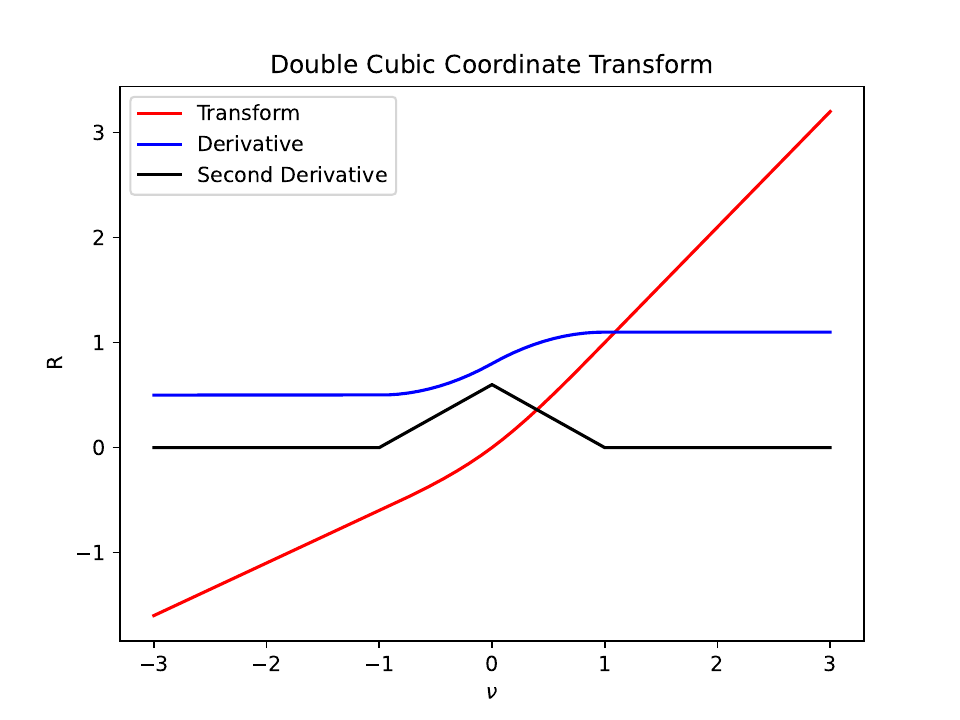}
    \includegraphics[width=0.49\textwidth]{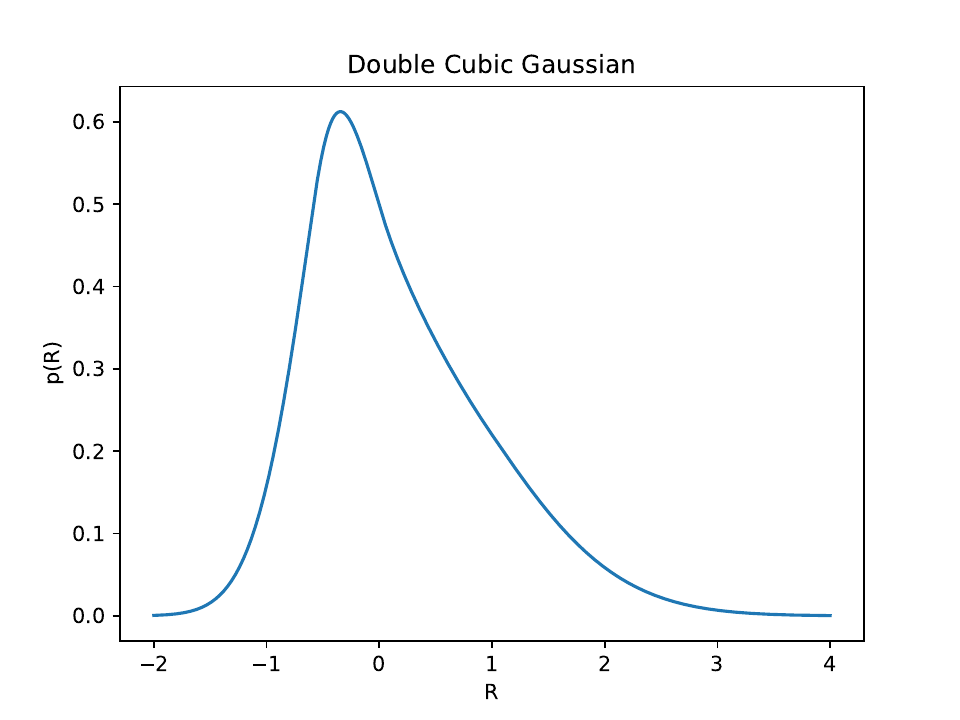}
    \caption{\label{fig:doublecGaussian} The Double Cubic Gaussian.
        The panel on the left illustrates the coordinate transformation used to construct a double cubic Gaussian with parameters $M = 0$, $\sigma^+ = 1$, $\sigma^- = 0.6$.  The panel on the right shows the corresponding density.}
    \end{center}
\end{figure}

To construct the double cubic Gaussian, the transformation generating the railway Gaussian is simplified by eschewing the parabolic section in the middle. The natural choice $h_l = h_r = 1$ then leads to
\begin{equation}
    \label{eq:doublectransform}
    R = M + \begin{cases}
    \frac{1}{4} (5 \sm - \sp) (\nu + 1) - \sm & \text{for } \nu \le -\!1 \\
    \nu \left[\frac{1}{4} \left(\nu^2+3 \nu+2\right) (\sp-\sm)+\sm\right] & \text{for } -\!1  \le \nu \le 0 \\
    \nu \left[\frac{1}{4} \left(\nu^2-3 \nu+2\right)
    (\sm-\sp)+\sp\right] & \text{for } 0 \le \nu \le 1 \\
    \frac{1}{4} (5 \sp - \sm) (\nu - 1) + \sp & \text{for } \nu \ge 1
    \end{cases}.
\end{equation}
The dependence of $R$ on $\nu$ is continuous together with its first two derivatives, while the third derivative exhibits discontinuities at $\nu = -1, 0, 1$. An example double cubic coordinate transformation and the corresponding density are shown in Figure~\ref{fig:doublecGaussian}.

\subsection{The symmetric beta Gaussian}
\label{sec:symbetaGauss}

The symmetric beta Gaussian distribution is constructed using the following transformation:
\begin{equation}
    \label{eq:symbetatransform}
    R(\nu) = A \int_0^\nu dy \int_{0}^{y} b(x, p, h) dx + k \nu + M,
\end{equation}
where
\begin{equation}
    \label{eq:symbeta}
    b(x, p, h) = \frac{1}{A} \frac{d^2 R(\nu)}{d \nu^2} = \begin{cases}
    0 & \text{for } |x| \ge h \\
    \left[1 - \left(\frac{x}{h}\right)^2\right]^p & \text{for } |x| < h
    \end{cases}
\end{equation}
is a shifted and scaled (as well as unnormalized) density of the symmetric beta distribution. $h$~is a~positive width parameter. In order to maintain the continuity of the transformation second derivative and to enable efficient numerical calculation of the cumulants, the power parameter $p$ is expected to be a~small positive integer (the C++/Python software is restricted to $p \in \{1, ..., 20\}$; it is not yet implemented in the R software). For given values of $p$, $h$, and $M$, parameters $A$ and $k$ are chosen to satisfy the conditions $R(-1) = M - \sm$ and $R(1) = M + \sp$, while the condition $R(0) = M$ is satisfied by Equation~\ref{eq:symbetatransform} automatically.

The transformation defined by Equation~\ref{eq:symbetatransform} is continuous together with its $p + 1$ leading derivatives.  This is reflected in the smoothness of the $R$ distribution whose density will possess at least $p$ continuous derivatives at all inner points of its support. (The Jacobian spike, if present, lies at the support boundary.)  In the limit $h \rightarrow 0$ the distribution of $R$ tends to the dimidiated Gaussian, and in the limit $h \rightarrow \infty$ it becomes the distorted Gaussian. Practically usable values of $h$ are between 0.1 and 10.0 or so; values outside this range can lead to numerical instabilities in the software.

\begin{figure}
    \begin{center}
    \includegraphics[width=0.4\textwidth]{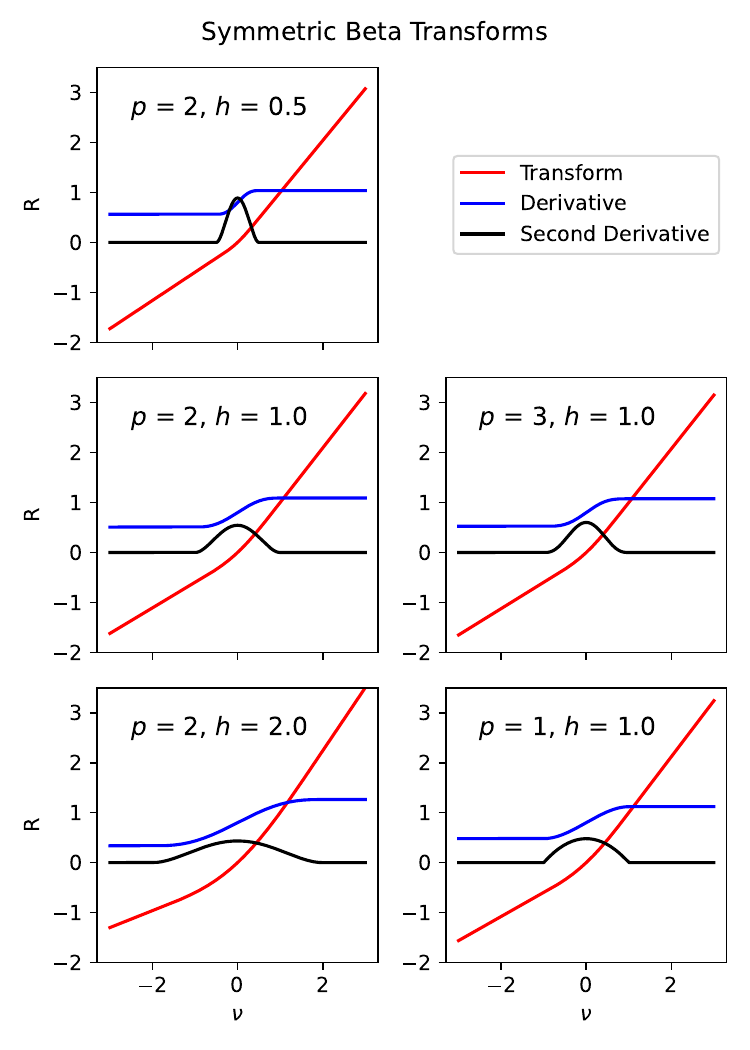}
    \includegraphics[width=0.59\textwidth]{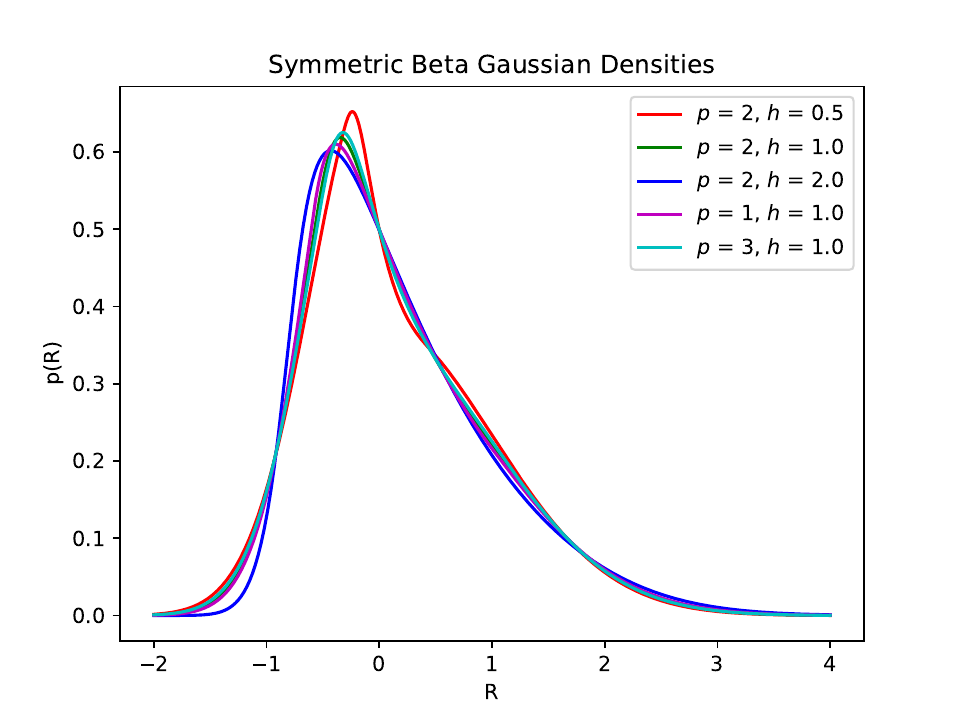}
    \caption{\label{fig:symbetaGaussian} 
        The Symmetric Beta Gaussian. The panel on the left illustrates the coordinate transformations used to construct 
        densities shown on the right. In all cases, $M = 0$, $\sigma^+ = 1$, $\sigma^- = 0.6$.}
    \end{center}
\end{figure}

Examples of the transformations generated by Equation~\ref{eq:symbetatransform}, together with the corresponding densities, are illustrated in Figure~\ref{fig:symbetaGaussian} for a number of different $p$ and $h$ settings.

\subsection{The quantile variable width Gaussian}

The quantile variable width (QVW) Gaussian distribution is a special case of the simple Q-normal distribution proposed in~\cite{ref:keelin2011}.  It is based on the following idea. Suppose, $Q(z)$ is the quantile function (i.e., the inverse cumulative distribution function) of the standard normal. Then $\mu + \sigma \, Q(z)$ is the quantile function of the Gaussian distribution with mean $\mu$ and standard deviation $\sigma$.  We can also define a quantile function which looks like $\mu + \sigma(z) \, Q(z)$. In a wider sense, the distribution defined by this quantile function can be interpreted as a Gaussian with variable width $\sigma(z)$. The width in this case depends on the cumulative probability rather than on the coordinate.  The QVW Gaussian distribution is employing the following simple parameterization:
\begin{equation}
    \sigma(z) = \sigma_0 \left[1 + a \left(z - \frac{1}{2}\right)\right],
\end{equation}
where $a$ is a shape parameter governing the asymmetry.  The complete quantile function is then
\begin{equation}
    q(z) = \mu_0 + \sigma_0 \left[1 + a \left(z - \frac{1}{2}\right)\right] Q(z).
\end{equation}
The density that corresponds to the cdf value $z$ is, of course, $\left(\frac{d q(z)}{d z}\right)^{-1}$.  The $z$ value corresponding to a given density argument $x$ has to be determined by solving the equation $q(z) = x$ numerically.

The first three moments of the QVW Gaussian distribution are
\begin{align}
    \begin{split}
    \mu & = \mu_0 + a \sigma_0 M_{11} \\
    \sigma^2 & = \sigma_0^2 \left[1 + a^2 \left(M_{22} - M_{11}^2 - \frac{1}{4}\right)\right] \\
    \gamma & = \frac{a \sigma_0^3}{4} \left[8 a^2 M_{11}^3 - 3 M_{11} (4 + a^2 (4 M_{22} - 1)) + 
    3 (a - 2)^2 M_{31} - 6 (a - 2) a M_{32} + 4 a^2 M_{33}\right],
    \end{split}
\end{align}
where
\begin{equation}
    M_{kn} = \int_{-\infty}^{\infty} z^k \Phi(z)^n \phi(z) dz.
\end{equation}
Here, $\phi(z)$ is the standard normal density, and $\Phi(z)$ is the standard normal cdf.  In particular, $M_{11} = \frac{1}{2 \sqrt{\pi}}$, $M_{22} = \frac{\sqrt{3} + 2 \pi}{6 \pi}$, $M_{31} = M_{32} = \frac{5}{4 \sqrt{\pi}}$. Multiple precision numerical evaluation of $M_{33}$ gives 0.6751064260945980674284983.

\begin{figure}
    \begin{center}
    \includegraphics[width=0.49\textwidth]{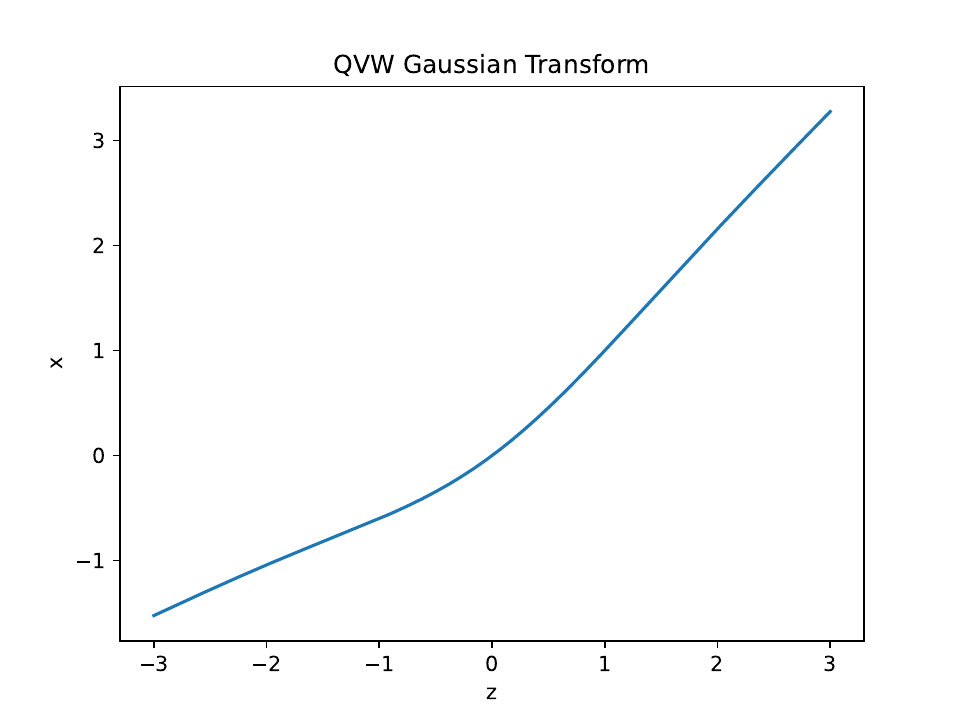}
    \includegraphics[width=0.49\textwidth]{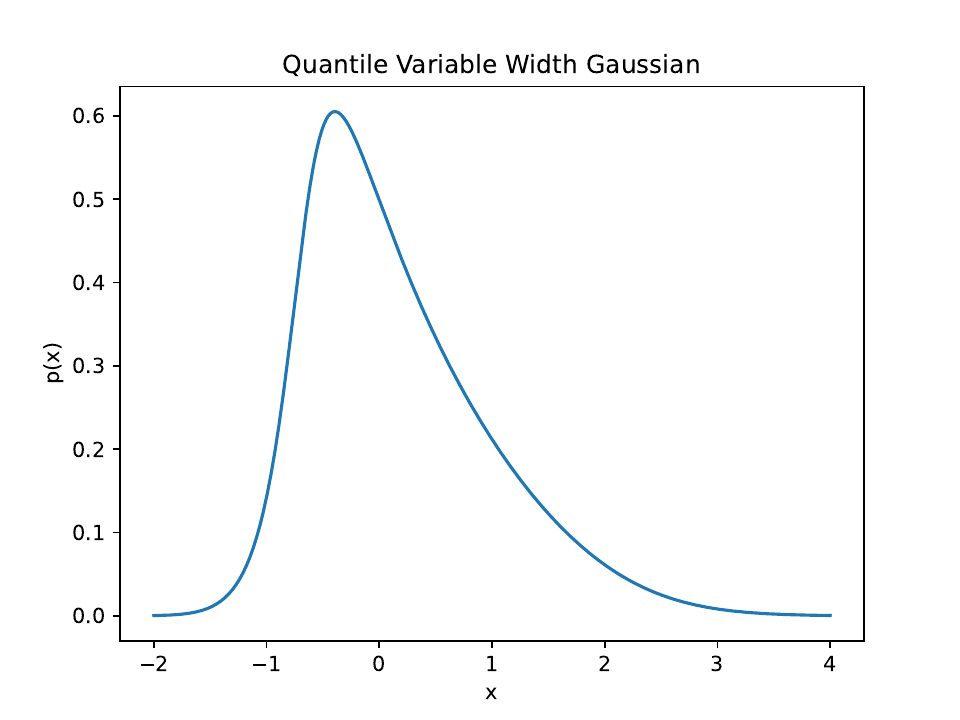}
    \caption{\label{fig:QVWGaussian} 
    The QVW Gaussian.
        The panel on the right illustrates the density which corresponds to the asymmetric uncertainty result $0^{+1.0}_{-0.6}$.  The panel on the left shows the corresponding coordinate transformation given by $x = q(\Phi(z))$.}
    \end{center}
\end{figure}

An example QVW Gaussian density and the corresponding transformation of the standard normal are shown in Figure~\ref{fig:QVWGaussian}.

\subsection{The Fechner distribution}

The probability density function of the Fechner (also known as the split-normal) distribution is given by
\begin{equation}
    \label{eq:fechnerpdf}
    p(x) = \begin{cases}
    \sqrt{\frac{2}{\pi}} \frac{1}{\sigma_1 + \sigma_2} \exp \left( -\frac{1}{2} \left( \frac{x - m}{\sigma_1}\right)^2 \right) & \text{for } x \le m \\
    \sqrt{\frac{2}{\pi}} \frac{1}{\sigma_1 + \sigma_2} \exp \left( -\frac{1}{2} \left( \frac{x - m}{\sigma_2}\right)^2 \right) & \text{for } x \ge m \\
    \end{cases},
\end{equation}
where $m$ is the distribution mode, while non-negative parameters $\sigma_1$ and $\sigma_2$ define characteristic distribution widths below and above the mode, respectively.

The first three moments of the Fechner distribution are given by
\begin{align}
    \label{eq:fechnermoments}
    \begin{split}
    \mu & = m + \sqrt{\frac{2}{\pi}} (\sigma_2 - \sigma_1) \\
    \sigma^2 & = \left(1 - \frac{2}{\pi}\right) (\sigma_2 - \sigma_1)^2 +  \sigma_1 \sigma_2 \\
    \gamma & = \sqrt{\frac{2}{\pi}} (\sigma_2 - \sigma_1) \left[ \left(\frac{4}{\pi} - 1\right) (\sigma_2 - \sigma_1)^2 +  \sigma_1 \sigma_2 \right]
    \end{split}.
\end{align}
The largest possible normalized skewness, $\gamma/\sigma^{3/2}$, is reached for $\sigma_1 = 0$ and $\sigma_2 > 0$. It corresponds to the skewness of the half-Gaussian, $\sqrt{2} (4 - \pi) (\pi - 2)^{-3/2}$.  This limits the asymmetry that can be represented by the Fechner distribution, $\frac{\sigma^+ - \sigma^-}{\sigma^+ + \sigma^-}$, to about 0.21564027.

\begin{figure}
    \begin{center}
    \includegraphics[width=0.49\textwidth]{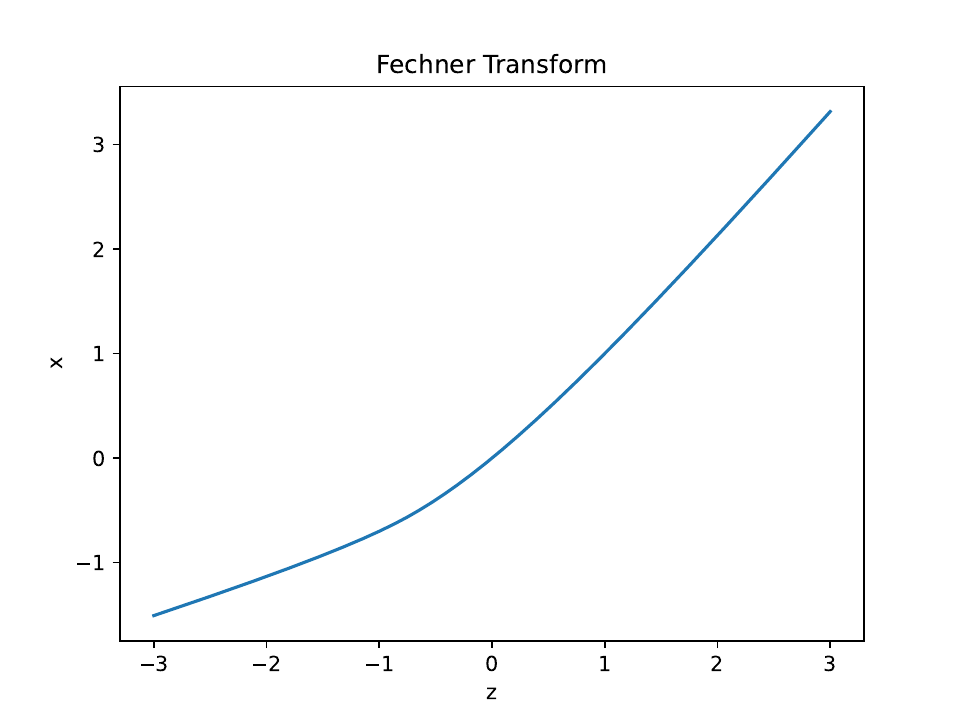}
    \includegraphics[width=0.49\textwidth]{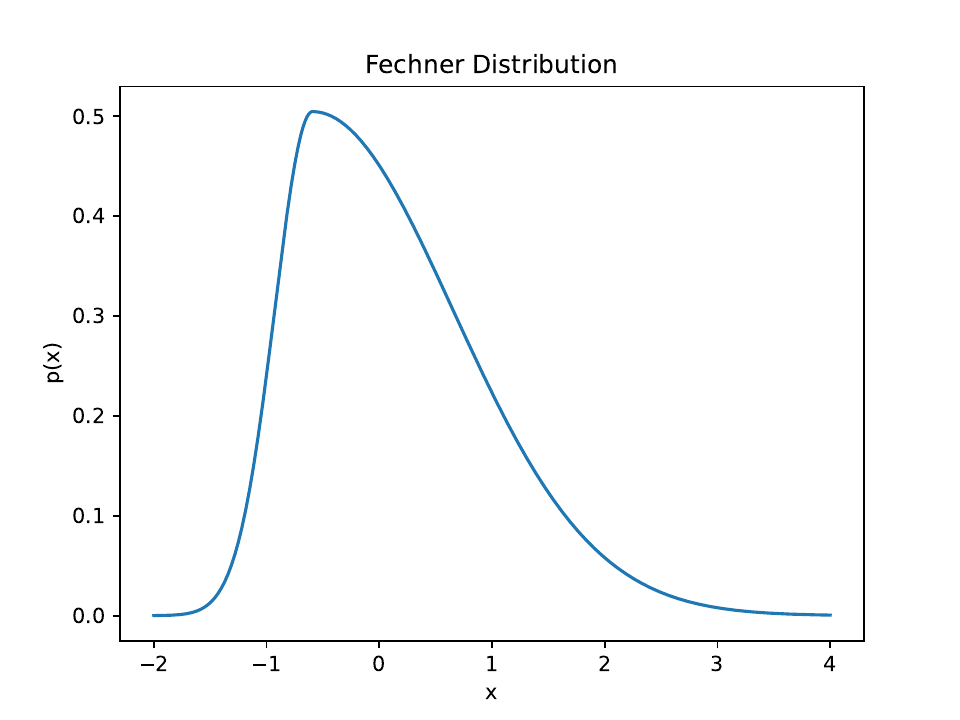}
    \caption{\label{fig:FechnePDF} 
        The Fechner distribution. The panel on the right illustrates the probability density function which corresponds to the asymmetric uncertainty result $0^{+1.0}_{-0.7}$. (The result $0^{+1.0}_{-0.6}$ for which a number of other distributions are plotted in this appendix can not be represented by Fechner.)
        The panel on the left shows the corresponding coordinate transformation given by $x = q(\Phi(z))$.}
    \end{center}
\end{figure}

Construction of the Fechner distribution from moments or from quantiles has to be performed numerically.  An example Fechner probability density function and the corresponding transformation of the standard normal are shown in Figure~\ref{fig:FechnePDF}.

\subsection{The Edgeworth expansion}
\label{sec:edgeworth}

The Edgeworth expansion \cite{Edgeworth} is appropriate with an asymmetry ascribable to the fact that the sample size is finite and the Central Limit Theorem has not had a chance to make the distribution accurately Gaussian. This is not a probability transform so there is no equivalent to Figures~\ref{fig:OPAT1} or \ref{fig:OPAT2}.  The pdf (taking only the first term after the Gaussian)  is given by
\begin{equation}
    p(x)={1 \over \sigma \sqrt{2 \pi}} e^{-\half(x-\mu)^2 / \sigma^2 } \left[ 1+{\gamma \over 6 \sigma^3}He_3({x-\mu \over \sigma}) \right]
,\label{eq:edgeworth0}
\end{equation}
where $He_3(z)=z^3-3z$ is the third order Hermite polynomial.  The Edgeworth model has the nice feature that the moments provide the function parameters for Equation~\eqref{eq:edgeworth0}, with $V=\sigma^2$.

The cumulative equivalent of this $p(x)$, the cumulative distribution function,  is
\begin{equation}
    F(z)=\Phi(z) -{\gamma \over 6 \sigma^3} \phi(z) (z^2-1),
\end{equation}
where $z\equiv (x-\mu)/\sigma$ and $\phi(z)$ and $\Phi(z)$ are the Gaussian density and distribution functions.  The second term vanishes at the 1-sigma points, $z=\pm 1$, so the width of the 68\% central interval is the same for all values of $\gamma$. As $\mu$ is also independent of $\gamma$, this means that the choice of parameters $\AE {\mu}  {\sp}  {\sm} $ cannot be used with this form, as mentioned in Section~\ref{sec:problems}.

If the quantile parameters $\AE M {\sp} {\sm}  $ are given, the moments are
\begin{eqnarray}
    \mu &=& M+\half(\sp-\sm) \nonumber \\
    \sqrt{V}\equiv \sigma &=& \half(\sp+\sm) \\
    \gamma &=& {6 \sigma^3 \over (z^2-1) G(z) } \left(\Phi(z)-\half\right) {\rm\  with\  } z={M-\mu \over \sigma}. \nonumber
    \label{eq:edgeworth1}
\end{eqnarray}
The reverse process again needs a numerical solution. Starting from $z=0$ one can iterate
\begin{equation}
    z=\Phi^{-1}\left[\half+ {\gamma \over 6 \sigma^3} (z^2-1) G(z)\right],
    \label{eq:edgeworth2}
\end{equation}
where $\Phi^{-1}$ is the inverse of the Gaussian cdf. This seems to work better than Newton's method, thanks to the constantly changing gradient. From $z$ one obtains $M=\sigma z+\mu$, and thence $\sp=\sigma +\mu- M$ and $\sm=\sigma -\mu+ M$.

The Edgeworth form, as one soon discovers, has the disadvantage that unless the asymmetry is quite small, the pdf goes negative, which is clearly disallowed.

\subsection{The skew normal}
\label{sec:azzalini}

Azzalini \cite{azzalini} suggests using a distribution whose density can be elegantly written as
\begin{equation}
    p(z)=2 \phi(z) \Phi(\alpha z).
    \label{eq:azzalini}
\end{equation}
In addition to the asymmetry parameter $\alpha$ one introduces a scale parameter $\omega$ and a location parameter $\xi$, so $z \equiv {x-\xi \over \omega}$. Writing for convenience $\delta = {\alpha \over \sqrt{1+\alpha^2}}$, the moments are given by
\begin{eqnarray}
    \mu&=&\xi+ \omega \delta \sqrt{2 \over \pi} \nonumber \\
    V &=& \omega^2 \left(1-{2 \delta^2 \over \pi} \right) \\
    \gamma &=& \left( {4-\pi \over 2} \right) (\omega \delta \sqrt{2/\pi}) ^3 \nonumber
    \label{eq:azzalini1}.
\end{eqnarray}
To convert in the reverse direction
\begin{eqnarray}
    \omega \delta &=& \sqrt{\pi \over 2} \sqrt[3]{2 \gamma \over 4-\pi} \nonumber \\
    \omega &=& \sqrt{V+{2 (\omega \delta)^2 \over \pi}} \\
    \xi &=& \mu - \omega \delta \sqrt{2 \over \pi} \nonumber
    .\label{eq:azzalini2}
\end{eqnarray}
For the quantile parameters we need the cumulative distribution, which is not so elegant,
\begin{equation}
    F(z)=\Phi(z)-2 T(z,\alpha),
\end{equation}
where $T(z,\alpha)$ is Owen's T function
\begin{equation}
    T(z,\alpha)={1 \over 2 \pi} \int_0^\alpha {e^{-\half z^2(1+x^2)} \over 1+x^2 }\, dx.
\end{equation}
From a value of $\alpha$ one can determine the appropriate values of $z$ to get the required quantiles $F(z)$= 0.16, 0.5, and 0.84, iterating using Newton's method, and then use $\xi$ and $\omega$ to convert these to $\AE {M} {\sp} {\sm} $.

To convert in the other direction, when the distribution is specified by $\AE M {\sp} {\sm} $, the parameter $\alpha$ can be found numerically by adjusting it to get the desired value of the asymmetry ${\sp-\sm \over \sp+\sm}$.  $\omega$ is then found by scaling to get the desired value of $\sp+\sm$, and $\xi$ from the desired value for $M$.  If the distribution is specified by  $\AE {\mu} {\sp} {\sm} $ then to obtain the other forms is not impossible (as it is for the Edgeworth distribution), but is so impractical as to be essentially useless.

This distribution does not describe large asymmetries, like the Edgeworth distribution, but for a different reason.  In the large $\alpha$ limit it becomes a half Gaussian, and higher asymmetries cannot be accommodated.

\subsection{The Johnson system}

The combination of Johnson distributions $S_B$ (bounded), $S_U$ (unbounded), together with their limiting cases (Gaussian and log normal), form a very flexible system of four-parameter densities~\cite{ref:johnson, ref:johnsonbook, ref:hahnshap}.  It turns out that there is a Johnson distribution for every possible combination of mean, standard deviation, skewness, and kurtosis. It is in fact convenient to think of the set as if it was a single distribution whose density is given by $j(R, \mu, \sigma, s, \kappa)$. (In this parameterization, skewness $s$ and kurtosis $\kappa$ are normalized.)

Four parameters are one too many to represent a central value with two asymmetric uncertainties. There is, however, a rather natural way to restrict the system to three parameters as follows: $j_r(R, \mu, \sigma, s) = j(R, \mu, \sigma, s, \hat{\kappa})$, where the kurtosis parameter $\hat{\kappa}$ is chosen for every $s$ according to the maximum entropy principle: $\hat{\kappa} = \underset{\kappa}\argmax[S(s, \kappa)]$.  The standardized entropy $S(s, \kappa)$ is given by
$$
    S(s, \kappa) = -\int_{-\infty}^{\infty} j(u, 0, 1, s, \kappa) \ln j(u, 0, 1, s, \kappa) \,du.
$$
The entropy does not depend on $\mu$. As $\kappa$ is normalized, $\hat{\kappa}$ does not depend on $\sigma$ either.

It turns out that, for $|s| > 0$, the maximum entropy principle always selects the $S_U$ distribution (and the Gaussian for $s = 0$).  Therefore, the $S_U$ distribution will be discussed here in more detail. Its density is given by
\begin{equation}
    p(x) = \frac{\delta}{\lambda \sqrt{2 \pi \left[1 + 
    \left(\frac{x - \xi}{\lambda}\right)^{2}\right]}} \,e^{-\frac{1}{2}
    \left[\gamma + \delta\,\mbox{\scriptsize sinh}^{-1}\left(\frac{x-\xi}
    {\lambda}\right)\right]^{2}},
    \label{eq:johnsu}
\end{equation}
where the set of parameters $\{\xi, \lambda, \gamma, \delta\}$ is in one-to-one correspondence to the set $\{\mu, \sigma, s, \kappa\}$.  Variable $z$ distributed according to the standard normal can be obtained from Johnson's $S_U$ variate $x$ by the transformation $z = \gamma + \delta \,\mbox{sinh}^{-1}\left(\frac{x-\xi}{\lambda}\right)$.  Alternatively,
\begin{equation}
    \label{eq:johnsontransform}
    x = \xi + \lambda \,\mbox{sinh}\left( \frac{z - \gamma}{\delta} \right)
\end{equation}
transforms the standard normal into Johnson's $S_U$.

With the notation $\Omega = \gamma/\delta$ and $\omega = \exp(\delta^{-2})$, the moments of the $S_U$ distribution are given by~\cite{ref:johnsonbook}
\begin{align}
    \begin{split}
    \mu & = \xi - \lambda \sqrt{\omega} \,\mbox{sinh}\, \Omega \\
    \sigma^2 & = \frac{\lambda^2}{2}  (\omega - 1) (\omega \,\mbox{cosh}\, 2 \Omega + 1 ) \\
    s^2 & = \frac{\omega (\omega - 1) \left[\omega (\omega + 2) \,\mbox{sinh}\, 3 \Omega + 3 \,\mbox{sinh}\, \Omega\right]^2}{2 (\omega \,\mbox{cosh}\, 2 \Omega + 1 )^3} \\
    \kappa & = \frac{\omega^2 (\omega^4 + 2 \omega^3 + 3 \omega^2 - 3) \,\mbox{cosh}\, 4 \Omega + 4 \omega^2 (\omega + 2) \,\mbox{cosh}\, 2 \Omega + 3 (2 \omega + 1)}{2 (\omega \,\mbox{cosh}\, 2 \Omega + 1 )^2}.
    \end{split}
\end{align}
The sign of the skewness $s$ is opposite to that of $\gamma$. The mapping from the parameter set $\{\mu, \sigma, s, \kappa\}$ back to $\{\xi, \lambda, \gamma, \delta\}$ can be efficiently performed numerically, according to the algorithm presented in~\cite{ref:AS99}.  The calculation of $\hat{\kappa}$ for the given $s$ also has to be performed numerically, as well as the determination of $\mu$, $\sigma$, and $s$ from quantiles.

\begin{figure}
    \begin{center}
    \includegraphics[width=0.49\textwidth]{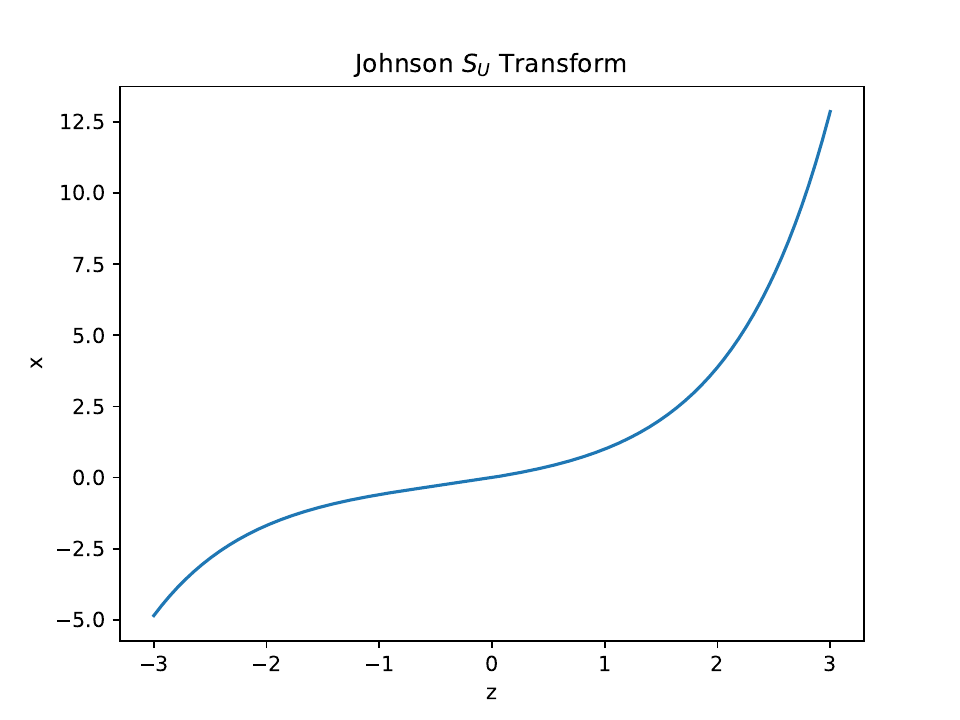}
    \includegraphics[width=0.49\textwidth]{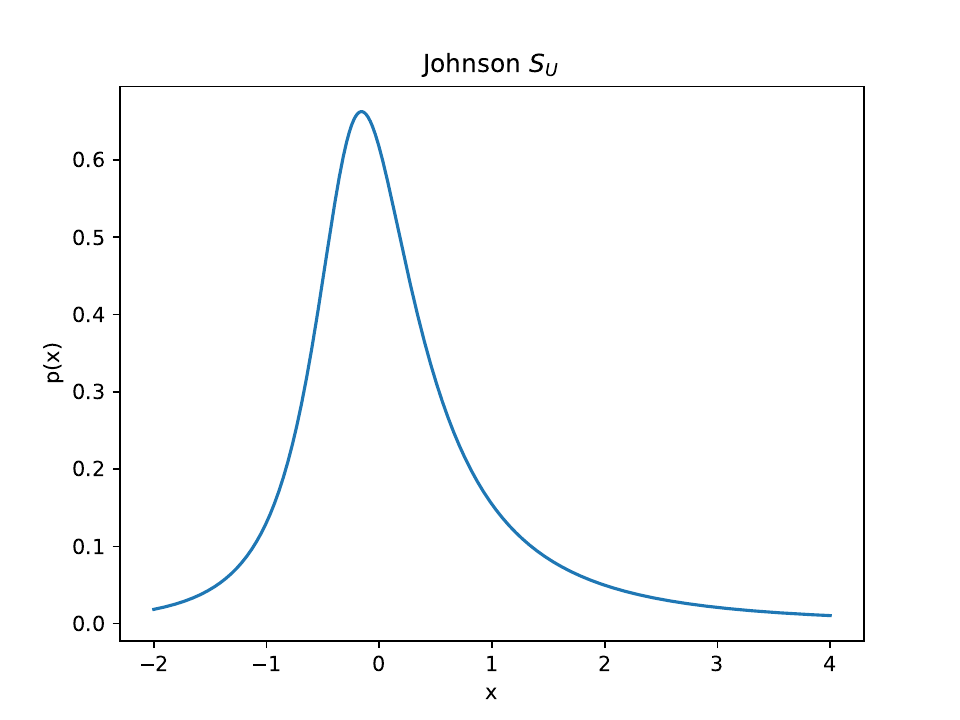}
    \caption{\label{fig:johnsonSU} 
        The Johnson distribution. The panel on the right illustrates the Johnson $S_U$ density which corresponds to the asymmetric uncertainty result $0^{+1.0}_{-0.6}$.  The panel on the left shows the matching coordinate transformation calculated according to Equation~\ref{eq:johnsontransform}.  As expected, this transformation maps $-1$ to $-0.6$, $0$ to $0$, and $1$ into $1$.}
    \end{center}
\end{figure}

An example $S_U$ density and the corresponding transformation are shown in Figure~\ref{fig:johnsonSU}.

\subsection{The log normal} 
\label{sec:lognormal}

The log normal distribution is a special three-parameter member of the Johnson system. For a fixed skewness, the kurtosis of the log normal is larger than the kurtosis of all $S_B$ distributions, but smaller than the kurtosis of all $S_U$. It is thus the limiting distribution for $S_U$ with the smallest possible kurtosis and for $S_B$ with the largest possible one.

A variable distributed according to the log normal can be obtained from the standard normal variate $z$ by applying the transformation
\begin{equation}
    \label{eq:lognormaltransform}
    x = \xi + \exp \left(\frac{z - \gamma}{\delta}\right).
\end{equation}
This also means that the variable $z = \gamma + \delta \ln (x - \xi)$ is distributed according to the standard normal if $x$ is distributed according to the log normal.  With this parameterization\footnote{The parameterization described here is consistent with~\cite{ref:johnsonbook}, but differs from the standard one.}, for positive values of skewness, the log normal density is given by
\begin{equation}
    \label{eq:lognormal}
    p(x) = \begin{cases}
    0, & x \le \xi \\
    \frac{\delta}{\sqrt{2 \pi} (x - \xi)} e^{-\frac{1}{2}\left[\gamma + \delta \ln (x - \xi)\right]^2}, & x > \xi
    \end{cases}.
\end{equation}
Distributions with negative values of skewness can be obtained by utilizing $p(-x)$ and adjusting the parameters $\xi$ and $\gamma$ as necessary.

With the notation $\omega = \exp(\delta^{-2})$, the moments of the log normal distribution (with normalized $s$ and $\kappa$) are given by~\cite{ref:johnsonbook}
\begin{align}
    \begin{split}
    \mu & = \xi + \sqrt{\omega} \exp \left(-\frac{\gamma}{\delta}\right) \\
    \sigma^2 & = \omega (\omega - 1) \exp \left(-\frac{2 \gamma}{\delta}\right) \\
    s^2 & = (\omega - 1) (\omega + 2)^2 \\
    \kappa & = \omega^4 + 2 \omega^3 + 3 \omega^2 - 3
    \end{split}.
\end{align}
The mapping from the parameter set $\{\mu, \sigma, s\}$ back to $\{\xi, \gamma, \delta\}$ can be performed by first solving the equation $s^2 = (\omega - 1) (\omega + 2)^2$ for $\omega$, calculating $\delta$, and then finding all other parameters. See~\cite{ref:johnsonbook} for details. The mapping from the quantiles to the distribution parameters has to be constructed numerically.

\begin{figure}
    \begin{center}
    \includegraphics[width=0.49\textwidth]{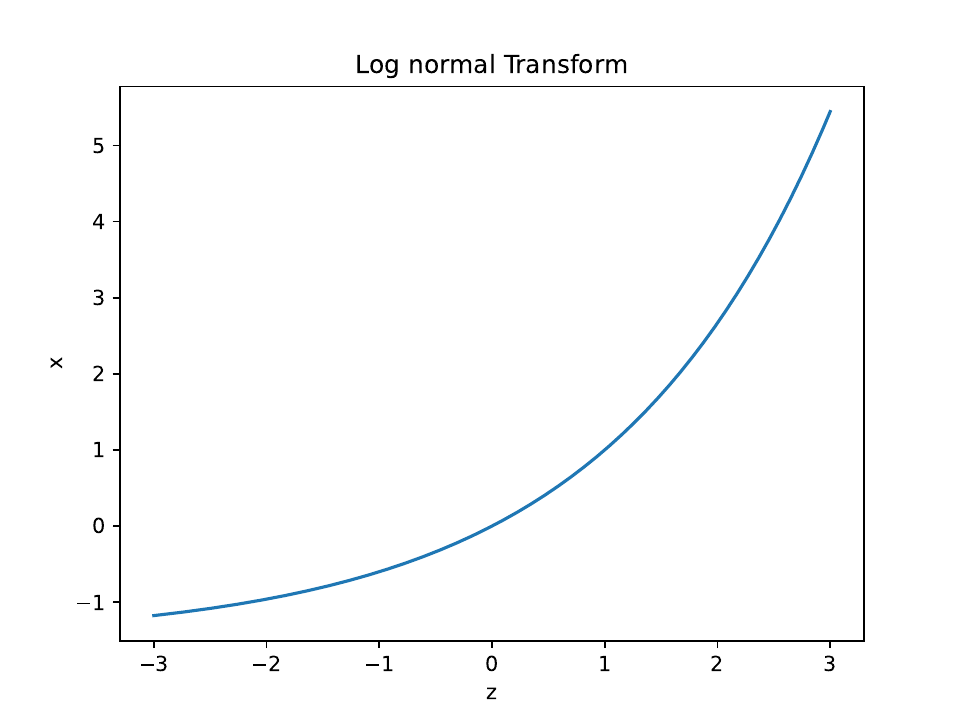}
    \includegraphics[width=0.49\textwidth]{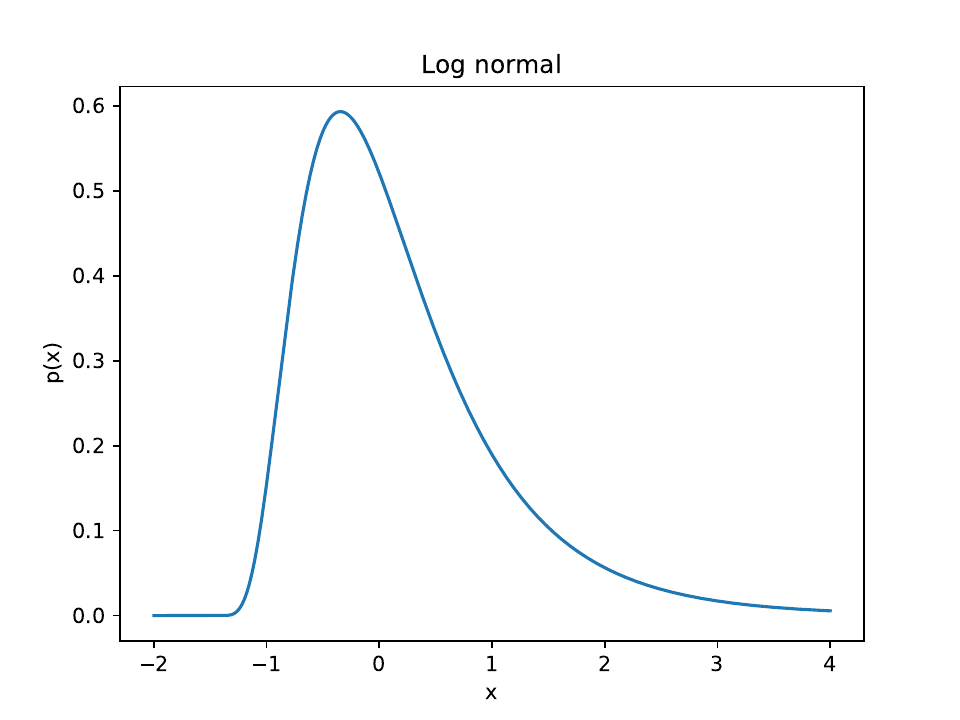}
    \caption{\label{fig:lognormal} 
        The Log Normal distribution. The panel on the right illustrates the density which corresponds to the asymmetric uncertainty result $0^{+1.0}_{-0.6}$.  The panel on the left shows the matching coordinate transformation calculated according to Equation~\ref{eq:lognormaltransform}.}
    \end{center}
\end{figure}

An example log normal density and the corresponding transformation are shown in Fig.~\ref{fig:lognormal}.




\section{Modelling likelihood errors}
\label{app:loglikelihoods}

Here we describe several models for likelihood curves that
are approximately parabolic.
There are many other possible forms (for example, the `rotated parabola'), and
both implementations of software provided are flexible so that more models (including models with more parameters) can be added without needing to re-write all the code.

\subsection{The cubic}

The obvious extension to a parabola is the cubic
\begin{equation}
    \ln L(a)=-\half \left( \alpha a^2 + \beta a^3 \right)
\end{equation}
with
\begin{equation}
    \alpha={\sp^3 + \sm^3 \over \sp^2\sm^2(\sp+\sm)} \qquad{\rm and} \qquad \beta={\sm^2-\sp^2 \over \sp^2\sm^2(\sp+\sm)}.
\end{equation}
If $\sp$ and $\sm$ differ by less than a factor of 2, this gives curves which behave sensibly in the range $[\hat a - \sm, \hat a + \sp]$,  but outside that range the cubic term gives an unwanted turning point. (If the asymmetry is large, the curve will have a minimum inside the range in addition to the maximum at $\hat a = 0$.) Also the curve does not go to $-\infty$ at both large positive and negative $a$. So although the cubic is the obvious extension of the quadratic, it often gives problems.

\subsection{The broken parabola}

Another obvious simple solution which gives $\Delta \ln L = - \half$ at $a=\sp$ and $a=- \sm$ is just
\begin{equation}
    \label{eq:brokenParabola}
    \ln L(a) = g(a) = \begin{cases}
        -\frac{1}{2} \frac{a^2}{\sm^2} & \text{for } a \le 0 \\
        -\frac{1}{2} \frac{a^2}{\sp^2} & \text{for } a \ge 0 \\
    \end{cases}.
\end{equation}
(Here, we explicitly assume $\hat a = 0$.) Unfortunately, this solution suffers from the unphysical discontinuity of the second derivative at $a = 0$.

\subsection{The symmetrized parabola}

The symmetrized parabola model is constructed in a Bayesian manner. It starts by assuming a~flat prior for the parameter and then treats Equation~\ref{eq:brokenParabola} as a logarithm of the (unnormalized) parameter probability density.  The resulting distribution is Fechner with density  given by Equation~\ref{eq:fechnerpdf} in which $m = \hat a$, $\sigma_1 = \sm$, and $\sigma_2 = \sp$.  This distribution is then exchanged for the Gaussian distribution with the same mean and standard deviation as the Fechner where the moments are calculated according to Equation~\ref{eq:fechnermoments}.  The symmetrized parabola is the logarithm of this Gaussian density.

Note that the peak position of the symmetrized parabola is no longer at $\hat a$; it is shifted to the right if $\sp > \sm$ or to the left in case $\sp < \sm$.

\subsection{The constrained quartic}

A quartic curve can be constrained to give only one maximum by making the second derivative a perfect square:
\begin{equation}
    (\ln L)''(a)=-\half(\alpha + \beta a)^2 \qquad{\rm which \ yields} \qquad  \ln L(a)=-\half \left( {\alpha^2 a^2 \over 2 } + {\alpha \beta a^3 \over 3} + {\beta^2 a^4 \over 12} \right) .
\end{equation}
The parameters are given by
\begin{equation}
    \beta = {1 \over \sp \sm} \sqrt{ {12(\sm+\sp)^2 \pm 24 \sqrt{4 \sp \sm^3 + 4 \sm \sp^3 - 2 \sp^4 - 2 \sm^4} \over  3 \sm^2 + 2 \sp \sm + 3 \sp^2}}.
\end{equation}
The $\pm$ term should be taken as negative to give a small quartic term. In very asymmetric cases, where $\sp$ and $\sm$ differ by more than a factor of $(1 + \sqrt[4]{12} + \sqrt{3})/2 \approx 2.29663$, the inner root is negative, indicating that there is no solution in the desired form.

Having found $\beta$, $\alpha$ is found from the two $\Delta \ln L=-\half$ points.  These specify that both
\begin{equation}
    {\alpha^2 \sp^2 \over 2}+{\alpha \beta \sp^3 \over 3} + {\beta^2 \sp^4 \over 12}=1
    \qquad {\rm and }
    \qquad
    {\alpha^2 \sm^2 \over 2}-{\alpha \beta \sm^3 \over 3} + {\beta^2 \sm^4 \over 12}=1
\end{equation}
which have respective solutions
\begin{equation}
    \alpha = - {\beta \sp \over 3} \pm {\sqrt{72-2 \beta^2 \sp^4} \over 6 \sp}
    \qquad {\rm and }
    \qquad
    \alpha =  {\beta \sm \over 3} \pm {\sqrt{72-2 \beta^2 \sm^4} \over 6 \sm}.
    \label{eq:quarticalpha}
\end{equation}
The two pairs of results have one solution in common, which is the one to be taken. (The roots in Equation~\ref{eq:quarticalpha} do not go negative.) 

This gives better behaviour than the cubic for large $a$, but is not always so satisfactory in the central $[\hat a - \sm,\hat a +\sp]$ region.

\subsection{The molded quartic}

To address the discontinuity problem of the broken parabola given by Equation~\ref{eq:brokenParabola}, one can make a~quartic polynomial $q(a)$which looks as close as possible to the broken parabola $g(a)$ on the interval $[- \sm, \sp]$ but, of course, has continuous derivatives at $0$. The {\it molded quartic} $\ln L(a) \equiv q(a)$ is constructed by minimizing $\int_{-\sm}^{\sp} (q(a) - g(a))^2 da$ subject to the constraints $q(-\sm) = q(\sp) = -1/2$, $q(0) = 0$, $q'(0) = 0$, and $q''(0) < 0$. One can imagine that, in this construction, $g(a)$ serves as a~mold into which $q(a)$ is stamped.

The solution of this constrained minimization problem is $q(a)= -\frac{1}{2} (\alpha a^4 + \beta a^3 + \gamma a^2)$, where the parameters are given by 
{\footnotesize
\begin{align}
    \begin{split}
    \alpha & = 3 (\sm - \sp)^2 (5 \sm^6 + 8 \sm^5 \sp + 5 \sm^4 \sp^2 + 8 \sm^3 \sp^3 + 5 \sm^2 \sp^4 + 8 \sm \sp^5 + 5 \sp^6)/\eta \\
    \beta & = (\sm - \sp) [25 (\sm^8+\sp^8) + 14 (\sm^7 \sp - \sm^6 \sp^2 + \sm^5 \sp^3 - \sm^4 \sp^4 + \sm^3 \sp^5 - \sm^2 \sp^6 + \sm \sp^7)]/\eta \\
    \gamma & = (10 \sm^{10} - 5 \sm^9 \sp + 30 \sm^7 \sp^3 - 6 \sm^6 \sp^4 + 6 \sm^5 \sp^5 - 6 \sm^4 \sp^6 + 30 \sm^3 \sp^7 - 5 \sm \sp^9 + 10 \sp^{10})/\eta
    \end{split}.
\end{align}
}
Here, $\eta = 2 \sm^2 \sp^2 (\sm + \sp)^4 (5 \sm^4 - 10 \sm^3 \sp + 12 \sm^2 \sp^2 - 10 \sm \sp^3 + 5 \sp^4)$.

With these parameters, the constraint $q''(0) < 0$ (equivalently, $\gamma > 0$) is satisfied automatically for all $\sm > 0$ and $\sp > 0$. The requirement that $q(a)$ has a single extremum (the maximum at 0) leads to the condition that $\sp$ and $\sm$ must not differ by more than a factor of about 3.40804.

\subsection{The matched quintic}

The matched quintic log likelihood model assumes that, outside of the interval $[-\sm, \sp]$, the log likelihood behaves as a second order polynomial with the second derivative $-(1/\sp)^{2}$ for $a > \sp$ and $-(1/\sm)^{2}$ for $a < -\sm$. Inside the interval $[-\sm, \sp]$ the polynomial takes the form
\begin{equation}
    \ln L(a)=-\half (\alpha a^5 + \beta a^4 + \gamma a^3 + \delta a^2).
\end{equation}
The coefficients are determined from the condition $\Delta \ln L(a) = -\half$ at the interval boundaries and from the requirement that the second derivative remains continuous at $-\sm$ and $\sp$. This leads to
\begin{align}
    \begin{split}
    \alpha & = -10 (\sm - \sp)/\eta \\
    \beta & = -18 (\sm - \sp)^2/\eta \\
    \gamma & = 45 \sm \sp (\sm - \sp)/\eta \\
    \delta & = (8 \sm^4 + 19 \sm^3 \sp - 19 \sm^2 \sp^2 + 19 \sm \sp^3 + 8 \sp^4)/\eta
    \end{split}.
\end{align}
Here, $\eta = \sm^2 \sp^2 (8 \sm^2 + 19 \sm \sp + 8 \sp^2)$.

This log likelihood model behaves similarly to Equation~\ref{eq:brokenParabola} for large values of $|a|$ but avoids the discontinuity in the second derivative. The third derivative is still discontinuous at $a = -\sm$ and $a = \sp$.

The requirement that $\ln L(a)$ has a single extremum (the maximum at 0) leads to the condition that $\sp$ and $\sm$ must not differ by more than a factor of about 2.426419.

\subsection{The interpolated 7th degree polynomial}

Another way of avoiding the second derivative problem of Equation~\ref{eq:brokenParabola} consists in assuming that the log likelihood curve should take the form of Equation~\ref{eq:brokenParabola} for $a < -\sm$ and $a > \sp$, while on the interval $[-\sm, \sp]$ the curve is constructed by matching the values of the broken parabola $g(a)$ and its derivatives $g'(a)$, and $g''(a)$ at the interval boundaries: $a = -\sm$ and $a = \sp$. Together with the conditions $\ln L(0) = 0$ and $\left. \frac{d}{d a} \ln L(a) \right|_{a = 0} = 0$, we have eight equations that can be satisfied by a 
7\supers{th} degree polynomial. This polynomial takes the form
\begin{equation}
    \ln L(a)=-\half (\alpha a^7 + \beta a^6 + \gamma a^5 + \delta a^4 + \epsilon a^3 + \zeta a^2)
\end{equation}
with the coefficients given by
\begin{align}
    \begin{split}
    \alpha & = 6 (\sm - \sp)/\eta \\
    \beta & = 15 (\sm - \sp)^2/\eta \\
    \gamma & = 10 (\sm - \sp) (\sm^2 - 4 \sm \sp + \sp^2)/\eta \\
    \delta & = -30 \sm \sp (\sm - \sp)^2/\eta \\
    \epsilon & = 30 \sm^2 \sp^2 (\sm - \sp)/\eta \\
    \zeta & = (\sm^6 + 4 \sm^5 \sp + 6 \sm^4 \sp^2 - 6 \sm^3 \sp^3 + 6 \sm^2 \sp^4 + 4 \sm \sp^5 + \sp^6)/\eta
    \end{split}.
\end{align}
Here, $\eta = \sm^2 \sp^2 (\sm + \sp)^4$.

The requirement that $\ln L(a)$ has a single extremum (the maximum at 0) leads to the condition that $\sp$ and $\sm$ must not differ by more than a factor of about 2.744405.

In combination with Equation~\ref{eq:brokenParabola} for $a < -\sm$ and $a > \sp$, this model has a very reasonable behavior for large values of $|a|$ but can be somewhat ``wavy" on the interval $[-\sm, \sp]$. It also has a discontinuous third derivative at $a = -\sm$ and $a = \sp$.

\subsection{The double quartic}

This curve is constructed as follows:
\begin{equation}
    \ln L(a) = \begin{cases}
        P_{2}^-(a) & \text{for } a \le - \sm \\
        P_{4}^-(a) & \text{for } - \sm \le a \le 0 \\
        P_{4}^+(a) & \text{for } 0 \le a \le \sp \\
        P_{2}^+(a) & \text{for } \sp \le a
    \end{cases},
\end{equation}
where $P_{2}^-(a)$ and $P_{2}^+(a)$ are quadratic polynomials and $P_{4}^-(a)$ and $P_{4}^+(a)$ are quartics defined by the following conditions:
\begin{eqnarray}
P_{2}^-(-\sm) = P_{4}^-(-\sm) &=& P_{2}^+(\sp) = P_{4}^+(\sp) = -\frac{1}{2},\nonumber\\
\left.
\frac{d}{d a} P_{2}^-(a)\right|_{a = -\sm}&=&   \left.\frac{d}{da} P_{4}^-(a) \right|_{a = -\sm},\nonumber\\
 \left. \frac{d^2}{d a^2} P_{2}^-(a) \right|_{a = -\sm} &=&  \left. \frac{d^2}{d a^2} P_{4}^-(a) \right|_{a = -\sm} = -\frac{1}{\sm^2},\nonumber\\
 \left. \frac{d}{d a} P_{2}^+(a) \right|_{a = \sp} &=&  \left. \frac{d}{da} P_{4}^+(a) \right|_{a = \sp},\\
 \left. \frac{d^2}{d a^2} P_{2}^+(a) \right|_{a = \sp} &=&  \left. \frac{d^2}{d a^2} P_{4}^+(a) \right|_{a = \sp} = -\frac{1}{\sp^2},\nonumber\\
 P_{4}^-(0) = P_{4}^+(0) = 0\\
 \left. \frac{d}{da} P_{4}^-(0) \right|_{a = 0} &=&  \left. \frac{d}{da} P_{4}^+(0) \right|_{a = 0} = 0,\nonumber\\
 \left. \frac{d^2}{d a^2} P_{4}^-(0) \right|_{a = 0} &=&  \left. \frac{d^2}{d a^2} P_{4}^+(0) \right|_{a = 0} = -\frac{1}{\sigma_0^2}.\nonumber 
\end{eqnarray}
 $\sigma_0$ is a somewhat arbitrary parameter describing the second differential at the peak. When it is selected, these conditions define the polynomial coefficients uniquely.  The resulting curve is continuous together with its first two derivatives.
Two choices of $\sigma_0$ have been investigated.  

The {\it simple double quartic} is obtained by setting $\sigma_0 = \sqrt{\sm \sp}$.  For this choice of $\sigma_0$, the requirement that $\ln L(a)$ has a single extremum (the maximum at 0) leads to the condition that $\sp$ and $\sm$ must differ by less than a factor of $\frac{68}{11}$.

The {\it molded double quartic} minimizes the ``molding" functional $\frac{1}{\sm} \int_{-\sm}^{0} \left[\ln L(a) - g(a)\right]^2 da + \frac{1}{\sp} \int_{0}^{\sp} \left[\ln L(a) - g(a)\right]^2 da$, with $g(a)$ defined by Equation~\ref{eq:brokenParabola}.  This expression is minimized by setting $\sigma_0 = \sqrt{\frac{\sm^4 + \sp^4}{\sm^2 + \sp^2}}$.  This choice results in a curve exhibiting reasonable behavior even for
very large asymmetries.

\subsection{The double quintic}

This curve is constructed as follows:
\begin{equation}
    \ln L(a) = \begin{cases}
        P_{2}^-(a) & \text{for } a \le - \sm \\
        P_{5}^-(a) & \text{for } - \sm \le a \le 0 \\
        P_{5}^+(a) & \text{for } 0 \le a \le \sp \\
        P_{2}^+(a) & \text{for } \sp \le a
    \end{cases},
\end{equation}
where $P_{2}^-(a)$ and $P_{2}^+(a)$ are quadratic polynomials and $P_{5}^-(a)$ and $P_{5}^+(a)$ are quintics defined by the following conditions:

\begin{eqnarray}
P_{2}^-(-\sm) =  P_{5}^-(-\sm) &=& P_{2}^+(\sp) = P_{5}^+(\sp) = -\frac{1}{2},\nonumber\\
\left. \frac{d}{d a} P_{2}^-(a) \right|_{a = -\sm} &=& \left. \frac{d}{da} P_{5}^-(a) \right|_{a = -\sm} = \frac{1}{\sm}
,\nonumber\\
\left. \frac{d^2}{d a^2} P_{2}^-(a) \right|_{a = -\sm} &=&  \left. \frac{d^2}{d a^2} P_{5}^-(a) \right|_{a = -\sm} = -\frac{1}{\sm^2}
,\nonumber \\
\left. \frac{d}{d a} P_{2}^+(a) \right|_{a = \sp} &=&  \left. \frac{d}{da} P_{5}^+(a) \right|_{a = \sp} = -\frac{1}{\sp}
,\nonumber\\
\left. \frac{d^2}{d a^2} P_{2}^+(a) \right|_{a = \sp} &=&  \left. \frac{d^2}{d a^2} P_{5}^+(a) \right|_{a = \sp} = -\frac{1}{\sp^2}
,\\
P_{5}^-(0) = P_{5}^+(0) = 0,\nonumber\\ 
\left. \frac{d}{da} P_{5}^-(0) \right|_{a = 0} &=&  \left. \frac{d}{da} P_{5}^+(0) \right|_{a = 0} = 0
,\nonumber \\
\left. \frac{d^2}{d a^2} P_{5}^-(0) \right|_{a = 0} &=&  \left. \frac{d^2}{d a^2} P_{5}^+(0) \right|_{a = 0} = -\frac{1}{\sigma_0^2}.\nonumber
\end{eqnarray}
The resulting curve is continuous together with its first two derivatives.  It coincides with Equation~\ref{eq:brokenParabola} outside of the interval $[-\sm, \sp]$.

Again, when the value of the parameter $\sigma_0$ is selected, these conditions define the polynomial coefficients uniquely. Two choices of $\sigma_0$ have been investigated. 

The {\it simple double quintic} is obtained by setting $\sigma_0 = \sqrt{\sm \sp}$.  For this choice of $\sigma_0$, the requirement that $\ln L(a)$ has a single extremum (the maximum at 0) leads to the condition that $\sp$ and $\sm$ must differ by less than a factor of $13.5$.

The {\it molded double quintic} minimizes the ``molding" functional
$\frac{1}{\sm} \int_{-\sm}^{0} \left[\ln L(a) - g(a)\right]^2 da + \frac{1}{\sp} \int_{0}^{\sp} \left[\ln L(a) - g(a)\right]^2 da$,
with $g(a)$ defined by Equation~\ref{eq:brokenParabola}.  This expression is minimized by setting $\sigma_0 = \sqrt{\frac{\sm^4 + \sp^4}{\sm^2 + \sp^2}}$.  This choice results in a curve exhibiting reasonable behavior even for very large asymmetries.

\subsection{The conservative spline}

This curve is constructed as follows:
\begin{equation}
    \label{eq:conservspline}
    \ln L(a) = \begin{cases}
        P_{2}^-(a) & \text{for } a \le a\sub{left} \\
        \alpha a^3 + \beta a^2 & \text{for } a\sub{left} \le a \le a\sub{right} \\
        P_{2}^+(a) & \text{for } a \ge a\sub{right}
    \end{cases},
\end{equation}
where $P_{2}^-(a)$ and $P_{2}^+(a)$ are quadratic polynomials.  The values, derivatives, and second derivatives of these polynomials are matched to the values, derivatives, and second derivatives of  the central cubic at the points $a\sub{left}$ (with $a\sub{left}$ restricted to the interval $[-\sm, \ 0]$) and $a\sub{right}$ (with a\sub{right} in $[0, \ \sp]$), so that the whole curve is continuous together with its first two derivatives. The parameters $\alpha$ and $\beta$ are chosen so that the curve satisfies the condition $\ln L(-\sm) = \ln L(\sp) = -\frac{1}{2}$. Assuming, for the moment, that $\sp < \sm$, the parameters $a\sub{left}$ and $a\sub{right}$ are set in such a way that the magnitude of the second derivative of this curve does not exceed $\frac{\kappa}{\sp^2}$, and equals to $\frac{\kappa}{\sp^2}$ for $a \ge a\sub{right}$, and does not become less than $\frac{1}{\kappa \sm^2}$, and equals to $\frac{1}{\kappa \sm^2}$ for $a \le a\sub{left}$, where $\kappa \ge 1$ is some value specified by the user. Larger values of $\kappa$ result in a smoother second derivative (smaller $|\alpha|$ and larger interval $[a\sub{left}, \ a\sub{right}]$), while the choice $\kappa = 1$ results in $a\sub{left} = a\sub{right} = 0$ thus reducing Equation~\ref{eq:conservspline} to Equation~\ref{eq:brokenParabola}.

This log-likelihood construction mitigates the discontinuity of the second derivative afflicting Equation~\ref{eq:brokenParabola} while maintaining user-controlled lower and upper bounds on the Fisher information of the log-likelihood in the parameter space. If a linear log-likelihood curve is added to this one, the uncertainties of the result determined from the $\Delta \ln L(a) = -\frac{1}{2}$ equation will be bounded by the range $[\mbox{min}(\sm, \sp)/\sqrt{\kappa},\ \sqrt{\kappa} \,\mbox{max}(\sm, \sp)]$ no matter how large is the slope of the added curve.

\subsection{The log logistic-beta}

The density of the logistic-beta distribution (a.k.a. the type IV generalized logistic) is given by~\cite{ref:johnsonkotzv2}:
\begin{equation}
    f(x, \alpha, \beta) = \frac{\sigma^{\alpha}(x) \sigma^{\beta}(x)}{B(\alpha, \beta)},
\end{equation}
where $B(\alpha, \beta)$ is the beta function, $\sigma(x) = 1/(1 + e^{-x})$ is the logistic function, and $\alpha, \beta > 0$ are the shape parameters. After adding a scale parameter and shifting the distribution so that its mode is at 0, the logarithm of this density can be parameterized as
\begin{equation}
    \label{eq:loglogbeta}
    \ln L(a) = g c^2 \left((A-1) \ln \left[\frac{1}{2} \left\{A
        \left(e^{\frac{a}{c}}-1\right)+e^{\frac{a}{c}}+1\right\}\right]-(A+1) \ln
        \left[\frac{1}{2} e^{-\frac{a}{c}} \left\{A
        \left(e^{\frac{a}{c}}-1\right)+e^{\frac{a}{c}}+1\right\}\right]\right).
\end{equation}
Here, the logarithm of the logistic-beta density (with the $x \rightarrow a$ replacement) has been reparameterized in terms of the new parameters $g > 0$, $c > 0$, and $|A| < 1$, and a constant term has been added so that $\ln L(0) = 0$. The role of the new parameters can be appreciated as follows:
\begin{itemize}
    \item {
    $A$ is the asymmetry parameter. The limiting behavior of $\ln L(a)$ for $a \rightarrow -\infty$ is an asymptote with the positive slope $g c (A + 1)$. For $a \rightarrow +\infty$, the asymptote has the negative slope $g c (A - 1)$. In case $A = 0$, these slopes are the same in magnitude and $\ln L(a)$ becomes an even function.
    }
    \item {
    $g$ is the parameter regulating the magnitude of the $\frac{d^2 \ln L(a)}{d a^2}$ maximum. Indeed, it can be shown that $\ln L(a)$ is strictly concave, while $\left|\frac{d^2 \ln L(a)}{d a^2}\right|$ is maximized at $a = c  \ln\left(\frac{1 - A}{1 + A}\right)$. At that point, $\frac{d^2 \ln L(a)}{d a^2} = -\frac{g}{2}$.
    }
    \item {
    $c$ is the shape parameter regulating how quickly the asymptotic behavior is approached. For small values of $c$ the curve looks like two straight lines joined by a narrow transition section. For large values of $c$ the curve looks more parabolic.
    }
\end{itemize}
Equation~\ref{eq:loglogbeta} satisfies the conditions $\ln L(0) = 0$ and $\left. \frac{d \ln L(a)}{d a} \right|_{a = 0} = 0$ automatically. It also needs to satisfy the conditions $\ln L(-\sm) = \ln L(\sp) = -\frac{1}{2}$. While with three parameters $A$, $c$, and $g$ the latter conditions can be satisfied in a multitude of ways, we choose the parameterization that minimizes~$g$, i.e., $A$ and $c$ are chosen in such a way that $|\ln L(\sp)/g|$ is maximized subject to the constraint $\ln L(-\sm) = \ln L(\sp)$.  This results in a conservative log-likelihood model with limited slopes and the second derivative exceeding $\frac{1}{\min (\sm, \sp)^2}$ in magnitude only on a short interval.

\subsection{The logarithmic}

One can also use a logarithmic form
\begin{equation}
    \ln L(a)=-\half \left({\ln (1 + \gamma a) \over \ln \beta }\right) ^2
\end{equation}
with
\begin{equation}
    \beta ={\sp \over \sm} \qquad {\rm and } \qquad  \gamma={\sp-\sm \over \sp \sm}.
\end{equation}
This is easy to implement, and has some motivation as it describes a parabola which has been modified by the expansion/contraction of $a$ at a constant rate. Its unpleasant features are that it is undefined for $a$ beyond some point in the direction of the smaller error, as $1+\gamma a$ goes negative,
and it does not give a parabola in the $\sp=\sm$ limit.

\subsection{The generalised Poisson}

Starting from the Poisson likelihood, $\ln L(a)=-a + N \ln a -\ln N!$, which has a positive skewness, one can generalise to
\begin{equation}
    \ln L(a)=-\alpha(a+\beta) + \N \ln(a+\beta) + constant,
\end{equation}
where $\N$ is a continuous parameter which produces the skewness, and $\alpha$ and $\beta$ are scale and location constants. To get the maximum in the right place requires $\N=\alpha\beta$ and, adjusting the constant to make $\ln L(0)=0$, this becomes
\begin{equation}
    \ln L = -\alpha a + \N \ln \left(1+{\alpha a \over \N} \right).
\end{equation}
Writing $\gamma = \alpha/\N$ the equations at $\sm$ and $\sp$ lead to
\begin{equation}
    {1 - \gamma \sm \over 1 + \gamma \sp}=e^{-\gamma(\sp +\sm)}.
\end{equation}
This has to be solved numerically for $\gamma$. The solution lies between 0 and $1/\sm$, and can be found by repeatedly taking the midpoint\footnote{Attempts to use more sophisticated algorithms were unsuccessful.}. $\N$ is then found from
\begin{equation}
    \N={1 \over 2(\gamma \sp-\ln(1+\gamma \sp))}.
\end{equation}
This does fairly well, though the need for a numerical solution makes it inelegant.  If the curve to be fitted has a negative skewness ($\sm > \sp)$, then the sign of $a$ has to be flipped.

\subsection{The linear sigma}
\label{sec:linearsigma}

Bartlett suggests (and justifies) that the likelihood function is described by a Gaussian whose width varies as a function of the parameter \cite{bartlett1,bartlett2,deltalnL}
\begin{equation}
    \label{eq:dependentsigma}
    \ln L(a)=-\half \left( {a - \hat a \over \sigma(a)}\right)^2.
\end{equation}
This does not include the $-\ln \sigma$ term from the denominator of the Gaussian, however omitting this term  improves the accuracy of $\Delta \ln L=-\half$ errors\cite{deltalnL}.

If we suppose that the variation of $\sigma$  is linear, at any rate in the  region of interest
\begin{equation}
    \sigma(a)=\sigma+\sigma' (a-\hat a),
\end{equation}
\begin{equation}
    \ln L(a)=-\half \left( {a - \hat a \over \sigma+\sigma'(a-\hat a)}\right)^2
    \label{eq:linearsigma},
\end{equation}
then the requirement that this go through the two $-\half$ points gives
\begin{equation}
    \sigma={2 \sp \sm \over \sp+\sm} \qquad {\rm and } \qquad \sigma'={\sp-\sm \over \sp+\sm}.
\end{equation}
This form does well in most cases, and the parameters are easy to find.

\subsection{The linear variance}
\label{sec:linearvariance}

As an alternative to Equation~\eqref{eq:linearsigma}, we could equally plausibly assume that the
variance has a linear variation
\begin{equation}
    V(a)=V+V'(a-\hat a)
\end{equation}
giving
\begin{equation}
    \ln L(a)=-\half {(a-\hat a)^2 \over V + V'(a-\hat a)}
    \label{eq:linearV}.
\end{equation}
The parameters are again easy to find:
\begin{equation}
    V=\sp \sm \qquad V'=\sp-\sm,
\end{equation}
and this also does very well. Non-physical results can occur in principle when $V+V'(a-\hat a)$ goes negative,
but only for large deviation and large asymmetries and this does not present a problem in practice.

\subsection{The linear sigma in the log space}

Equation~\eqref{eq:dependentsigma} works for any positive, monotonic function $\sigma(a)$ as long as $\sigma(\hat a - \sm) = \sm$ and  $\sigma(\hat a + \sp) = \sp$.  As the values of $\sigma$ are supposed to be positive, it seems natural to interpolate these values in the log space. However, simple linear interpolation of $\ln \sigma$ as a function of  $a$ is not going to work: for large values of $|a|$ the exponent of a linear function will grow faster than $|a - \hat a|$, and the magnitude of $\ln L(a)$ will decrease, creating at least one additional unphysical extremum.

Instead, it can be useful to perform linear interpolation of $\ln \sigma$ as a function of some variable $u = Q(a)$, where $Q(a)$ is a monotonic transform that maps the range of $a$ from $[-\infty, \infty]$ to some compact interval $[u_{\text{min}}, u_{\text{max}}]$. Without loss of generality, this interval can be chosen to be $[0, 1]$. In this case,  $Q(a)$ becomes a cumulative density function of some distribution supported on $[-\infty, \infty]$.

Here, we choose $Q(a)$ to be the cdf of the Fechner distribution whose density is given by Equation~\ref{eq:fechnerpdf} in which we set $m = \hat{a}$, $\sigma_1 = \sm$, and $\sigma_2 = \sp$.  The equation for $\ln \sigma$ is then 
\begin{equation}
    \ln \sigma(u) = \alpha u + \beta,
\end{equation}
where the coefficients $\alpha$ and $\beta$ are chosen to satisfy the conditions
\begin{align}
    \begin{split}
        \ln \sigma(Q(\hat a - \sm)) & = \ln \sm\\
        \ln \sigma(Q(\hat a  + \sp)) & = \ln \sp. \\
    \end{split}
\end{align}
Then, of course, $\sigma(a) = e^{\alpha Q(a) + \beta}$.

The requirement that $\ln L(a)$ has a single extremum (the maximum at 0) leads to the condition that $\sp$ and $\sm$ must not differ by more than a factor of about 5.338453.

\subsection{The double cubic sigma in the log space}

Note that Equation~\eqref{eq:dependentsigma} results in a curve with a continuous second derivative if $\sigma(a)$ is continuous with its first and second derivatives everywhere except the point $a = \hat a$. At that point it is sufficient for $\sigma(a)$ just to be continuous, while the continuity of the derivatives is not required (as long as the first and the second derivative are finite on both sides of $\hat a$). This allows us to construct a useful log likelihood model as follows:
\begin{equation}
    \ln \sigma(a) = \begin{cases}
        \ln \sm & \text{for } a \le \hat a - \sm \\
        P_{3}^-(a) & \text{for } \hat a - \sm \le a \le \hat a \\
        P_{3}^+(a) & \text{for } \hat a \le a \le \hat a + \sp \\
        \ln \sp & \text{for } \hat a + \sp \le a
    \end{cases},
\end{equation}
where $P_{3}^-(a)$ and $P_{3}^+(a)$ are cubic polynomials satisfying the following conditions: 
\begin{eqnarray}
P_{3}^-(\hat a - \sm) &=&  \ln \sm
,\nonumber\\
\left. \frac{d}{d a} P_{3}^-(a) \right|_{a = \hat a - \sm} &=&  0
,\nonumber \\
\left. \frac{d^2}{d a^2} P_{3}^-(a) \right|_{a = \hat a - \sm} &=&  0
,\nonumber\\
P_{3}^-(\hat a) = P_{3}^+(\hat a) &=&  \ln \sigma_0
,\\
P_{3}^+(\hat a + \sp) &=&  \ln \sp
.\nonumber\\
\left. \frac{d}{d a} P_{3}^+(a) \right|_{a = \hat a + \sp} &=&  0
,\nonumber\\
\left. \frac{d^2}{d a^2} P_{3}^+(a) \right|_{a = \hat a + \sp} &=&  0.\nonumber
\end{eqnarray}
 When the value of the overall  parameter $\sigma_0$ is selected, these conditions define the polynomial coefficients uniquely.
We choose $\sigma_0$ by numerically minimizing the following ``molding" functional: 
$\frac{1}{\sm} 
\int_{\hat a - \sm}^{\hat a} \left\{\ln L(a) - g(a - \hat a)\right\}^2 da
+ \frac{1}{\sp} \int_{\hat a}^{\hat a + \sp} \left\{\ln L(a) - g(a - \hat a)\right\}^2 da$,
with $g(a)$ defined by Equation~\ref{eq:brokenParabola}.  The resulting curve has two continuous derivatives and exhibits reasonable behavior even for very asymmetric errors.

\subsection{The quintic sigma in the log space}

This model is defined by the following dependence of $\sigma$
on the parameter:
\begin{equation}
    \ln \sigma(a) = \begin{cases}
        \ln \sm & \text{for } a \le \hat a - \sm \\
        P_{5}(a) & \text{for } \hat a - \sm \le a \le \hat a + \sp \\
        \ln \sp & \text{for } \hat a + \sp \le a
    \end{cases}.
\end{equation}
The coefficients of the quintic polynomial $P_5(a)$ are chosen in such a way that the $\ln \sigma(a)$ function remains continuous together with its first and second derivatives at $a = \hat a - \sm$ and $a = \hat a + \sp$.

The requirement that $\ln L(a)$ has a single extremum (the maximum at 0) leads to the condition that $\sp$ and $\sm$ must not differ by more than a factor of about 4.107184572.

\subsection{The PDG method}

In combining results with asymmetric errors the Particle Data Group~\cite{PDG} uses $\sp$ for $a > \hat a + \sp$, $\sm$ for $a < \hat a - \sm$, and a linearly varying sigma, as given by Equation~\eqref{eq:linearsigma}, 
for the intermediate region. Note that the PDG log likelihood has discontinuous derivatives at $a = \hat a - \sm$ and $a = \hat a + \sp$.

\subsection{The Edgeworth expansion}

The Edgeworth expansion of Equation~\eqref{eq:edgeworth0} gives
\begin{equation}
    \ln L(a)=-\half\left({a - a_0 \over \sigma}\right)^2 + \ln \left(1+\alpha He_3\left({a-a_0 \over \sigma}\right)\right)
\end{equation}
with $He_3(z)=z^3-3z$.  

Note that the peak of the distribution is not at $a_0$, and the peak value is not 0.  The parameter $\alpha$ can be found numerically from the asymmetry $(\sp-\sm)/(\sp+\sm)$ as that is independent of scale and location.  Once this is found, the values of $\sigma$ and $a_0$ are just found by scaling and shifting.

\subsection{The skew normal}

The skew normal pdf, Equation~\eqref{eq:azzalini}, takes the likelihood form
\begin{equation}
\ln L(a) = -\half\left({a-\xi \over \omega}\right)^2+\ln{\Phi\left(\alpha \left[{a-\xi)\over \omega}\right]\right)}.
\end{equation}
Again the location parameter, here $\xi$, is not the peak position, and the log likelihood at the peak is not zero but must be determined.  Given parameters $\xi,\omega, \alpha$ the log likelihood can be mapped out numerically to find the peak and the $\Delta \ln L=-\half$ errors.  For the reverse procedure $\alpha$ is first found from the asymmetry, $\sp-\sm \over \sp+\sm$, which is independent of $\omega$ and $\xi$, then the scale $\omega$ from $(\sp+\sm)$, and finally the location $\xi$ from the peak position.







\section{Implementation in C++/Python}
\label{sec:Cpp}

Open source C++/Python code for modeling asymmetric errors is available on GitHub.  The code is split into two packages, one implemented in C++ and the other mostly in Python. The reason for dual language implementation is that C++ is better suited for number crunching while Python environment allows for rapid development of small programs and provides convenient plotting facilities.

The first package, ``ase", contains a C++ library implementing probability distributions described in Appendix~\ref{app:modelling-pdf-errors} and log-likelihood curves described in Appendix~\ref{app:loglikelihoods}, as well as a number of requisite numerically intensive algorithms (minimization, root finding, special functions, quadratures, convolutions, etc).  The URL for this package is \textcolor{blue}{\href{https://github.com/igvgit/AsymmetricErrors}{https://github.com/igvgit/AsymmetricErrors}}

The second package, ``asepy", contains a Python API generated by SWIG~\cite{ref:SWIG} for the C++ classes and functions from the first package.  It also includes a number of additional high-level utilities and programs for combining errors and results, as well as for plotting distributions and log-likelihoods. This package comes with its own documentation and a sizeable collection of examples.  The URL for this package is \textcolor{blue}{\href{https://github.com/igvgit/AsymmetricErrorsPy}{https://github.com/igvgit/AsymmetricErrorsPy}}


\section{Implementation in R}
\label{sec:R}

The implementation in R is available as a package, which can be installed, from within an R session, by
%
{\tt \small install.packages("https://barlow.web.cern.ch/programs/AsymmetricErrors.tar.gz")}. Alternatively, if preferred, the file can be downloaded and installed with {\tt R CMD install {\sl downloadfilename}}.  (The case required for {\tt install} appears to vary between systems.)  Once installed, when required it can be loaded and attached whenever needed by {\tt library(AsymmetricErrors)}. 

The package comprises a set of functions that use S3 classes to distinguish between different models.  Not all models have been implemented (yet). A list of likelihood models can be found by {\tt methods(lnL)}.  

There are 4 functions to handle likelihoods:

\begin{itemize}
    \item []
    \leftline{\tt getlnLpars({\sl spec}, {\sl type})} 
    where {\tt \sl spec} is a named vector containing {\tt val}, the position of the peak, and the positive and negative errors {\tt sp} and {\tt sm}.  {\tt \sl type } is the model to be used. It returns a named vector with the appropriate model parameters added to {\tt \sl spec}, with class {\sl \tt \sl  type}.  If there is no set of parameters that can be found for a model to match the required specification then a {\tt warning} is raised and a {\tt NULL} is returned. The {\tt warning} can be suppressed (if appropriate) by a standard R {\tt suppressWarning} call, and robust code will check for any {\tt NULL} returns.

    \item[] \leftline{\tt lnL(p, a)}
    evaluates the log likelihood at {\tt a} (which can be a vector) according to the model and parameters {\tt p}. It returns a vector of the same length as {\tt a}.
    \item [] \leftline{\tt combinelnLresults(r)} takes a list of parameter vectors (which must all be of the same model) and gives the combined result as a parameter vector.
    \item[] \leftline{\tt combinelnLerrors(r)} takes a list of parameter vectors, which must all be of the same model, and gives the combined result as a parameter vector. (The central value of the result, which is set to zero, is meaningless.)
\end{itemize}

These use named vectors as parameters and return values.  The indexing of parameters within these vectors is not standardised: access should always be by name. 

Pdfs are handled similarly.  A list of models can be found by {\tt methods(Pdf)}.  (Notice the uppercase P, which avoids confusion with existing R graphics-related functions.)  They are a little more complicated (but this is hidden from the user) as the specification in {\tt getPdfpars} can be given either as the moments, {\tt mu, V, gamma} or by the quantiles {\tt M, sp, sm} (with {\tt M} the median and {\tt M-sp} and {\tt M+sp} the 68\% central interval).  The function will take whichever set it is given and evaluate the model parameters.  It will then use those to calculate the other set and include these in the returned vector.  This may be inefficient as they will not always be needed, but it makes the handling easier: any parameter vector produced by {\tt getPdfpars} will contain both the moments and the quantile parameters, as well as the specific model parameters.

\begin{itemize}
    \item[]\leftline{\tt getPdfpars({\sl spec}, {\sl type})} 
    where {\tt \sl spec} is a named vector containing either the moments  {\tt mu, V, gamma}, or the quantile parameters {\tt M, sp, sm}.  {\tt \sl type } is the model to be used.  It returns a named vector with the appropriate model parameters and the alternative specifiers added to {\tt \sl spec}, with class {\sl  type}.  Again, if there is no set of parameters for the required specification then a {\tt warning} is raised and {\tt NULL} is returned.
    \item[]\leftline{\tt Pdf(p, a) } 
    evaluates the pdf at {\tt a} according to the model and parameters {\tt p}.
    \item[]\leftline{\tt combinePdferrors(r)} 
    takes a list of parameter values (which must all be of the same model), adds their moments to get the total $\mu,V, \gamma$ and returns the parameter set that give this combined result.
    \item[]\leftline{\tt combinePdfresults(r)} 
    takes a list of parameter values (which must all be of the same model) and gives the combined result.
    \item[]\leftline{\tt getflipPdfpars({\sl spec, type, direction)}} 
    returns the model parameters for a `flipped' OPAT result.  {\tt direction} is $+$1 (the default) when both deviations are positive and $-$1 when both are negative (the given values of both {\tt sp} and {\tt sm} are always positive).  Only implemented for the dimidiated model.
\end{itemize}

All these functions have an associated {\tt help} giving further details.

Some classes (such as  {\tt edgeworth} and {\tt azzalini})  are used both  for the likelihoods and for the pdfs, but the two are distinct, with no connection. The function {\tt Title(p)}, which returns the name of the model of the parameters in printable form, with spaces and upper-case letters, may be useful.

\paragraph{\bf Examples}
~ \\[2ex]
We give some simple illustrative examples. They are very basic, just to illustrate simple uses,  and do not contain the checks and elaboration required for `good practice'.  They assume that the library has been installed on your system and 
loaded and attached in this R session by {\tt library(AsymmetricErrors)}.

\begin{itemize}
    \item Likelihood errors
    \begin{enumerate}
        \item To combine two results for the Higgs width, using the linear variance model:
        \begin{verbatim}
    ATLAS <- getlnLpars(c(val=4.5,sp=3.3,sm=2.5),"linearvariance")
    CMS <- getlnLpars(c(val=3.2,sp=2.4,sm=1.7),"linearvariance")
    p <- combinelnLresults(list(ATLAS,CMS))
    print(p)   
        \end{verbatim}
    \item To plot the lnL curves for the result $\AE 5.3 1.2 0.9 $ using all available models:
        \begin{verbatim}
    result <- c(val=5.2,sp=1.2,sm=0.9)
    a <- seq(0,10,.01)
    plot(0,0,type='n',xlim=range(a),ylim=c(-5,0))
    for (t in methods(lnL)) {
      p <- getlnLpars(result,substring(t,5))
      lines(a,lnL(p,a))
    }
        \end{verbatim}

    The substring call is to remove the leading {\tt lnL.} from the model names.  The 
    program should be elaborated to give titles, colours, and a legend, as desired.
    \item The near end of an object is $\AE 100 3 4 $ km away, and the far end is $\AE 200 4 5 $ km away. To calculate the error on the length of the object:
        \begin{verbatim}
    near <- getlnLpars(c(val=100,sp=3,sm=4),"linearsigma")
    far <- getlnLpars(c(val=200,sp=4,sm=5),"linearsigma")
    print(combinelnLerrors(list(near,far)))
        \end{verbatim}
    \item The combination of 3 results shown in Figure~\ref{fig:comb1} can be obtained by:
        \begin{verbatim}
    t <- "linearvariance"
    v1 <- getlnLpars(c(sp=0.7,sm=0.5,val=1.9),t)
    v2 <- getlnLpars(c(sp=0.6,sm=0.8,val=2.4),t)
    v3 <- getlnLpars(c(sp=0.5,sm=0.4,val=3.1),t)
    a <- seq(1,4,.01)
    plot(a,lnL(v1,a),type='l',ylab="lnL")
    lines(range(a),-.5*c(1,1))
    lines(a,lnL(v2,a))
    lines(a,lnL(v3,a))
    res <- combinelnLresults(list(v1,v2,v3))
    lines(a,lnL(res,a),col='red')
    print(res)
        \end{verbatim}
    \end{enumerate}
    \item Pdf Errors
    \begin{enumerate}
        \item The left hand plot curves of Figure~\ref{fig:simplepdfs} were produced by the code:
        \begin{verbatim}
    x <- seq(0,10,.01)
    p1 <- getPdfpars(c(M=5,sp=1.1,sm=0.9),"dimidiated")
    p2 <- getPdfpars(c(M=5,sp=1.1,sm=0.9),"distorted")
    plot(x,Pdf(p1,x),type='l',lwd=2,col='red',ylab="P(x)")
    lines(x,Pdf(p2,x),col='green',lwd=2)
    legend("topright",col=c("red","green"),lwd=2, 
           legend=c("Dimidiated","Distorted")) 
        \end{verbatim}
        and the right hand plot is the same, with {\tt sp=0.85, sm=1.15} in place of {\tt sp=1.1, sm=0.9}.
        \item If a table of OPAT-style systematic errors has been saved on a file {\tt errors.txt} as bare numbers, one set per line, for $\sp$ and $\sm$, then to find the total, using the dimidiated model:
        \begin{verbatim}
    df <- read.table("errors.txt")
    N <- dim(df)[1]
    errorlist <- list()
    for(i in 1:N) errorlist[[i]] <- 
      getPdfpars(c(M=0,sp=df[i,1],sm=df[i,2]),"dimidiated")
      print(combinePdferrors(errorlist))
        \end{verbatim}
        \item To find the first three moments for a pdf specified as $\AE 5.0 1.1 0.9 $ using all available models:
        \begin{verbatim}
    models <- methods(Pdf)
    for(m in models){
    m <- substring(m,5)
    p <- getPdfpars(c(M=5,sp=1.1,sm=0.9),m)
    print(paste("model",m," mean ",p['mu'],
       " variance ",p['V']," skewness ",p['gamma']))
    }
        \end{verbatim}
         (The {\tt substring} call is to remove the {\tt Pdf.} at the start of each name.)
        \item To combine two measurements $\AE 12.34 0.56 0.78 $  and $\AE 12.43 0.65 0.87 $  using the railway Gaussian:
        \begin{verbatim}
    p1=getPdfpars(c(M=12.34,sp=0.56,sm=0.78),"railway")
    p2=getPdfpars(c(M=12.43,sp=0.65,sm=0.87),"railway")
    print(combinePdfresults(list(p1,p2)))
        \end{verbatim}
    \end{enumerate}
    \item Likelihood and Pdf Errors
    \begin{enumerate}
        \item The combined results of Example~\ref{final.result} were produced by the code:
        {\scriptsize 
        \begin{verbatim}
poissonerrors <- function(x){
    L0 <- x*log(x)-x
    g <- function(y) { return (-y+x*log(y)- L0 +.5)}
    hi <- uniroot(g,c(x,x+2*sqrt(x)),tol=1.e-8)
    lo <- uniroot(g,c(x,max(0.0001,x-2*sqrt(x))),tol=1.e-8)
    return(c(sp=hi$root-x,sm=x-lo$root))
    }
phi <- function(x){ # seek Delta ln L = -1/2 upwards
   xx <- x[1]
   L0 <- x[2]
   g <- function(y){  return(-y+xx*log(y)-L0+.5)}
   return(uniroot(g,c(xx,xx+2*sqrt(xx)),tol=1.e-8)$root-xx)
   }
plo <- function(x){ # seek Delta ln L = -1/2 downwards
   xx <- x[1]
   L0 <- x[2]
   g <- function(y){  return(-y+xx*log(y)-L0+.5)}
   return(xx-uniroot(g,c(xx,max(0.0001,xx-2*sqrt(xx))),tol=1.e-8)$root)
   }
model1 <- "distorted"
model2 <- "linearvariance"
t <- 1 # recording time of 1 second
s <- as.numeric(readline("Error on time in sec - suggest 0.1 "))
sp <- 1/(t-s)-1/t
sm <- 1/t-1/(t+s)
rt <- 1/t
print(paste("result for time factor ",rt," + ",sp, " - ",sm))

A <- 100 # acceptance factor

n <- as.numeric(readline("Expected number of events - suggest 50 "))
pp <- poissonerrors(n)
print(paste(" Number with poisson errors is ",n,"+",pp['sp'],"-",pp['sm']))
r <- n*A*rt # the result
sp1 <- as.numeric(pp['sp']*A*rt)
sm1 <- as.numeric(pp['sm']*A*rt)
sp2 <- n*A*sp
sm2 <- n*A*sm
print(paste("Quoted result ", r, "+",sp1,"-",sm1," (stat) +",sp2,"-",sm2," (syst)"))

p1p <- getPdfpars(c(M=r,sp=sp1,sm=sm1),model1)
p2p <- getPdfpars(c(M=r,sp=sp2,sm=sm2),model1)
p3p <- combinePdferrors(list(p1p,p2p))
print("Combine (wrongly) as two pdfs"); print(p3p)
p1L <- getlnLpars(c(val=r,sp=sp1,sm=sm1),model2)
p2L <- getlnLpars(c(val=r,sp=sp2,sm=sm2),model2)
p3L <- combinelnLerrors(list(p1L,p2L))
print("combine(wrongly) as two likelihoods"); print(p3L)
        \end{verbatim}
        }
     \end{enumerate}
\end{itemize}


\bibliographystyle{unsrt}


\begin{thebibliography}{10}

\bibitem{ATLAShiggs}
S~Manzoni.
\newblock Higgs boson mass and width measurement with the {ATLAS} detector.
\newblock In {\em Proc. EPS-HEP 2023, Venice}, 2023.

\bibitem{CMShiggs}
F~Errico.
\newblock Higgs boson properties (mass/width) at {CMS}.
\newblock In {\em Proc. EPS-HEP 2023, Venice}, 2023.

\bibitem{NOvA}
M~Frank.
\newblock The latest three-flavor neutrino oscillation results from {NOvA}.
\newblock In {\em Proc. EPS-HEP 2023, Venice}, 2023.

\bibitem{Belle}
M~Reif.
\newblock Recent {B}elle {II} results on hadronic {B} decays.
\newblock In {\em Proc. EPS-HEP 2023, Venice}, 2023.

\bibitem{LHCb}
V~Chobanova.
\newblock Time-dependent cp violation measurements in b decays.
\newblock In {\em Proc. EPS-HEP 2023, Venice}, 2023.

\bibitem{Schmelling}
M~Schmelling.
\newblock Averaging measurements with hidden correlations and asymmetric
  errors.
\newblock arXiv:hep-ex/0006004v1, 2000.

\bibitem{Systematic}
R~J Barlow.
\newblock Asymmetric systematic errors.
\newblock arXiv:physics/0306138v1, 2003.

\bibitem{Statistical}
R~J Barlow.
\newblock Asymmetric statistical errors.
\newblock arXiv:physics/0406120v1, 2004.

\bibitem{dagostini}
G~d'Agostini.
\newblock Asymmetric uncertainties: Sources, treatment and potential dangers.
\newblock arXiv:physis/0403086.v2, 2004.

\bibitem{Possolo}
C~Merkat A~Possolo and O~Biodnar.
\newblock Asymmetrical uncertainties.
\newblock {\em Metrologia}, 56:045009, 2019.

\bibitem{Demortier}
L Demortier and L Lyons.
\newblock Everything you always wanted to know about pulls.
\newblock {\em CDF Note}, CDF/ANAL/PUBLIC/5776  (v3), 2008.


\bibitem{PDGhandbook}
S Navas et al. (Particle Data Group)
\newblock Review of Particle Properties.
\newblock Phys. Rev. D 110, 030001 (2024)



\bibitem{deltalnL}
R~J Barlow.
\newblock A note on ${\Delta} \ln l=-{1 \over 2}$ errors.
\newblock arXiv/physics/0403046, 2004.


\bibitem{Heinrich}
J~G Heinrich.
\newblock Coverage of Error Bars for Poisson Data
\newblock  CDF report CDF-6438 2003.

\bibitem{Garwood}
F~Garwood. 
\newblock Fiducial limits for the Poisson distribution.
\newblock {\em Biometrika} 28 437–442, 1936.

\bibitem{bartlett1}
M~S Bartlett.
\newblock On the statistical estimation of mean lifetimes.
\newblock {\em Phil. Mag}, 44:244, 1953.

\bibitem{bartlett2}
M~S Bartlett.
\newblock Estimation of mean lifetimes from multiple plate cloud chamber
  tracks.
\newblock {\em Phil. Mag}, 44:1407, 1953.

\bibitem{LHCbLambdab}
The LHCb Collaboration (R.~Aaj et~al).
\newblock Amplitude analysis of the decay {$\Lambda^0_b \to pK^-\gamma$}.
\newblock arXiv:2403.03710, 2024.

\bibitem{Edgeworth}
P~Hall.
\newblock {\em The Bootstrap and Edgeworth Expansion}.
\newblock Springer, 1992.

\bibitem{hall}
P~Hall.
\newblock Theoretical comparison of bootstrap confidence intervals.
\newblock {\em Ann. Stat.}, 16:927, 1988.

\bibitem{ref:keelin2011}
T.W. Keelin and B.W. Powley.
\newblock Quantile-parameterized distributions.
\newblock {\em Decision Analysis}, 8:206, 2011.

\bibitem{azzalini}
A~Azzalini.
\newblock A class of distributions which includes the normal ones.
\newblock {\em Scandinavian Journal of Statistics}, 12:181, 1985.

\bibitem{ref:johnson}
N.L. Johnson.
\newblock Systems of frequency curves generated by methods of translation.
\newblock {\em Biometrika}, 36:149, 1949.

\bibitem{ref:johnsonbook}
W.P. Elderton and N.L. Johnson.
\newblock {\em Systems of Frequency Curves}.
\newblock Cambridge University Press, 1969.

\bibitem{ref:hahnshap}
G.J. Hahn and S.S. Shapiro.
\newblock {\em Statistical Models in Engineering}.
\newblock Wiley, 1994.

\bibitem{ref:AS99}
R.~Hill I.D.~Hill and R.~L. Holder.
\newblock Algorithm {AS} 99: fitting {J}ohnson curves by moments.
\newblock {\em Applied Statistics}, 25:180, 1976.

\bibitem{ref:johnsonkotzv2}
S.~Kotz N.L.~Johnson and N.~Balakrishnan.
\newblock {\em Continuous Univariate Distributions, Vol. 2}.
\newblock Wiley, 1995.

\bibitem{PDG}
Wei-Ming Jao.
\newblock Private communication, 2004.

\bibitem{ref:SWIG}
{Simplified Wrapper and Interface Generator}.
\newblock \url{https://www.swig.org}.


\end{thebibliography}

\end{document}